\documentclass[11pt,tightenlines,eqsecnum,aps,amsmath,amssymb,nofootinbib,prd]{revtex4}
\usepackage{graphicx, wrapfig}
\usepackage{amssymb}
\usepackage{bm}
\usepackage{color}
\usepackage{mathrsfs}
\usepackage{subfigure}
\usepackage[normalem]{ulem}

\def\v{\vec}

\def\f{\frac}

\def\d{\textrm{d}}

\def\t{\tilde}

\def\t{\tilde}
\def\h{\hat}

\def\pp{p_{\phi}}
\def\dpp{\delta p_{\phi}}
\def\dph{\delta\phi}

\def\hz{\mathring{h}}
\def\gz{\mathring{g}}

\def\pz{\mathring{\pi}}

\def\l{\left}
\def\r{\right}

\def\A1IJ{{\cal A}_1^{ij}}
\def\A2IJ{{\cal A}_2^{ij}}
\def\A3IJ{{\cal A}_3^{ij}}
\def\A4IJ{{\cal A}_4^{ij}}
\def\A5IJ{{\cal A}_5^{ij}}
\def\A6IJ{{\cal A}_6^{ij}}
\def\A1ij{{\cal A}^1_{ij}}
\def\A2ij{{\cal A}^2_{ij}}
\def\A3ij{{\cal A}^3_{ij}}
\def\A4ij{{\cal A}^4_{ij}}
\def\A5ij{{\cal A}^5_{ij}}
\def\A6ij{{\cal A}^6_{ij}}

\usepackage{enumerate}

\newcommand{\be}{\nopagebreak[3]\begin{equation}}
\newcommand{\ee}{\end{equation}}
\newcommand{\bfig}{\nopagebreak[3]\begin{figure}}
\newcommand{\efig}{\end{figure}}
\newcommand{\bea}{\nopagebreak[3]\begin{eqnarray}}
\newcommand{\ea}{\end{eqnarray}}

\newcommand{\bmult}{\nopagebreak[3]\begin{multline}}
\newcommand{\emult}{\end{multline}}

\usepackage{colordvi}
\usepackage{xcolor}

\begin{document}
\title{Hamiltonian theory of classical and quantum gauge invariant perturbations in Bianchi I spacetimes}

\author{Ivan Agullo${}^{(1)}$}
\email{agullo@lsu.edu}
\author{Javier Olmedo${}^{(1)}$}
\email{jolmedo1@lsu.edu}
\author{V.~Sreenath${}^{(2)}$}
\email{sreenath@nitk.edu.in}
\affiliation{
${}^{(1)}$Department of Physics and Astronomy, Louisiana State University, Baton Rouge, Louisiana  70803, USA
}
\affiliation{${}^{(2)}$Department of Physics, National Institute of Technology Karnataka, Surathkal, Mangalore 575025, India}

\begin{abstract}
We derive a Hamiltonian formulation of the theory of gauge invariant, linear perturbations in anisotropic Bianchi I spacetimes, and describe how to quantize this system. The matter content is assumed to be a minimally coupled scalar field with potential $V(\phi)$. We show that a Bianchi I spacetime generically induces both anisotropies and quantum entanglement on cosmological perturbations,  and provide the  tools to compute the details of these features. We then apply this formalism to a scenario in which the inflationary era is preceded by an anisotropic Bianchi I phase,  and discuss the  potential imprints in observable quantities. The formalism developed here  paves the road to a simultaneous canonical quantization of both the homogeneous degrees of freedom and the perturbations, a task that we develop in a companion paper. 

\end{abstract}

\maketitle
%%%%%%%%%%%%%%%%%%%%%%%%%%%%%%%%%%%%%%%%%%%%%%%%%%%%%%%%%%%%%%%%%%%%%%%%%%%%%%%
\section{Introduction}
\label{sec:introduction}
%%%%%%%%%%%%%%%%%%%%%%%%%%%%%%%%%%%%%%%%%%%%%%%%%%%%%%%%%%%%%%%%%%%%%%%%%%%%%%%

One of the attractive features of the cosmic inflationary scenario  is that it helps to explain why our Universe looks so simple at large scales. This is the case, in particular, if one pays attention to  anisotropies. According to the Belinskii-Khalatnikov-Lifshitz conjecture \cite{bkl}, the anisotropies are  expected to dominate the expansion close to the big bang, and  could have left some traces in the present Universe. But in the absence of anisotropic sources, the contribution of shears to Einstein's equations fall off with the expansion  significantly faster than the contributions from radiation, matter, or a cosmological constant. Consequently, an inflationary phase of exponential expansion is very efficient in washing anisotropies away (see \cite{wald,turner,moss,ppu-BI2,pu-BI,gs,guptsingh} and references therein). This fact  simplifies enormously the analysis of the generation of the primordial perturbations during inflation, since one can safely neglect anisotropic aspects of the spacetime and work in the much simpler Friedmann-Lema\^itre-Robertson-Walker (FLRW) scenario. However, the analysis of perturbations requires one to specify  the quantum state describing them at the onset of inflation, and it is  common to choose this state to be isotropic too (e.g. the  Bunch-Davies vacuum). This is a stronger assumption. Contrary to the anisotropies in the spacetime geometry,  anisotropic features in perturbations do not dilute with the expansion \cite{agulloparker}.  The best  inflation can do to wash  anisotropies in perturbations away is to red-shift them out of the observable patch of the Universe. But red-shift is different from dilution; red-shift is inversely proportional to the scale factor, while dilution scales with its inverse cube. Therefore, red-shift is  efficient only if inflation lasts significantly longer than the minimum amount required. These arguments, together with the detection of anomalous anisotropic features in the large-angle temperature correlation functions in the cosmic microwave background (CMB) by WMAP \cite{wmap} and PLANCK \cite{Planck2015Isotropy}, have boosted the motivation to study  primordial anisotropies.

The best studied anisotropic spacetimes are the ones with Bianchi I-type geometries, the simplest family of spacetimes containing anisotropies. They are spatially flat and  reduce to the flat FLRW universe in the shear-free limit. A special subfamily of  Bianchi I spacetimes characterized by containing an extra spatial rotational symmetry was analyzed in \cite{bb-bi-lrs,fls-bi-lrs,gcp-lrs,paban1,paban2,paban3}, where predictions for the inflationary power spectrum and non-Gaussianity were made. Another type of anisotropic  models, the so-called shear-free spacetimes, have been  studied in \cite{s-free1,s-free2}. For the more general Bianchi I  geometries sourced by a scalar field, a complete and detailed analysis of the classical theory of gauge invariant perturbation was provided in \cite{ppu-BI1}. The power spectrum for scalar and tensor perturbations was also analyzed in \cite{ppu-BI2}, although in a less rigorous manner. These works correctly pointed out that the main observational features of an anisotropic phase are expected  for large angular scales in the CMB in the form of anisotropic power spectra and cross-correlations between scalar and tensor perturbations (see  Ref. \cite{pu-BI} for a  recent summary.)

The goal of this paper is, on the one hand, to introduce a Hamiltonian or phase space analysis of classical and quantum gauge invariant perturbations  in a  Bianchi I spacetime 
(for a Hamiltonian analysis in FLRW, see e.g: \cite{Langlois:1994ec, Nandi:2015ogk, Nandi:2016pfr,abs}). 
At the classical level, our final result is equivalent to the outcome of \cite{ppu-BI1,ppu-BI2}, and in this respect our analysis provides a complementary viewpoint from a purely canonical perspective. More precisely, rather than starting from Einstein equations, expanding them in perturbations, and identifying what combinations of perturbations remain invariant under changes of coordinates that are linear in the perturbations  \cite{ppu-BI1,ppu-BI2}, we start from the linearized phase space of general relativity around Bianchi I geometries, and use canonical methods to isolate the gauge invariant degrees of freedom at leading order in perturbations. This procedure elegantly reduces  the problem of finding gauge invariant fields and their equations of motion to solving a Hamilton-Jacobi equation for the generating function of a canonical transformation. Our approach  provides a  geometric and tractable approach to deal with the complexities of cosmological perturbations in the  presence of anisotropies and, in particular, makes it possible  to implement the mathematical framework in a user-friendly computational algorithm written in {\tt Mathematica}, that we have made publicly available in \cite{ntbk}.

On the other hand, the quantum theory of cosmological perturbations presented in this paper differs  from previous treatments. The quantization of the gauge invariant perturbations in Bianchi I spacetimes offers extra challenges compared to the FLRW counterpart, arising from the fact that scalar and tensor perturbations are {\em coupled} in the presence of anisotropies (see \cite{peloso,kioto} for previous analyses). Interacting field theories are known to be significantly less tractable than free ones, and perturbative techniques are often required to derive physical predictions. In this paper, we provide a complete and exact (i.e.\ nonperturbative in anisotropies) formulation of the quantum field theory of gauge invariant fields. The key observation is that, although these fields are coupled, at leading order in perturbations the theory is still {\em linear}. It is therefore possible to use rigorous quantization techniques for linear fields in curved spacetimes \cite{waldbook}. We follow a canonical (or Hamiltonian) viewpoint and quantize the theory starting from the classical phase space. This strategy has several advantages, particularly in the formulation of the Schr\"odinger picture, which contains  important subtleties in curved spacetimes \cite{aa}. This picture is particularly illuminating to show how anisotropies in the spacetime geometry induce   quantum entanglement between scalar and tensor perturbations.  

Another fact that motivates our analysis is the extension of the theory presented here to scenarios of quantum cosmology, where the Bianchi I geometry itself is also quantized, together with the perturbations. Many of the approaches to quantum cosmology are formulated in a Hamiltonian language, and therefore  one  needs the canonical description of perturbations introduced in this paper to simultaneously quantize  the Bianchi I  background together with  the gauge invariant  perturbations. We illustrate this point in  detail in a companion paper \cite{aos2}, where we study this problem in a scenario where the big bang singularity is replaced by a cosmic bounce, which connects two classical branches of the Universe, one contracting and one expanding. The Universe  isotropizes in the past and future, but it is  anisotropic around the time of the bounce. One can then analyze the evolution of gauge invariant  perturbations that start in an adiabatic vacuum state in the remote past, propagate  across the anisotropic bounce, and  continue the evolution until the inflationary phase of the Universe. This is a neat example that shows the way cosmic perturbations retain memory of the anisotropic phase of the Universe and leave an imprint in the CMB, even though  anisotropies in the background spacetime are  relevant only during a short period of time around the cosmic bounce \cite{aos2}. 

This paper is organized as follows. In Sec. \ref{sec.2} we formulate the canonical theory of Bianchi I geometries. Section \ref{sec.3} describes the classical theory of linear perturbations thereon, and  the way to isolate  the gauge invariant degrees of freedom of these perturbations. Section \ref{sec.4} is devoted to the formulation of the quantum kinematics, i.e.\ the construction of the Hilbert space and a representation of field and momentum operators on it. Dynamics on this Hilbert space is introduced in Sec. \ref{sec.5}, both in the Heisenberg and Schr\"odinger  pictures. These two viewpoints illuminate  complementary aspects of the time evolution, particularly regarding quantum entanglement between scalar and tensor perturbations. Section \ref{sec.6} illustrates our theoretical construction with a concrete example of a Bianchi I phase of the Universe followed by a period of inflation. Appendixes A, B and C, contain some  details and calculations  that have been omitted in the main body of this article.

%%%%%%%%%%%%%%%%%%%%%%%%%%%%%%%%%%%%%%%%%%%%%%%%%%%%%%%%%%%%%%%%%%%%%%%%%%%%%%%
\section{Hamiltonian formulation of Bianchi I spacetimes}\label{sec.2}

We are interested in general relativity minimally coupled to a scalar field $\Phi$ that evolves under the influence of a potential $V(\Phi)$. We assume the spacetime  manifold to be $M=\mathbb{R} \times M_3$, with $M_3$ having the $\mathbb{R}^3$ topology. In the Arnowitt-Deser-Misner (ADM) formulation \cite{Arnowitt:1962hi}, the phase space ${\bf{V}}_{\rm GR}$ of general relativity  is characterized by two  couples of fields defined on  $M_3$,    $(\Phi(\vec{x}),P_{\Phi}(\vec{x}); h_{ij}(\vec{x}),\pi^{ij}(\vec{x}))$, where $P_{\Phi}(\vec{x})$ is the conjugate momentum of $\Phi(\vec{x})$, $h_{ij}(\vec{x})$ is a Riemannian metric that describes the intrinsic spatial geometry of $M_3$, and its conjugate momentum  $\pi^{ij}(\vec{x})$ describes  its extrinsic geometry (Latin indices $i,j$ run from 1 to 3). Recall that fields in phase space do not depend on time---time will appear below as the parameter along the flow generated by the Hamiltonian. The nonvanishing Poisson brackets between these fields are
\be \label{PB} \{ \Phi(\vec{x}),P_{\Phi}(\vec{x}')\}=\delta^{(3)}(\vec{x}-\vec{x}')\, ,\hspace{1cm}
\{ h_{ij}(\vec{x}),\pi^{k l}(\vec{x}')\}=\delta_{(i}^k\delta_{j)}^{l}\delta^{(3)}(\vec{x}-\vec{x}')\, .\ee
where $\delta_{(i}^k\delta_{j)}^{l}\equiv \frac{1}{2} (\delta_{i}^k\delta_{j}^{l}+\delta_{j}^k\delta_{i}^{l})$.  These canonical fields are subject to the four constraints of general relativity: The scalar and diffeomorphism (or vector) constraints
\bea \label{scons} \mathbb S (\vec{x})&=&\f{2\kappa}{\sqrt{h}}\l( \pi^{ij}\pi_{ij} -\f{1}{2}\pi^2 \r) - \f{\sqrt{h}}{2\kappa} ~^{(3)}R + \f{1}{2\sqrt{h}} P_{\Phi}^2 + \sqrt{h} \, V(\Phi) + \f{\sqrt{h}}{2} D_i \Phi D^i \Phi \approx 0 \, ,  \\ \label{vcons}
\mathbb V_i(\vec{x})&=&  -2 \sqrt{h}\,  h_{ij} \, D_k(h^{-1/2} \pi^{kj}) +P_{\Phi} \, D_i \Phi \approx 0\, , \ea
where $\kappa=8\pi G$, and $h$, $^{(3)}R$, and $D_i$ are the determinant, the Ricci scalar, and the covariant derivative associated with the metric $h_{ij}$, respectively.

Time evolution in ${\bf{V}}_{\rm GR}$ is generated by the Hamiltonian $\mathcal H$, which  is a combination of constraints 
\be \label{ham} \mathcal H=\int \d^3x \, \Big[N(\vec{x}) \,  \mathbb S (\vec{x})+N^i(\vec{x}) \, \mathbb V_i (\vec{x})\Big]\, .\ee
 $N(\vec{x})$ and $N^i(\vec{x})$ are called the lapse and shift functions, respectively, and they play the role of Lagrange multipliers. See \cite{ttbook} for  details of the ADM formulation omitted here.

We are interested in geometries  that are ``close'' to a homogeneous, anisotropic Bianchi I spacetime. In the Hamiltonian language, this means that we will restrict our attention  to a subset of the phase space ${\bf{V}}_{\rm GR}$ made of Bianchi I-type spacetimes ${\bf V}_{\rm BI}\in {\bf{V}}_{\rm GR}$ together with purely inhomogeneous linear perturbations around it.  In that neighborhood, we can write the canonical fields as
\bea \label{pert} \Phi(\vec{x})&=&\phi+\dph(\vec{x}) \, , \nonumber \\
P_{\Phi}(\vec{x})&=&\pp+\dpp(\vec{x}) \, , \nonumber \\ 
h_{ij}(\vec{x})&=&\hz_{ij}+\delta h_{ij}(\vec{x}) \, , \nonumber \\
\pi^{ij}(\vec{x})&=&\pz^{ij}+\delta \pi^{ij}(\vec{x})\, ,
\ea
where $\dph(\vec{x}), \dpp(\vec{x}),\delta h_{ij}(\vec{x}),\delta \pi^{ij}(\vec{x})$ describe small perturbations around  the homogeneous variables $\phi, \pp, \hz_{ij}, \pz^{ij}$ (From now on, all the indices $i,j,k,\ldots $ will be raised and lowered with $\hz^{ij}$ and $\hz_{ij}$, respectively). The background variables are defined as the homogeneous part of the canonical fields, in the sense that $\phi\equiv 1/\mathcal{V}_0\, \int_{M_3}d^3x\,  \Phi(\vec{x})$, and similarly for the other variables.\footnote{Because in the canonical treatment of Bianchi I geometries we have to deal  with homogeneous fields,  and because $M_3$ is noncompact, the spatial integrals involved in the definition of the Hamiltonian and the Poisson brackets diverge. This spurious infrared divergence can be eliminated by restricting the  integrals to a fiducial coordinate volume $\mathcal{V}_0$, arbitrarily large but finite, that can be understood as an infrared regulator. Physical predictions will not depend on $\mathcal{V}_0$, and we can take $\mathcal{V}_0\to \infty$ at the end of the calculation.}
 In Fourier space, the background variables encode the $\vec k=0$ mode of the canonical fields.  This automatically implies that perturbations  are purely inhomogeneous, in the sense that $\int_{M_3}d^3x\,\dph(\vec{x})=0$. Equivalently, they have Fourier components with $\vec k\neq 0$ only.

We now discuss the dynamics of the background variables and postpone the study  of perturbations for the next section. The  variables $\phi, \pp, \hz_{ij}, \pz^{ij}$ are chosen to describe a Bianchi I geometry.  The  nonzero canonical Poisson brackets are
\be \label{bpb}  \{\phi, \pp\}=\frac{1}{\mathcal{V}_0}\, ,  \hspace{0.5cm} \{\hz_{ij}, \pz^{kl} \}=\frac{1}{\mathcal{V}_0} \, \delta_{(i}^k\delta_{j)}^{l}\, ,\ee 
Next, as it is customary, we  restrict ourselves to spatial coordinates    $(x_1,x_2,x_3)$ for which the canonical variables take a diagonal form (this is always possible for Bianchi I metrics when the matter content is a perfect fluid \cite{hawkingellis})
\be \label{back} \hz_{ij}={\rm diag}(a_1^2,a_2^2,a_3^2) \, ,  \hspace{1cm} \pz^{ij}={\rm diag}\left(\f{\pi_{a_1}}{2\, a_1},\f{\pi_{a_2}}{2\, a_2},\f{\pi_{a_3}}{2\, a_3}\right)  \, .\ee
With this choice of numerical factors in (\ref{back}), the Poisson brackets (\ref{bpb}) translate to $\{a_i,\pi_{a_j}\}=\frac{1}{\mathcal{V}_0} \delta_{ij}$. Note that the subscripts  $i,j$ in $a_i$ and $\pi_{a_j}$ are just labels, and not tensorial indices. The scalar constraint, when  restricted to ${\bf V}_{\rm BI}$, takes the form 
\bea  \label{Fcons} \mathbb S^{(0)}&=& \f{1}{2\sqrt{\hz}} \biggl[ 
 \kappa \l( \f{a_1^2 \pi_{a_1}^2}{2} + \f{a_2^2 \pi_{a_2}^2}{2}  + \f{a_3^2 \pi_{a_3}^2}{2} 
 - a_1 \pi_{a_1} a_2 \pi_{a_2} - a_2 \pi_{a_2} a_3 \pi_{a_3} - a_3 \pi_{a_3} a_1 \pi_{a_1} \r)\nonumber\\
 && + p_{\phi}^2 + 2 \hz V({\phi}) \biggr] \approx 0\, ,
\ea%
where  $\hz=(a_1a_2a_3)^2=a^6$ is the determinant of $\hz_{ij}$, and we have defined the average scale factor as $a\equiv (a_1a_2a_3)^{1/3}$.  The vector constraint vanishes identically due to the homogeneity (and, as it is standard in the literature of Bianchi models, we set the shift $N^i$ equal to zero\footnote{This condition yields a spacetime metric invariant under parity (spatial inversions). The converse is also true: imposing invariance under spatial inversion implies $N^i=0$. This symmetry will  play an important role in the quantum theory of gauge invariant perturbations discussed below.}).
Then, the Hamiltonian (\ref{ham})  reduces to
\be \label{backH0} \mathcal H_{_{\rm BI}}=\int \d^3x \, N \, \mathbb S ^{(0)}\, .
\ee
Since $ \mathbb S ^{(0)}$ is homogeneous, only homogeneous lapses $N$ contribute to (\ref{backH0})---this is because the integral $\int_{M_3}d^3x$ of any purely inhomogeneous function vanishes identically---and then the spatial integral produces simply the total coordinate volume, $\mathcal H_{_{\rm BI}}=\mathcal{V}_0 \, N \, \mathbb S ^{(0)}$. Choosing $N=1$ corresponds to using  the familiar cosmic time $t$, and $N=a$ to conformal time $\eta$. The equations of motion are then given by Hamilton's equations  (we use cosmic time)
\bea \label{eoma} \dot a_i&=&\{a_i, \mathcal H_{_{\rm BI}}\} ;\hspace{0.5cm} \dot \pi_{a_i}=\{\pi_{a_i}, \mathcal H_{_{\rm BI}}\} \, ; \\ \nonumber 
\label{eomphi} \dot \phi&=&\{\phi, \mathcal H_{_{\rm BI}}\}= \f{\pp}{a^3}  \, ; \hspace{1cm} \dot{p}_\phi=\{\pp, \mathcal H_{_{\rm BI}}\}=- a^3 \, \f{\d V(\phi)}{\d\phi} \, .\ea
All aspects about dynamics can be extracted from these equations. Recall that under a rescaling of the three spatial coordinates $x_i\to \alpha_i \, x_i$ (no sum in repeated indices), the directional scale factors change as $a_i\to \alpha_i\, a_i$. Therefore, the scale factors $a_i$ are not physical observables---only ratios $a_i(t)/a_i(t')$ are. Hence, a  solution to these equations is uniquely characterized by specifying the value of $\pi_{a_1}(t_0)$, $\pi_{a_2}(t_0)$, $\pi_{a_3}(t_0)$, $\phi(t_0)$ and  $\pp(t_0)$ at some instant $t_0$ [the choice of $a_i(t_0)$ does not alter the physical content of the solution]. But since these degrees of freedom are subject to the constraint (\ref{Fcons}), a dynamical trajectory can be singled out, for instance, by specifying the first four and the sign of $\pp(t_0)$ [the constraint only determines $\pp^2(t_0)$, and not its sign]. 

It is common and  convenient to rewrite Eqs. (\ref{eoma}) in a different form. Namely, the dynamical degrees of freedom can be separated into those describing the evolution of a  spatial physical volume element, and those describing anisotropies. The equations of motion associated with the former take a form similar to the Friedmann equations of isotropic cosmology, while the dynamics of the anisotropies is determined by another set of differential equations. In order to obtain these equations, let us first define appropriate variables. Consider the timelike vector field $t^a\equiv (\partial_t)^a$ (where $a,b,\ldots$ are  spacetime tensor indexes). Let us decompose the tensor $\nabla_a t_b$ in its acceleration, expansion, shear, and twist \cite{waldGR}, where $\nabla_a$ is the covariant derivative compatible with the spacetime metric $g_{ab}$. The acceleration $a_b\equiv t^a \nabla_a t_b$ is zero, since $t^a$ is geodesic. The twist $w_{ab}$, that is given by the antisymmetric part of $\nabla_a t_b$, also vanishes, since $t^a$ is hypersurface orthogonal. The expansion is defined by the trace of $\nabla_a t_b$, and it is given by
\be \Theta\equiv \hz^{ab}\nabla_a t_b=\frac{\dot a_1}{a_1}+\frac{\dot a_2}{a_2}+\frac{\dot a_3}{a_3} \, ,\ee
with $\hz_{ab}=g_{ab}+n_an_b$, and  $n^a$ the unit  vector field normal to $M_3$ (with our choice $N^i=0$ for the shift, we have $t^a=n^a$). The average Hubble rate $H=\frac{\dot a}{a}$ is related to the expansion by  $H= \frac{1}{3} \Theta=\frac{1}{3}\left(H_1+H_2+H_3\right)$,  where  $H_i\equiv \frac{\dot a_i}{a_i}$ are the directional Hubble rates. The shear is defined as the symmetric, trace-free part of $\nabla_a t_b$
\be \sigma_{ab}=\nabla_{(a} t_{b)}-\frac{1}{3}\,  \Theta\, \hz_{ab} = {\rm diag}(0,a_1^2\, \sigma_1,a_2^2\, \sigma_2,a_3^2\, \sigma_3)\,,\ee
where $\sigma_i=(H_i-H)$, $i=1,2,3$. The pullback of this spacetime tensor to the spatial hypersurface $M_3$ is therefore
\be \sigma_{ij}={\rm diag}(a_1^2\, \sigma_1,a_2^2\, \sigma_2,a_3^2\, \sigma_3)\,.\ee
Since $\sigma_{ij}$ is traceless with respect to $\hz_{ij}$, its components are not independent, but they are  constrained by  $\sigma_1+\sigma_2+\sigma_3=0$. For later use, it is convenient to define the shear squared
\be\sigma^2=\sigma_{ij}\sigma^{ij}=\sigma_1^2+\sigma_2^2+\sigma_3^2= (H_1-H)^2+(H_2-H)^2+(H_3-H)^2 \, ,
\ee
with $\sigma^{ij}=\hz^{ik}\hz^{jl}\sigma_{kl}$. The relation of the canonical momenta $\pi_{a_i}$ with $H$ and  $\sigma_i$ can be obtained from the familiar relation between momenta and velocities, and it reads 
\be \pi_{a_i}=\frac{1}{\kappa}\frac{a^3}{a_i}\, (\sigma_i-2\, H)\, . \ee
With these definitions at hand, we can now extract from (\ref{eoma}) the equations of motion for the degrees of freedom that describe the evolution of the spatial volume element. They take the form
\be \label{aver} \frac{\ddot a}{a}=-\frac{\kappa}{6}\, [\rho+3\, P]-\frac{\sigma^2}{3} \, ; \hspace{1cm} \ddot \phi+3\f{\dot a}{a}\, \dot\phi+\frac{\d V(\phi)}{\d\phi}=0 \, . \ee
These variables are   subject to the scalar constraint (\ref{Fcons}), which can be written as
\be  \label{cons} H^2=\frac{\kappa}{3}\, \rho +\frac{\sigma^2}{6}\, , \ee
where we have defined the energy and pressure densities of $\phi$, $\rho\equiv \f{1}{2}\dot \phi^2+V(\phi)$ and $P\equiv \f{1}{2}\dot \phi^2-V(\phi)$, respectively.  Note that these expressions contain information about the anisotropies, via $\sigma^2$, and therefore the evolution of the mean scale factor is coupled to the dynamics  of  anisotropies. But as we will shortly see, the evolution of  $\sigma^2$ is remarkably simple, and it is given   by $\sigma^2=\frac{\Sigma^2}{a^6}$, where $\Sigma^2$ is a constant.\footnote{The factor $1/a^6$ implies that the contribution of anisotropies dilutes as stiff matter, faster than cold matter or radiation in an expanding universe. But note that this evolution for $\sigma^2$ is true only in  the absence of anisotropic sources in the matter sector, as it is the case if matter  is made of  a scalar field. In the more general case where the matter source is given by a perfect fluid with stress-energy tensor containing a nonzero anisotropic stress  $t_{ab}$, $T_{ab}=\rho \, n_an_b+P\, (\gz_{ab}+n_an_b)+t_{ab}$, the Eqs.   (\ref{eqanis})  describing the evolution of anisotropies acquire a source term proportional to $t_{ab}$, $\dot \sigma^a_{\  b}=-3\, H\, \sigma^a_{\ b}+\kappa \, t_{ab}$, and the evolution of $\sigma^2$ becomes more complicated.} Adding this piece of information makes Eqs.  (\ref{aver})--(\ref{cons}) a complete system for $a$ and $\phi$, which can be solved independently of other details in the anisotropies. Equations  (\ref{aver}) and (\ref{cons}), which we have derived from Hamilton's equations,  are equivalent to the diagonal components of Einstein's equations, and for $\Sigma^2=0$ they reduce to the  familiar FLRW theory.

On the other hand,  (\ref{eoma}) provides the following equations of motion for the anisotropies
\be \label{eqanis} \dot \sigma^i_{\  j}=-3\, H\, \sigma^i_{\ j} \, .\ee
These equations  are equivalent to the traceless components of Einstein's equations. The solutions to (\ref{eqanis}) are simply $\sigma_i=\Sigma_i/a^3$, where $\Sigma_i$ are three constants, constrained to satisfy $\Sigma_1+\Sigma_2+\Sigma_3=0$; hence, only two of them are independent. From this solution we immediately see that $\sigma^2=\frac{\Sigma^2}{a^6}$, where $\Sigma^2=\Sigma_1^2+\Sigma_2^2+\Sigma_3^2$. 

It is convenient to parametrize the freedom in the $\Sigma_i$'s in terms of $\Sigma^2$ and another constant of motion, $\Psi$, as
\be \label{shearsev} \sigma_1= \sqrt{\frac{2}{3}}\, \frac{\Sigma}{a^3}\, \sin \Psi\, , \hspace{0.3cm} \sigma_2= \sqrt{\frac{2}{3}}\, \frac{\Sigma}{a^3}\, \sin \left(\Psi+\frac{2\pi}{3}\right)\, , \hspace{0.3cm}\sigma_3= \sqrt{\frac{2}{3}}\, \frac{\Sigma}{a^3}\, \sin \left(\Psi+\frac{4\pi}{3}\right)\, ,\ee
where $\Sigma\equiv \sqrt{\Sigma^2}$. The relevant values of $\Psi$ fall in the range  $[\pi/6,\pi/2]$. Values outside this interval  only add a physically  irrelevant permutation of the values of the $\sigma_i$'s. 

To summarize,  by specifying  $H(t_0)$, $\phi(t_0)$, and the sign of $\dot \phi(t_0)$ at some instant $t_0$, together with $\Sigma^2$,  Eqs. (\ref{aver})--(\ref{cons}) provide a unique solution for $a(t)$ and $\phi(t)$ that completely describes the  evolution of the scalar field and the spatial volume element. Furthermore, a choice of $\Psi$ completely specifies the evolution of anisotropies by means of Eqs. (\ref{shearsev}).

\section{Perturbations}\label{sec.3}

Perturbation  fields $\dph (\vec{x}),\, \dpp (\vec{x}), \, \delta h_{ij}(\vec{x}),\, \delta \pi^{k l}(\vec{x})$ were defined in Eqs. (\ref{pert}), and their canonical Poisson brackets  can be obtained from  (\ref{PB}) and (\ref{bpb}). They are
\be \label{pertPB} \{ \dph (\vec{x}),\dpp (\vec{x}')\}=\delta^{(3)}(\vec{x}-\vec{x}')-\frac{1}{\mathcal{V}_0}\, 
;\hspace{0.5cm} \{ \delta h_{ij}(\vec{x}),\delta \pi^{k l}(\vec{x}')\}=\delta_{(i}^k\delta_{j)}^{l}\, \Big(\delta^{(3)}(\vec{x}-\vec{x}')-\frac{1}{\mathcal{V}_0}\Big)\, .\ee
Equations (\ref{PB}) and (\ref{bpb}) also imply that all Poisson brackets between background variables and perturbations vanish. The distribution $\delta^{(3)}(\vec{x}-\vec{x}')-\frac{1}{\mathcal{V}_0}$ is  the Dirac delta  on the space of purely inhomogeneous fields. Perturbations are subject to the four constraints (\ref{scons}) and (\ref{vcons}). It is convenient to expand them as
\bea \mathbb S(\vec x)&=&\mathbb S^{(0)}+\mathbb S^{(1)}(\v x)+\mathbb S^{(2)}(\v x)+\mathbb S^{(3)}(\v x)+\cdots\,  ,\nonumber \\
 \mathbb V_i(\v x)&=&\mathbb V_i^{(0)}+\mathbb V_i^{(1)}(\v x)+\mathbb V_i^{(2)}(\v x)+\mathbb V_i^{(3)}(\v x)+\cdots\, ,\ea
where the superscripts in parentheses denote the number of perturbation fields contained in each term. In this paper we will work at the lowest order in perturbations, that corresponds to keeping only linear terms   in the equations of motion. This is equivalent to truncate the constraints at {\em second} order, i.e.\ to disregard $\mathbb S^{(3)}(\v x)$, $\mathbb V^{(3)}(\v x)$ and higher order terms. 

Next, we expand  the lapse and shift as $N+\delta N(\vec x)$ and $N^i+\delta N^i(\vec x)$, where $N$ and $N^i$ are homogeneous, and for consistency with the gauge  used for the Bianchi I background metric, we take $N^i=0$. On the other hand, the perturbations $\delta N(\v x)$ and $\delta N^i(\v x)$ are the inhomogeneous parts of the lapse and shift, respectively. 

Recall that in a Bianchi I spacetime $\mathbb V_i^{(0)}$ identically vanishes, and $\mathbb S^{(0)}$ only constrains background degrees of freedom. Hence the physics of perturbations needs to be extracted from the constraints that are linear and quadratic in the perturbations. It is both natural and  convenient to interpret $\mathbb S^{(1)}(\v x)$ and $\mathbb V_i^{(1)}(\v x)$ as constraints on perturbations, and  define their Hamiltonian evolution from the quadratic contributions in the perturbations to the constraints. This is what we do in the next two subsections. 

\subsection{Gauge invariant perturbations\label{sec.3.A}}

We have a total of seven degrees of freedom (per point of space) in configuration variables---six from gravity, $\delta h_{ij}(\vec{x})$, and  one from the matter sector $ \dph (\vec{x})$---and seven  conjugate momenta. But they are subject to four constraints, $\mathbb S^{(1)}(\v x)\approx 0$, $\mathbb V_i^{(1)}(\v x)\approx 0$. In Dirac's terminology, these are first class constraints,  meaning that they are generators of gauge transformations. This is to say, the flow they generate in phase-space  relates configurations that must be identified as physically equivalent. Hence, each of these four constraints reduces the number of physical degrees of freedom by two, one due to the restriction they impose to the hypersurface where they vanish, and another arising from the identification of points along the gauge orbits they generate. Therefore, we are left with $14-8=6$ physical degrees of freedom (per point of space) in the phase space of perturbations. The goal of this section is to isolate these degrees of freedom. Their dynamics will be studied in the next section. 

To isolate the physical degrees of freedom we will extract out of the seven canonical pairs of perturbations three pairs that are {\em gauge invariant}, i.e.\ that remain invariant under the gauge flows or, equivalently, that  Poisson-commute with the four gauge generators $\mathbb S^{(1)}(\v x)$ and $\mathbb V_i^{(1)}(\v x)$. There exists an elegant and simple procedure to do this \cite{GENR},  consisting in finding a new set of canonical variables in which these four constraints are a subset of the new momenta. This is of course possible because these constraints are first class, i.e.\ they Poisson-commute among themselves.\footnote{While the Poisson brackets between any of the three vector constraints $\mathbb V_i^{(1)}(\v x)$ vanish, the vector constraints do not Poisson commute with $\mathbb S^{(1)}(\v x)$ off shell. However, these Poisson brackets are proportional to the zeroth-order scalar constraint $\mathbb S^{(0)}$, that vanishes on solutions of the background equations of motion (i.e.\ on-shell).}
For linear systems such as the one we are considering here, it is always possible to  achieve this globally in the phase space of  perturbations. The canonically conjugate variables of those four momenta are obviously pure gauge fields. On the other hand, the canonical Poisson brackets guarantee that the other three  canonical pairs are automatically gauge invariant. Furthermore, two facts make this strategy  useful. On the one hand,  the problem of finding this canonical transformation reduces to solving a simple Hamilton-Jacobi equation for a generating function and, on the other hand, the dynamics of gauge invariant and pure gauge fields decouple, allowing us to write a theory solely in terms of gauge invariant  (unconstrained) fields. There are however  multiple solutions to this problem (obviously, since  linear combinations of gauge invariant fields are also gauge invariant). We will choose the gauge invariant fields that in the isotropic limit reduce to the familiar scalar comoving curvature perturbations and tensors modes, that are commonly  used in FLRW cosmologies. 

In order to meet our goal, we  start by applying the standard  scalar-vector-tensor (SVT) decomposition to the metric perturbations. 
This decomposition is based on the property of perturbations under rotations that leave $\vec k$ invariant, and it is particularly useful in  spacetimes that are symmetric under rotations, such as FLRW geometries, since it guarantees that SVT modes evolve independently of each other.  Bianchi I metrics do not have any rotational symmetry, and therefore the SVT decomposition does not offer any clear advantage compared to other choices---none of the decompositions available in the literature decouples the different components in $\delta h_{ij}(\vec{x})$ \cite{ppu-BI2,s-free2}. But we still find the SVT decomposition the most useful choice, since in scenarios of interest for cosmology the spacetime isotropizes at late times. 
We begin by Fourier-expanding the metric perturbations\footnote{Here we use the Fourier expansion of fields in a box of fiducial volume $\mathcal{V}_0$. But one must keep in mind that we will  take the limit $\mathcal{V}_0\to\infty$ at the end of the calculation. Working in a box only changes the calculations in that the wave numbers $\vec k$ are restricted to  a discrete lattice $\left(\frac{\mathcal{V}_0^{1/3}}{2\pi}\,\vec k\right)\in  \mathbb{Z}^3$.}
\be \label{Fexph} \delta h_{ij}(\vec{x})=\sum_{\vec{k}\neq \v 0} \delta \t{h}_{ij}(\vec{k})\, e^{i\, \v{k}\cdot\v{x}}\, ; \hspace{1cm} \delta \pi^{ij}(\vec{x})=\sum_{\vec{k}\neq \v 0} \delta \t{\pi}^{ij}(\vec{k})\, e^{i\, \v{k}\cdot\v{x}}\, .\ee 
Here, $\v{k}\cdot\v{x}=k_i\,x^i$, and $k_i$ is time independent (the so-called  comoving wave vector). Note also that the ``zero-mode" $\vec k=\v 0$  has been excluded from the sum; this is because perturbations are purely inhomogeneous fields  and  do not have any homogeneous components. Similarly, we Fourier-expand  the perturbations of the scalar field and its conjugate  momentum
 \be \label{Fexph2}\delta \phi(\vec{x})=\sum_{\vec{k}\neq \v 0} \delta \t{\phi}(\vec{k})\, e^{i\, \v{k}\cdot\v{x}}\, ; \hspace{1cm} \delta p_{\phi}(\vec{x})=\sum_{\vec{k}\neq \v 0} \delta \t{p}_{\phi}(\vec{k})\, e^{i\, \v{k}\cdot\v{x}}\, .\ee 
The Poisson brackets (\ref{pertPB}) imply
 \be\label{cpb} \{ \delta  \t \phi(\vec{k}),\delta \t p_{\phi}(\vec{k'})\}=\mathcal{V}_0^{-1}\, \delta_{\vec{k},-\vec{k}'}\,; \quad \{ \delta  \t h_{ij}(\vec{k}),\delta \t \pi^{k l}(\vec{k'})\}=\mathcal{V}_0^{-1}\, \delta_{(i}^k\delta_{j)}^{l}\,  \delta_{\vec{k},-\vec{k}'}\, .\ee
Note that  the conjugate variables of $\delta  \t \phi(\vec{k})$ and $\delta  \t h_{ij}(\vec{k})$ are $\delta \t p_{\phi}(-\vec{k})$ and $\delta \t \pi^{ij}(-\vec{k})$, respectively, rather than $\delta \t p_{\phi}(\vec{k})$ and $\delta \t \pi^{ij}(\vec{k})$.

The scalar-vector-tensor decomposition is obtained by writing $ \delta  \t h_{ij}(\vec{k})$
 in a convenient basis in the vector space of $3\times 3$ symmetric matrices
 \be \label{eq:dh-to-gamma}\delta \tilde h_{ij}(\vec{k})=\sum_{n=1}^{6}  \gamma_n(\vec{k}) \,  {A}^{{(n)}}_{ij}(\hat k)\, ; \hspace{1cm} \delta \tilde \pi^{ij}(\vec{k})=\sum_{n=1}^{6}  \pi_n(\vec{k}) \,  {A}_{{(n)}}^{ij} (\hat k)\, ,  \ee
 where
\begin{align}
 {A}^{{(1)}}_{ij}\, &=\, \f{\hz_{ij}}{\sqrt{3}}, \hspace{0.5in} & {A}^{(4)}_{ij}\, &=\,\f{1}{\sqrt{2}}\, \l(\, \hat{ k}_i\, \h y_j\, +\, \hat{ k}_j\, \h y_i \,\r),\nonumber\\
 {A}^{(2)}_{ij}\, &=\,\sqrt{\f{3}{2}}\,\l(\hat{ k}_i\,\hat{ k}_j - \f{\hz_{ij}}{3}\r), \hspace{.5in}  &{A}^{(5)}_{ij}\,& = \, \f{1}{\sqrt{2}}\, \l(\, \hat x_i\, \h x_j\, -\, \hat y_i\, \h y_j \,\r), \nonumber\\
{A}^{(3)}_{ij}\, &=\, \f{1}{\sqrt{2}}\, \l(\, \hat{ k}_i\, \h x_j\, +\, \hat{ k}_j\, \h x_i \,\r),
 \hspace{.5in} &{A}^{(6)}_{ij}\, &=\,\f{1}{\sqrt{2}}\, \l(\, \hat x_i\, \h y_j\, +\, \hat x_j\, \h y_i \,\r), \label{matrixbases}
\end{align}
and ${A}_{{(n)}}^{ij} (\hat k)$ are obtained from ${A}^{{(n)}}_{ij}$ by  raising the indices with $\hz_{ij}$. In these expressions $\hat{ k}$ is the unit vector (with respect to $\hz_{ij}$) in the direction of $\v{k}$. Together with $\h x$ and  $\h y$, they form a time-dependent orthonormal triad with orientation defined by $\hat x \times \hat y=\hat k$.\footnote{\label{invA} Under a parity transformation $\delta h_{ij}(\vec x)\to  \delta h_{ij}(-\vec x)$, we have $\delta \tilde{h}_{ij}(\vec k)\to  \delta \tilde{h}_{ij}(-\vec k)$. Consequently, the matrices ${A}^{{(n)}}_{ij}(\hat k)$ transform as follows:  ${A}^{{(n)}}_{ij}(\hat k)\to {A}^{{(n)}}_{ij}(-\hat k)={A}^{{(n)}}_{ij}(\hat k)$ for $n=1,2,4,5$, and  ${A}^{{(n)}}_{ij}(\hat k)\to {A}^{{(n)}}_{ij}(-\hat k)=-{A}^{{(n)}}_{ij}(\hat k)$ for $n=3,6$, where we have used that under $\hat k\to -\hat k$, the unit vectors $\hat x$ and $\hat y$ transform to $\hat x$ and $-\hat y$ respectively (since the three unit vectors must maintain their relative orientation). This implies that under parity, $\gamma_n(\vec{k})$ transforms as $\gamma_n(\vec{k})\to \gamma_n(-\vec{k})$ for $n=1,2,4,5$, and $\gamma_n(\vec{k})\to -  \gamma_n(-\vec{k})$  for $n=3,6$. On the other hand, the reality of $\delta h_{ij}(\vec x)$ implies that,  under complex conjugation, $\bar \gamma_n(\vec k)=\gamma_n(-\vec k)$  for $n=1,2,4,5$, and $\bar \gamma_n(\vec k)=-\gamma_n(-\vec k)$  for $n=3,6$. Therefore, a parity transformation can be implemented by changing $\gamma_n(\vec{k})\to\bar \gamma_n(\vec{k})$ for all $n$.} The dependence on time of these three vectors originates from the time dependence of the Bianchi I metric $\hz_{ij}$, and  it makes ${A}^{{(n)}}_{ij} $   also functions of time (see Appendix \ref{appA} for further details). The components $\gamma_n(\vec k)$ and $ \pi_n(\vec k)$ are called scalar modes for $n=1,2$,  vector modes for $n=3,4$, and tensor  modes for $n=5,6$. These names are motivated from the transformation properties of the matrices ${A}^{(n)}_{ij}$  under  rotations around the direction $\hat{k}$. We implement the decomposition \eqref{eq:dh-to-gamma} as a time-dependent canonical transformation between $ (\tilde h_{ij}(\vec{k}),\delta \t{\pi}^{ij}(\vec{k}))$ and $ (\gamma_n(\vec k), \pi_n(\vec k))$. The details can be found in Appendix \ref{appA}. The Poisson brackets (\ref{cpb}) become
\bea \label{gammacomm}\{\gamma_n(\v k), \pi_m (\v{k}')\}&=&
\mathcal{V}_0^{-1} \, \delta_{nm} \, \delta_{\vec{k},-\vec{k}'}\, , \nonumber \\
\{ \gamma_n(\v k),\gamma_m (\v{k}')\}&=&0
\, , \nonumber \\
\{ \pi_n(\v k), \pi_m (\v{k}')\}&=&0\, .
\ea
For later use, we also define  $\sigma_{(n)}(\hat k) \equiv \sigma_{ij} \, {A}_{(n)}^{ij}(\hat k)$, for $n=2,\ldots, 6$,  as the  projection of the  shear tensor $\sigma_{ij}$ on the basis elements ${A}_{(n)}^{ij}(\hat k)$  (there is no $\sigma_{(1)}$, because $\sigma_{ij}$ is traceless). It should be clear from this definition that $\sigma_{(n)}(\hat k)$ {\em are not} the Fourier components of the tensor $\sigma_{ij} $---this should be obvious since $\sigma_{ij} $ is position independent, and therefore its Fourier transform would contain only the $\vec k=\v 0$ mode. $\sigma_{(n)}(\hat k)$ is rather a compact way of writing the product of  $\sigma_{ij} $ and the basis tensors ${A}_{(n)}^{ij}(\vec k)$, a combination that will  repeatedly  appear in our expressions below.

Expressions \eqref{eq:S1}--\eqref{eq:Vy} in Appendix \ref{appA}  show the form of the scalar and vector constraints written in terms of $ \gamma_n$ and $ \pi_n$. From them, it is straightforward to check that none of these  variables, neither $ \delta \t \phi$ nor $ \delta \t p_{\phi}$, Poisson-commute with either the scalar $\mathbb{S}^{(1)}$ or any of the vector constraints $\mathbb{V}_i^{(1)}$. Therefore, they are not gauge invariant. In order to find gauge invariant variables, as explained above, we look for a canonical transformation
\be  \label{CT}  \gamma_{\alpha} (\vec k)\, ,\,   \pi_{\alpha}(\vec k) \longrightarrow  \Gamma_{\alpha}(\vec k),  \Pi_{\alpha} (\vec k) \, \ee
[where we have defined $ \gamma_0\equiv \sqrt{4\kappa}\  \delta \t\phi(\vec k)$ and  $ \pi_0\equiv \sqrt{1/4\kappa}\  \delta \t p_\phi(\vec k)$] to new canonical pairs $ \Gamma_{\alpha}(\vec k)$ and $  \Pi_{\alpha} (\vec k)$, $\alpha=0,\ldots,6$, such that four of the new momenta agree with the Fourier components of the constraints
\be \label{eq:new-Pi}  \Pi_{3} (\vec k)=\frac{1}{|\v k|}\, \tilde{\mathbb{S}}^{(1)}(\vec k)\, , \hspace{0.4cm}  \Pi_{4} (\vec k)=\frac{1}{i \, |\v k|}\, \h k^j\,\tilde{\mathbb{V}}^{(1)}_j(\vec k)\, , \hspace{0.4cm}  \Pi_{5} (\vec k)=\frac{1}{i\, |\v k|}\,  \h x^j\,\tilde{\mathbb{V}}^{(1)}_j(\vec k)\, , \hspace{0.4cm}  \Pi_{6} (\vec k)=\frac{1}{i\, |\v k|}\, \h y^j \,\tilde{\mathbb{V}}_j^{(1)}(\vec k)\, . \ee
Here, $|\v k|\equiv \sqrt{k_ik^i}$ is the norm of $\v k$. The factor $\frac{1}{|\v k|}\, $ has been introduced for dimensional reasons (recall that $\vec k \neq \v 0$), and the imaginary unit for convenience in the calculation. As mentioned above, this automatically implies that  $\Gamma_{\alpha}(\vec k)$ are gauge invariant for $\alpha=0,1,2$, and pure gauge for $\alpha=3,4,5,6$. This transformation can be obtained by finding a suitable generating function $G( \gamma_{\alpha}, \Pi_{\alpha})$, that we choose to be of type 2---i.e.,  it depends on old variables $ \gamma_{\alpha}$ and new momenta $ \Pi_{\alpha}$---and from which the rest of the variables are given by
\begin{equation}\label{eq:pi-Gamma-G}
  \pi_{\alpha} (\vec k)= \f{\partial G( \gamma_{\beta},\,  \Pi_{\beta})}{\partial \gamma_{\alpha}(\vec k)} ,
 \hspace{2cm}
  \Gamma_{\alpha} (\vec k)= \f{\partial G( \gamma_{\beta},\,  \Pi_{\beta})}{\partial \Pi_{\alpha}(\vec k)} .
\end{equation}
The generating function we are looking for is a solution of the following Hamilton-Jacobi-type equations:
 \begin{eqnarray}\label{HJeq}
 \Pi_3 (\vec k)&=& \frac{1}{|\v k|}\,\tilde{\mathbb{S}}^{(1)}( \gamma_{\alpha},\,  
 \pi_{\alpha}\,=\, \f{\partial G( \gamma_{\beta},\,  \Pi_{\beta})}{\partial  \gamma_{\alpha}} )\, , \nonumber \\
 \Pi_{4} (\vec k)&=&\frac{1}{i|\v k|}\, \h k^j\,\tilde{\mathbb{V}}_j^{(1)}( \gamma_{\alpha},\, 
 \pi_{\alpha}\,=\, \f{\partial G( \gamma_{\beta},\,  \Pi_{\beta})}{\partial  \gamma_{\alpha}} ) \, , \nonumber \\
 \Pi_{5} (\vec k)&=&\frac{1}{i|\v k|}\, \h x^j\,\tilde{\mathbb{V}}_j^{(1)}( \gamma_{\alpha},\, 
 \pi_{\alpha}\,=\, \f{\partial G( \gamma_{\beta},\,  \Pi_{\beta})}{\partial  \gamma_{\alpha}} ) \, , \nonumber \\
 \Pi_{6}(\vec k) &=&\frac{1}{i|\v k|}\, \h y^j\,\tilde{\mathbb{V}}_j^{(1)}( \gamma_{\alpha},\, 
 \pi_{\alpha}\,=\, \f{\partial G( \gamma_{\beta},\,  \Pi_{\beta})}{\partial  \gamma_{\alpha}} ) \, . \end{eqnarray}
These differential equations for $G( \gamma_{\alpha},\,  \Pi_{\alpha})$ can be converted into algebraic equations by noticing that, because we are working at linear order in perturbations, the generating function $G( \gamma_{\alpha},\,  \Pi_{\alpha})$ can only depend on $ \gamma_{\alpha}$ and $ \Pi_{\alpha}$ quadratically, and hence it must be of the form 
\be \label{Gform} G=\sum_{\vec{k}} \, (B^{\alpha \beta}\,  \Pi_{\alpha} \, \gamma_{\beta}+C^{\alpha \beta}\, \gamma_{\alpha} \gamma_{\beta}) \, , \ee
where $B^{\alpha\beta}$ and $C^{\alpha \beta}$ are matrices whose unknown components do not depend on perturbations, although they can depend on background variables, and $C^{\alpha \beta}$ is symmetric. The generating function contains therefore $77$ unknown coefficients.\footnote{We could have also included in $G$ a term  of the form $\displaystyle \sum_{\vec{k}\neq \v 0}\, D^{\alpha\beta}\,  \Pi_{\alpha} \Pi_{\beta}$. We have not done so simply  because (\ref{Gform}) is already general enough to meet our goals.} Equations (\ref{HJeq}) provide then a set of algebraic relations for the components of  $B^{\alpha\beta}$ and $C^{\alpha \beta}$. More precisely,  (\ref{HJeq}) contain 44 equations, out of which only 38 are independent. Hence, these equations have multiple solutions, and any of them will provide us with three independent pairs of gauge invariant variables that are equally legitimate; physical predictions are of course independent of the variables we use in our calculations. As mentioned before, we choose the solution for which the gauge invariant variables agree with the familiar scalar perturbations and the two tensor modes in the isotropic limit. They are%
\begin{eqnarray}
  \Gamma_0(\vec k)\, &=&   \gamma_0\, +\, 
{
\f{\sqrt{\kappa}\,p_{\phi}}{\sqrt{1/6}\, \kappa\,a \, p_a\, 
 +\, a^3\, \sigma_{(2)} }\l(\sqrt{2} \, \gamma_1\, -\,  \gamma_2 \r)\, ,
}
\label{eq:G0}
\\
  \Gamma_1(\vec k)\, &=& \gamma_5\, +
{
\ \f{\,a^2\,\sigma_{(5)}}{\sqrt{1/6}\, \kappa\,p_a\, 
 +\,a^2\, \sigma_{(2)} }\ \l(\sqrt{2}\,  \gamma_1\, -\,  \gamma_2 \r)\, ,
}\label{eq:G1}
\\
   \Gamma_2(\vec k)\, &=& \gamma_6\, +
{
\ \f{\,a^2\,\sigma_{(6)}}{\sqrt{1/6}\, \kappa\,p_a\, 
 +\, \,a^2\, \sigma_{(2)} }\ \l(\sqrt{2}\,  \gamma_1\, -\,  \gamma_2 \r)\, ,
}\label{eq:G2}
\end{eqnarray}
where  $p_a$ is the canonically conjugate variable of the average scale factor $a$, and it is related to the expansion by $p_a = -2 a^2 \Theta/\kappa$. Note that, choosing three gauge invariant variables fixes $21$  coefficients,  leaving $18$ of them free, which can be fixed by demanding  their Hamiltonian to have a simple form.
Further details about this canonical transformation, such as the form of the conjugate momenta $\Pi_0$, $\Pi_1$, and $\Pi_2$, and of  the pure gauge fields, can be found in Appendix \ref{appA}. One can see there that $\Pi_0$, $\Pi_1$, and $\Pi_2$ also involve vector modes $\gamma_3$ and $\gamma_4$, and the components  $\sigma_{(3)}$ and $\sigma_{(4)}$ of the shear (recall that $\sigma_{(1)}=0$, because the shear tensor is traceless).   It is  straightforward to check that $ \Gamma_0$, $\Gamma_1$, and $\Gamma_2$ and their conjugate momenta Poisson-commute with the linear constraints.  Hence, they span the phase space of gauge invariant fields.
 
In the isotropic limit $\sigma_{(n)}\to 0$, $ \Gamma_1$ and $ \Gamma_2$  reduce to the familiar two polarizations of transverse and traceless tensor modes,  and
$ \Gamma_0$ becomes proportional to the comoving curvature perturbation $\mathcal{R}$,  i.e. $\Gamma_0= \sqrt{4\kappa}\, \frac{z}{a}\, \mathcal{R}$, where $z=-\frac{6}{\kappa}\frac{p_{\phi}}{p_a}=\frac{\dot \phi}{H}\, a$. But in presence of  anisotropies, there are no gauge invariant fields that are  combinations of tensor modes of the metric  only; mixture with scalar modes is needed to achieve gauge invariance.

\subsection{Dynamics: Physical Hamiltonian}\label{sec.3.B}

The strategy followed in the previous subsection guarantees that the dynamics of gauge invariant fields decouples from pure gauge ones \cite{GENR}.  The dynamics of the former is generated by the Hamiltonian (see Appendix \ref{B} for further details)
\be \label{Ham2}\mathcal{H_{\rm pert}}=\frac{N(t)\,{\cal V}_0}{2\, a(t)}\, \sum_{\vec k}\sum_{\mu,\mu'=0}^2\, \left[ \frac{4\kappa}{a^2(t)}\, \delta_{\mu,\mu'}\, |\Pi_{\mu}(\vec k)|^2+\, \frac{a^2(t)}{4\kappa}\, \Big( \delta_{\mu,\mu'}\, k^2+\, {\cal U}_{\mu \mu'}(t,\vec k)\Big)\, \Gamma_{\mu}(\vec k)\bar \Gamma_{\mu '}(\vec k)\right]\, ,\ee
where $k^2(t)\equiv \, a^2(t)k^ik_j=\, a^2(t)\left(\frac{k_1^2}{a_1^2(t)}+\frac{k_2^2}{a_2^2(t)}+\frac{k_3^2}{a_3^2(t)}\right)$, $\delta_{\mu,\mu'}$ is the Kronecker delta, and $N$ is the same lapse function adopted to evolve the background geometry in the previous section. If we choose $N=1$, this Hamiltonian generates evolution in cosmic time $t$, and in conformal time if $N=a$.   The (time-dependent) effective potentials $ {\cal U}_{\mu\mu'}$ are symmetric in $\mu$ and $\mu'$, and the off-diagonal terms vanish in the isotropic limit. In the presence of  anisotropies, these off-diagonal  components  describe the couplings between the different types of gauge invariant perturbations. They are given by
\bea
\, {\cal U}_{00}&=&a^2\, V_{\phi\phi}\, - \frac{2 \kappa \, \pp^2 {\cal F}_2}{a^3}  + 2 \kappa  \, {\cal F}_1  \left(-\frac{\kappa\,\pp^2\,p_a}{3a^{5}}\, + \,2 \,V_{\phi}\, \pp\right),\\ \nonumber
   {\cal U}_{01}&=&{\cal U}_{10} \,= \,\frac{2\sqrt{\kappa}}{a^2}\left(-a^2\,\pp\, \sigma_{(5)} \, {\cal F}_2  + a^5V_{\phi} \, \sigma_{(5)}\, {\cal F}_1 - a^2\,\pp \, {\cal G}_{5}\,  {\cal F}_1\, +\, \frac{\kappa}{6}\,\pp\,p_a\,\sigma_{(5)}\,{\cal F}_1\right)\,,\\ \nonumber
   {\cal U}_{02}&=&{\cal U}_{20}\,= \,\frac{2\sqrt{\kappa}}{a^2}\left(-a^2\,\pp\, \sigma_{(6)} \, {\cal F}_2  + a^5\,V_{\phi} \, \sigma_{(6)}\, {\cal F}_1 - a^2\,\pp \, {\cal G}_{6}\,  {\cal F}_1\, +\, \frac{\kappa}{6}\,\pp\,p_a\,\sigma_{(6)}\,{\cal F}_1\right)\,,\\ \nonumber
   {\cal U}_{12}&=&{\cal U}_{21}\,=\, 2\,\sigma_{(5)}\,\sigma_{(6)}\,\left( a^2 -\, a^3\,{\cal F}_2\,+\, \frac{2}{3}\,\kappa\,a\,p_a\,{\cal F}_1\right) - \left( \, 2\,a^3\,\sigma_{(6)}\,{\cal G}_{5}\, +\, 2\,a^3\,\sigma_{(5)}\,{\cal G}_{6}\right)\,{\cal F}_1\, \\ \nonumber
{\cal U}_{11}\, &=&\,- 2 \,a^2\, \sigma_{(6)}^2\, +\,\frac{\kappa p_a \,\sigma_{(2)}}{\sqrt{6}}\,-\,a^2\,\sqrt{\f{2}{3}}{\cal G}_2\, +\,\frac{4}{3}\,\kappa\,a\,p_a\,\sigma_{(5)}^2\,{\cal F}_1\,-\,4 \,a^3\, \sigma_{(5)}\,{\cal F}_1 \, {\cal G}_5\,-\,2\,a^3\, \sigma_{(5)}^2\,{\cal F}_2\, ,\\ \nonumber
  {\cal U}_{22}\, &=&\,- 2 \,a^2\, \sigma_{(5)}^2\, +\,\frac{\kappa p_a \,\sigma_{(2)}}{\sqrt{6}}\,-\,a^2\,\sqrt{\f{2}{3}}{\cal G}_2\, +\,\frac{4}{3}\,\kappa\,a\,p_a\,\sigma_{(6)}^2\,{\cal F}_1\,-\,4 \,a^3 \,\sigma_{(6)}\, {\cal F}_1\, {\cal G}_6\,-\,2\,a^3\, \sigma_{(6)}^2\, {\cal F}_2\,, \ea
 where $V_{\phi}\equiv dV/d\phi$, $V_{\phi\phi}\equiv d^2V/d\phi^2$, and
\begin{eqnarray}
    {\cal F}_1\, &=&\, \frac{-\frac{\kappa p_a}{2a^3}\, +\,\sqrt{\frac{3}{2}} \,\frac{\sigma_{(2)}}{a}}{
      2\kappa\rho\,+\,\sigma_{(3)}^2\,+\, \sigma_{(4)}^2+\, \sigma_{(5)}^2\,+\, \sigma_{(6)}^2},\\ \nonumber
  {\cal F}_2\, &=&\frac{\frac{3\kappa \, V}{a}\, - \,\frac{\kappa^2 p_a^2}{3a^5}\,+\,\frac{\kappa p_a\sigma_{(2)}}{2\sqrt{6}a^3}\, +\, \sqrt{\frac{3}{2}}\frac{{\cal G}_{2}}{a}\,-\,{\cal F}_1\left[\frac{\kappa^2 \pp^2p_a}{a^8}\,+\,2\,\sigma_{(3)}\, {\cal G}_3\,+\,2\,\sigma_{(4)}\, {\cal G}_4+\, 2\,\sigma_{(5)}\, {\cal G}_5\,+\, 2\,\sigma_{(6)}\, {\cal G}_6)\right]}{
    2\kappa\rho\,+\,\sigma_{(3)}^2\,+\, \sigma_{(4)}^2+\, \sigma_{(5)}^2\,+\, \sigma_{(6)}^2},\\ \nonumber
   {\cal G}_2 &=& \frac{\kappa p_a\sigma_{(2)}}{2\,a^2}\, -\, \sqrt{\frac{3}{2}}\left(\sigma_{(3)}^2 \,+\, \sigma_{(4)}^2\right),\\ \nonumber
   {\cal G}_3 &=& \frac{\kappa\, p_a\, \sigma_{(3)}}{2\,a^2}\,  +\, \frac{1}{\sqrt{2}} \left(\sqrt{3}\sigma_{(2)} \sigma_{(3)} - \sigma_{(3)} \sigma_{(5)} - \sigma_{(4)} \sigma_{(6)}\right),\\ \nonumber
   {\cal G}_4 &=&  \frac{\kappa p_a\sigma_{(4)}}{2\,a^2} + \frac{1}{\sqrt{2}} \left(\sqrt{3}\sigma_{(2)} \sigma_{(4)} + \sigma_{(4)} \sigma_{(5)} - \sigma_{(3)} \sigma_{(6)}\right)\,, \\ \nonumber
   {\cal G}_5 &=& \frac{\kappa p_a\sigma_{(5)}}{2\,a^2}\, +\, \frac{1}{\sqrt{2}}(\sigma_{(3)}^2 - \sigma_{(4)}^2),\\ \nonumber
   {\cal G}_6 &=& \frac{\kappa p_a\sigma_{(6)}}{2\,a^2} + \sqrt{2}\,\sigma_{(3)}\sigma_{(4)}.
  \end{eqnarray}

The dependence in $\vec k$ in the right-hand side of these expressions comes from $\sigma_{(n)}(\vec k)$ [defined below Eq.\ (\ref{gammacomm}) in Sec. \ref{sec.3.A}]. Time evolution is now given by Hamilton's equations, that are derived by using the Poisson brackets given in Eq.\ \eqref{Gamma-comm}.  In cosmic time, they read
\bea \label{perteqH} \dot \Gamma_{\mu}(\vec k)&=&\{\Gamma_{\mu}(\vec k),\mathcal{H_{\rm pert}}\}=\frac{4\kappa}{a^3}\, \Pi_{\mu}(\vec k)\, ,\nonumber \\ 
 \dot \Pi_{\mu}(\vec k)&=&\{\Pi_{\mu}(\vec k),\mathcal{H_{\rm pert}}\}=-\frac{a}{4\kappa}\,\sum_{\mu'=0}^2\,  (\delta_{\mu\mu'}\, k^2+ {\cal U}_{\mu \mu'})\, \Gamma_{\mu'}(\vec k)\, . \ea
As usual, we  obtain second-order differential equations for $ \Gamma_{\mu}(t,\vec k)$  by eliminating $\Pi_{\mu}$ 
\be \label{eqginper}\ddot \Gamma_{\mu}+3\, H\, \dot \Gamma_{\mu}+\frac{k^2}{a^2}\,  \Gamma_{\mu}+\frac{1}{a^2}\, \sum_{\mu'=0}^2\, {\cal U}_{\mu\mu'}\, \Gamma_{\mu'}=0\, ,  \ \ \ \ \mu=0,1,2\, . \ee
This is a set of three coupled, second-order, ordinary differential equations for each wave vector $\vec k$. Because the potentials ${\cal U}_{\mu \mu'}(t,\vec k)$ are time dependent, it is not possible to absorb these couplings by means of a local time-dependent redefinition of fields and time. In other words, it is not possible to simultaneously diagonalize  the matrix ${\cal U}_{\mu \mu'}(t,\vec k)$ with a local time-dependent transformation while keeping the other terms in these equations (including those containing time derivatives) diagonal. As mentioned above, in the isotropic limit, the potential ${\cal U}_{\mu \mu'}(t,\vec k)$ becomes diagonal and the equations  for $\Gamma_{0}$, $\Gamma_{1}$ and $\Gamma_{2}$ decouple and reduce to the familiar equations describing scalar and tensor gauge invariant perturbations in FLRW spacetimes. We have checked that Eqs.\ (\ref{eqginper})   are equivalent to the equations obtained from a Lagrangian approach, derived in \cite{ppu-BI1,ppu-BI2}. 

On the other hand, we have implemented the main steps of this analysis in a computer code written in the symbolic language of {\tt Mathematica}, and made publicly available in \cite{ntbk}. We have  also complemented this notebook with a computer code, based on the C programming language, and available in \cite{num-lib}, to solve Eqs. (\ref{eqginper}) and to compute observables in the CMB.

From a physical viewpoint, it is convenient to replace $\Gamma_1$ and $\Gamma_2$ by the combinations
\be \Gamma_{\pm2}(\vec k)\equiv \frac{1}{\sqrt{2}} \left(\Gamma_{1}(\vec k)\mp i\, \Gamma_{2}(\vec k)\right)\, . \ee
Under a rotation of angle $\theta$ around the direction $\hat k$, $ \Gamma_{\pm2}(\vec k)$ acquire a phase $e^{\pm i\, 2\, \theta}$;\ i.e. they transform as fields with spin weight $\pm2$. In the isotropic limit, these fields describe tensor modes with helicity $\pm2$ (i.e.\ circularly polarized radiation). Also, it is straightforward to check that  $\bar \Gamma_{\pm2}(\vec k)= \Gamma_{\pm2}(-\vec k)$, and under parity $\Gamma_{\pm2}(\vec k)\to \Gamma_{\mp2}(-\vec k)$.\footnote{This is to be contrasted with $\bar \Gamma_0(\vec k)= \Gamma_0(-\vec k)$,  $\bar \Gamma_1(\vec k)= \Gamma_1(-\vec k)$, and  $\bar \Gamma_2(\vec k)= -\Gamma_2(-\vec k)$. Note that $\Gamma_2$ is an ``anti-Hermitian" field; it is for this reason that in the quantum theory it is more convenient to work with the circularly polarized fields $\Gamma_{\pm 2}$. On the other hand, under a  parity transformation, $ \Gamma_0(\vec k)\to \Gamma_0(-\vec k)$,  $ \Gamma_1(\vec k)\to \Gamma_1(-\vec k)$, and  $ \Gamma_2(\vec k)\to -\Gamma_2(-\vec k)$.}
These properties will be useful in the next section. From now on, we will use these variables.

\section{Quantum theory: Kinematics}\label{sec.4}

In this section we discuss the quantum theory of the gauge invariant fields $\Gamma_{0}$, $\Gamma_{\pm2}$, again working in the canonical formalism. We focus here on the quantum kinematics, and leave the discussion of dynamics for the next section. The phase space ${\mathbb{V}}(\vec k)$ for a Fourier mode $\vec k$ of our system is made of three canonically conjugate pairs, that we will encode in a single element $v(\vec{k})=(\Gamma_{0}(\vec k),\,  \Gamma_{+2}(\vec k),\, \Gamma_{-2}(\vec k), \, \Pi_{0}(\vec k), \, \Pi_{+2}(\vec k), \, \Pi_{-2}(\vec k))\, \in\, {\mathbb{V}}(\vec k)$. The components of $v(\vec{k})$ will be denoted with the  index $S$, with $S$ running  from 0 to 5. We will reserve lower case indices $s=0,+2,-2$, to denote the three fields $\Gamma_{s}(\vec k)$ and momenta $\Pi_{s}(\vec k)$ individually. As we just discussed at the end of the previous section, if the spacetime were isotropic, the three fields $\Gamma_{s}$ would evolve independently, and the space of solutions to the equation of motion would acquire a product structure $\mathbb{S}=\mathbb{S}_{0}\times\mathbb{S}_{+2}\times \mathbb{S}_{-2}$. But in Bianchi I geometries, gauge invariant perturbations are coupled and we lose this product structure. However, the equations of motion are still {\em linear} in the fields, and consequently the space of solutions is a vector space (i.e.\ any linear combination of solutions is also a solution). It is precisely this vector space structure that allows us to formulate the quantum theory in an exact way, without the need of any perturbative treatment of the anisotropies.

The construction of the quantum theory for gauge invariant perturbations in Bianchi I spacetimes follows the same steps as the quantization of two harmonic oscillators with a linear, time-dependent  coupling between them. Appendix \ref{appB} describes that theory in some detail, and provides a pedagogical introduction to the Fock quantization of linear coupled systems. The analysis presented in this section differs from  Appendix \ref{appB} only in the fact that we are dealing here with fields, and hence with infinitely many degrees of freedom. 

The quantum theory is constructed as follows: 

\begin{enumerate}
\item The first step  is to ``complexify''  $\mathbb{V}(\vec k)$, in the sense that we must  extend the classical phase space to include arbitrary complex elements $v(\vec{k})$, and not only those satisfying the ``reality condition'' $\bar v(\vec{k})=v(-\vec{k})$. We call this larger phase space ${\mathbb{V}}_{\mathbb{C}}(\vec k)$. 

\item

The symplectic structure of the classical theory can be used to define  a natural Hermitian ``product" in ${\mathbb{V}}_{\mathbb{C}}(\vec k)$. Given any two elements  $v^{(1)}(\vec k)$ and $v^{(2)}(\vec k)$ in ${\mathbb{V}}_{\mathbb{C}}(\vec k)$, this product is
\be\label{prodFour}  \langle v^{(1)}(\vec k), v^{(2)}(\vec k)\rangle= i \, \mathcal{V}_0\, \sum_{s=0,\pm2}\,  \left(\bar \Gamma^{(1)}_{s}(\vec k)\,   \Pi^{(2)}_{ s} (\vec k)-\bar \Pi^{(1)}_{ s}(\vec k)\,  \Gamma^{(2)}_{ s} (\vec k) \right)\, .
 \ee
It satisfies all properties of a Hermitian inner product, except that it is not positive definite. 
 
\item The next step  is to choose a three-dimensional subspace of ${\mathbb{V}}_{\mathbb{C}}(\vec k)$ on which the product $\langle \cdot, \cdot\rangle$ is positive definite. We will denote it by  ${\mathbb{V}}_{\mathbb{C}}^+(\vec k)$. The properties of   $\langle\cdot , \cdot\rangle$ guarantee then that it is   negative definite on the complex conjugated subspace $\overline{\mathbb{V}}_{\mathbb{C}}^{+}(\vec k)$, and furthermore,  both subspaces are orthogonal to each other, and their sum equals  ${\mathbb{V}}_{\mathbb{C}}(\vec k)$. This means that
$$\mathbb{V}_{\mathbb{C}}(\vec k)={\mathbb{V}}_{\mathbb{C}}^+(\vec k)\oplus \overline {\mathbb{V}}_{\mathbb{C}}^{+}(\vec k)\, .$$
A choice of ${\mathbb{V}}_{\mathbb{C}}^+(\vec k)$ provides therefore a decomposition of ${\mathbb{V}}_{\mathbb{C}}(\vec k)$ in subspaces of positive and negative norm, with respect to (\ref{prodFour}). This decomposition is precisely the extra ingredient that one  needs in order to  quantize the classical theory. But note also that such decomposition is highly nonunique. There are infinitely many different choices of ${\mathbb{V}}_{\mathbb{C}}^+(\vec k)$ (see footnote \ref{Mink}). If the spacetime geometry has a timelike Killing vector field, like in flat spacetimes, a preferred choice of  ${\mathbb{V}}_{\mathbb{C}}^{+}(\vec k)$ is available, which corresponds to the familiar positive-frequency subspace. Such preferred structure is however absent in the  Bianchi I geometries under consideration (as it  is also absent in FLRW), and one needs to make a choice. The construction below---in particular the quantum state that we will call the Fock vacuum---depends on this choice. Now, the space ${\mathbb{V}}_{\mathbb{C}}^+(\vec k)$  equipped with the product $\langle \cdot,\cdot \rangle$ forms a three-dimensional Hilbert space $\mathfrak{h}(\vec k)$. The  (Cauchy completion of the)  sum for all $\vec k$, $\mathfrak{h}\equiv \oplus_{\vec k} \, \mathfrak{h}(\vec k)$, is known as the one-particle Hilbert space of the field theory. The Fock space is constructed by summing symmetric products of $\mathfrak{h}$ in the standard way (see e.g. Appendix A of \cite{waldbook} for a summary of this construction).

\item Next, we need a choice of  three basis vectors in  ${\mathbb{V}}_{\mathbb{C}}^+(\vec k)$, that we will denote by bold letters, ${\boldsymbol v}^{(\lambda)}(\vec k)$, where the index $\lambda=1,2,3$ labels each basis element. Together with  their conjugates $\bar{\boldsymbol  v}^{(\lambda)}(\vec k)$, they form a complete basis in ${\mathbb{V}}_{\mathbb{C}}(\vec k)$. One can intuitively think of ${\boldsymbol v}^{(\lambda)}(\vec k)$ as a generalization of  the ``normal modes" of the system. It is convenient for the calculations below to choose these vectors to be orthonormal. The orthonormality relations are
\bea \label{orth} \langle {\boldsymbol v}^{(\lambda)}(\vec k),\boldsymbol v^{(\lambda')}(\vec k)\rangle&=&\delta^{\lambda\lambda'}\, , \nonumber \\ \langle {\boldsymbol v}^{(\lambda)}(\vec k),\bar{\boldsymbol v}^{(\lambda')}(\vec k)\rangle&=&0\, . \ea 
Furthermore, one needs to impose  these additional conditions on the basis vectors 
\be \label{exctcond} {\cal V}_0\,\sum_{\lambda=1}^3 \left(\boldsymbol v^{(\lambda)}_{S}(\vec k)\bar{\boldsymbol v}^{(\lambda)}_{S'}(\vec k)-\bar{\boldsymbol v}^{(\lambda)}_{S}(-\vec k)\boldsymbol v^{(\lambda)}_{S'}(-\vec k)\right)=i\, \Omega_{SS'}\, , \ee where 
\be \Omega_{SS'}=\left( {\begin{array}{cc} 0& \mathbb{I}_{3\times3}   \\ -\mathbb{I}_{3\times3}& 0 \end{array} } \right)\, ,\ee
to ensure that the  canonical commutation relations  of fields and momenta  can be derived from  the algebra of creation and annihilation operators. Or in other words, to ensure that Eqs. (\ref{fopk})  provides an admissible  representation  of the field and momentum operators  in the Fock space.

\item We define now creation and annihilation operators. First, we will use the symbol $\hat V(\vec k)$ to encode all field and momentum operators in Fourier space. More explicitly, $\hat V (\vec k)=(\hat\Gamma_{0}(\vec k),\, \hat \Gamma_{+2}(\vec k),\, \hat\Gamma_{-2}(\vec k), \, \hat\Pi_{0}(\vec k), \, \hat\Pi_{+2}(\vec k), \, \hat\Pi_{-2}(\vec k))$. Each component of $\hat V(\vec k)$ will be denoted by $\hat V_S(\vec k)$, with $S$ running from 0 to 5. Now,  given a choice of positive-norm subspace  ${\mathbb V}_{\mathbb{C}}^+$ and a set ${\boldsymbol v}^{(\lambda)}(\vec k)$ of three basis vectors on it, the annihilation operators are defined as the ``projection" of the field operator on these basis elements
\be \label{creatopt} \hat a_{\lambda}(\vec k)\,\equiv \,\langle {\boldsymbol v}^{(\lambda)}(\vec k),\hat V(\vec k) \rangle \,  .\ee
The creation operators are obtained by Hermitian conjugation. The canonical commutation relations
\be
  [\hat V_S(\vec k),\hat V_{S'}(\vec k')]\,=\,i\, {\cal V}_0^{-1} \,\delta_{\vec k, -\vec{k}'}\, \Omega_{SS'},
\ee
then imply
\be [\hat a_{\lambda}(\vec k),\hat a_{\lambda'}(\vec k')]\,=\,0 \, ; \ \ \  [\hat a_{\lambda}(\vec k),\hat a_{\lambda'}^{\dagger}(\vec k')]\,=\,\delta_{\lambda\lambda'}\,\delta_{\vec k, \vec{k}'}\, , \ee
and vice versa. 

\item The Fock vacuum  is now defined as the  (normalized) state $|0\rangle$ that is annihilated by $\hat a_{\lambda}(\vec k)$ for all values of $\lambda$ and $\vec k$. It is  obvious  that, since the definition of $\hat a_{\lambda}(\vec k)$ rests on a choice of positive-norm subspace ${\mathbb{V}}_{\mathbb{C}}^+$, the notion of Fock vacuum depends  also  on that choice. %(All these vacua  are  invariant under spatial translations). As mentioned before, in situations in which the metric and the Hamiltonian are time dependent, there is no a preferred choice of vacuum. 

It     is straightforward to check that this construction guarantees that the vacuum state is  invariant under translations. The other isometry of the Bianchi I metric is parity, and it is natural to demand the vacuum to be parity invariant too. This will be the case if the one-particle Hilbert space $\mathfrak{h}$ remains invariant under parity.  This  can be translated to a condition on the basis vectors, as follows.  Under parity, the basis vectors transform as 
\be \label{paritytransf} {\boldsymbol  v}^{(\lambda)}(\vec k)= \left( {\begin{array}{c} {\boldsymbol  v}_0^{(\lambda)}(\vec k)  \\ {\boldsymbol  v}_1^{(\lambda)}(\vec k) \\ {\boldsymbol  v}_2^{(\lambda)}(\vec k) \\ {\boldsymbol  v}_3^{(\lambda)}(\vec k)  \\ {\boldsymbol  v}_4^{(\lambda)}(\vec k) \\ {\boldsymbol  v}_5^{(\lambda)}(\vec k) \end{array} } \right) \longrightarrow P[{\boldsymbol  v}^{(\lambda)}(\vec k)]=\left( {\begin{array}{c} {\boldsymbol  v}_0^{(\lambda)}(-\vec k)  \\ {\boldsymbol  v}_2^{(\lambda)}(-\vec k) \\ {\boldsymbol  v}_1^{(\lambda)}(-\vec k) \\ {\boldsymbol  v}_3^{(\lambda)}(-\vec k)  \\ {\boldsymbol  v}_5^{(\lambda)}(-\vec k) \\ {\boldsymbol  v}_4^{(\lambda)}(-\vec k) \end{array} } \right)\, , \ \ \ \lambda=1,2,3\, . \ee
Note that the components 1 and 2, as well as 4 and 5, have been interchanged in the  right-hand side---this is because parity interchanges  $\Gamma_{+2}$ and  $\Gamma_{-2}$. The vacuum state will be invariant under parity if $P[{\boldsymbol  v}^{(\lambda)}(\vec k)]$ remains within ${\mathbb{V}}_{\mathbb{C}}^+(\vec k)$, i.e. if $P[{\boldsymbol  v}^{(\lambda)}(\vec k)]$  has no component   on the negative-norm subspace $\overline {\mathbb{V}}_{\mathbb{C}}^+(\vec k)$. Or more explicitly, if $ P[{\boldsymbol  v}^{(\lambda)}(\vec k)]$ can be written as\footnote{This expression can also be derived by studying the effect of a parity transformation on the metric perturbations $\delta h_{ij}(\vec x)$ in position space.}
\be \label{parityvacuum}  P[{\boldsymbol  v}^{(\lambda)}(\vec k)]=\sum_{\lambda'} \alpha^{\lambda \lambda'} \,{\boldsymbol  v}^{(\lambda')}(\vec k)\, ,\ee
for some complex numbers $ \alpha^{\lambda \lambda'}$, satisfying $\sum_{\lambda''} \alpha^{\lambda \lambda''} \bar{\alpha}^{\lambda' \lambda''} =\delta^{\lambda \lambda'}$ (so the norm of  $P[{\boldsymbol  v}^{(\lambda)}(\vec k)]$ remains the same).  Condition (\ref{parityvacuum}) suffices to make all the two-point correlation functions defined below invariant under parity.

\item The field and momentum operators in Fourier space are represented  in the Fock space as
\be \label{fopk} \hat V_{S} (\vec k)=\sum_{\lambda} \,\left[{\boldsymbol v^{(\lambda)}_{S}}(\vec k)\, \hat a_{\lambda}(\vec k)\, +\bar {\boldsymbol v}^{(\lambda)}_{S}(-\vec k)\, \hat a^{\dagger}_{\lambda}(-\vec k)\right]\, . \ee 
Note that these operators trivially satisfy the ``reality condition'' $ \hat V^{\dagger}_{S} (\vec k)= \hat V_{S} (-\vec k)$.  From these expressions, we can easily compute the two-point correlation functions, and the result is
\be \label{psa} \langle 0|\{ \hat V_{S} (\vec k) , \hat V_{S'} (\vec k')\}|0\rangle={\mathcal V}_0^{-1}\ \frac{2\pi^2}{k^3}\, 2\,\mathcal{P}_{SS'}(\vec k) \, \delta_{\vec k,-\vec k'}\, , \ee
where $\mathcal{P}_{SS'}(\vec k)$ are  known as the  power spectra, and in terms of the basis vectors they read
\be \label{PSS} \mathcal{P}_{SS'}(\vec k)={\mathcal V}_0\, \frac{k^3}{2\pi^2} \, \sum_{\lambda}\,\frac{1}{2} \left[{\boldsymbol v}^{(\lambda)}_S(\vec k)\,\bar {\boldsymbol v}^{(\lambda)}_{S'}(\vec k)+\bar {\boldsymbol v}^{(\lambda)}_{S}(-\vec k)\, {\boldsymbol v}^{(\lambda)}_{S'}(-\vec k)\right] \, . \ee
The  brackets in (\ref{psa}) indicate  anticommutator $\{\hat V_{S} (\vec k),\hat V_{S'} (\vec k')\}\equiv \hat V_{S} (\vec k)\hat V_{S'} (\vec k')+\hat V_{S'} (\vec k')\hat V_{S} (\vec k)$, and we have focused only on the symmetric part of $ \langle 0|\hat V_{S} (\vec k)\hat V_{S'} (\vec k') |0\rangle$ because the antisymmetric part (the expectation value of the commutator) is state independent and completely determined by the canonical commutation relations. Note also that for all $S$ and $S'$, we have $\mathcal{P}_{SS'}(\vec k)=\mathcal{P}_{S'S}(-\vec k)$.
Equation   (\ref{psa})   defines the power spectra for all couples of field and/or momentum operators. In cosmology, we are interested in the spectra involving field operators alone, $\mathcal{P}_{s s'}(\vec k)$ with $s,s'=0,\pm2$, since this is what we can extract from observations of the CMB. So from now on we will focus on them. We now describe the most relevant properties of these spectra:

\begin{enumerate}[(i)]

\item For fields alone (and also for momenta alone) the two terms inside the square brackets in (\ref{PSS}) are equal to each other. This can be seen directly from (\ref{exctcond}), and it is a consequence of the fact that field operators commute among themselves. Then, the expression for $ \mathcal{P}_{ss'}(\vec k)$ reduces to
\be \label{Pssp} \mathcal{P}_{ss'}(\vec k)={\mathcal V}_0\, \frac{k^3}{2\pi^2} \, \sum_{\lambda}\, \left[{\boldsymbol v}^{(\lambda)}_s(\vec k)\,\bar {\boldsymbol v}^{(\lambda)}_{s'}(\vec k)\right]\, .\ee
\item  $\mathcal{P}_{ss'}(\vec k)$ is real and positive for $s=s'$, but  it can be  complex for $s\neq s'$, as it can be seen directly from (\ref{Pssp}).

\item $\mathcal{P}_{ss'}(\vec k)= \mathcal{P}_{s's}(-\vec k)$, for all $s$ and $s'$, as a consequence of the commutation relations of field operators.
\item $\overline{\mathcal{P}}_{ss'}(\vec k)= \mathcal{P}_{ss'}(-\vec k)$, for all $s$ and $s'$. This is a consequence of the reality condition satisfied by the fields, $\hat{\Gamma}^{\dagger}_s(\vec k)=\hat{{\Gamma}}_s(-\vec k)$. This implies that the real part of ${\mathcal{P}}_{ss'}(\vec k)$ remains invariant under inversion $\vec k \to -\vec k$ (do not confuse this operation with a parity transformation that also changes $s\to -s$; see below), while the imaginary part changes sign.  

\item Parity: because the fields $\hat \Gamma_s(\vec k)$ transform into $\hat  \Gamma_{-s}(-\vec k)$ under parity, we find that a parity transformation sends  $\mathcal{P}_{ss'}(\vec k)$ to  $\mathcal{P}_{-s-s'}(-\vec k)$. It is direct to check that condition (\ref{parityvacuum}) on the basis vectors guarantees that $\mathcal{P}_{ss'}(\vec k)=\mathcal{P}_{-s-s'}(-\vec k)$ for all $s$ and $s'$, i.e.\ all  spectra $\mathcal{P}_{ss'}(\vec k)$ are parity invariant.\footnote{In fact, it is straightforward to  check that condition (\ref{parityvacuum}) makes {\em all} power spectra $\mathcal{P}_{SS'}(\vec k)$ parity invariant, and not only those involving  field operators but no momenta. Since in a free theory the vacuum is completely characterized by the two-point functions $\langle 0|\{ \hat V_{S} (\vec k)\hat V_{S'} (\vec k')\}|0\rangle$, this proves that the vacuum state is invariant under parity.} Furthermore,  together with the  property (iii) this implies $\mathcal{P}_{ss'}(\vec k)=\mathcal{P}_{-s'-s}(\vec k)$, and in particular  $\mathcal{P}_{+2+2}(\vec k)=\mathcal{P}_{-2-2}(\vec k)$.

\item Rotations: because $\Gamma_s(\vec k)$ transform as fields of spin weight $s=0,\pm 2$ under rotations around ${\vec k}$, the power spectra $\mathcal{P}_{ss'}(\vec k)$ have spin weight $s-s'$. It is important to keep this in mind when expanding $\mathcal{P}_{ss'}(\vec k)$ in angular multipoles, because such expansion must be done using spin-weighted  spherical harmonics:
\be \mathcal{P}_{ss'}(\vec k)=\sum_{L=|s-s'|}^{\infty}\sum_{M=-L}^L \,   \mathcal{P}^{LM}_{ss'}(k)\ _{s-s'}Y_{LM}(\hat k) \, , \ee
where $ _{s-s'}Y_{LM}(\hat k)$ are  spherical harmonics with spin weight $s-s'$, normalized such that $\int d\Omega_{\h k} \, _{s}\bar Y_{L M}(\hat{k}){}_{s}Y_{L' M'}(\hat{k})=
\delta_{LL'}\delta_{MM'}$. Recall that $ _{s-s'}Y_{LM}(\hat k)$ are  zero for $L<|s-s'|$. This in turn implies that the isotropic  (i.e.\ $L=0$) part of $\mathcal{P}_{ss'}(\vec k)$  vanishes unless $s-s'=0$, and hence only $\mathcal{P}_{00}$, and $P_{+2+2}=P_{-2-2}$ can be  different from zero in the limit in which both the spacetime and the quantum  state of perturbations are isotropic. 

\end{enumerate}

In early-universe cosmology we are interested in the primordial power spectra {\em evaluated at the end of inflation}. Hence, we are ultimately interested in computing the time evolution of  $\mathcal{P}_{ss'}(\vec k)$, starting from some initial time and ending at the end of inflation.\footnote{In Ref.\ \cite{aos2} we provide a detailed analysis of the relation between the primordial power spectra $\mathcal{P}_{ss'}(\vec k)$ and the angular correlation functions for temperature and polarization in the CMB.} This will be the goal of the next section. 
\end{enumerate}

We close this section  by illustrating the construction explained above with a simple example. For the subspace of positive norm  ${\mathbb V}^+_{\mathbb{C}}(\vec k)$, we choose the space spanned by the three vectors
\bea \label{exp} {\boldsymbol v}^{(1)}(\vec k)&=&\sqrt{\frac{4\, \kappa }{ a^2\, \mathcal{V}_0}} \,\left(\frac{1}{\sqrt{2\, k}},0,0;\frac{a^2}{4\kappa}\,\frac{ -i\, k}{\sqrt{2\, k}},0,0\right) \, \nonumber \\
{\boldsymbol v}^{(2)}(\vec k)&=&\sqrt{\frac{4\, \kappa }{ a^2\, \mathcal{V}_0}} \,\left(0, \frac{1}{\sqrt{2\, k}} ,0;0,\frac{a^2}{4\kappa}\,\frac{ -i\, k}{\sqrt{2\, k}},0\right)\, \nonumber \\
{\boldsymbol v}^{(3)}(\vec k)&=&\sqrt{\frac{4\, \kappa }{ a^2\, \mathcal{V}_0}}  \, \left(0,0,\frac{1}{\sqrt{2\, k}};0,0,\frac{a^2}{4\kappa}\,\frac{ -i\, k}{\sqrt{2\, k}}\right) \, ,
\ea
where $k$ is the comoving wave number. It is straightforward to check that these elements satisfy the conditions (\ref{orth}) and (\ref{exctcond}), as well as (\ref{parityvacuum}).\footnote{Under parity,  $P[{\boldsymbol  v}^{(1)}(\vec k)]={\boldsymbol  v}^{(1)}(\vec k)$, $P[{\boldsymbol  v}^{(2)}(\vec k)]={\boldsymbol  v}^{(3)}(\vec k)$, $P[{\boldsymbol  v}^{(3)}(\vec k)]={\boldsymbol  v}^{(2)}(\vec k)$. Hence (\ref{parityvacuum}) is satisfied.} In the classical theory,  each element ${\boldsymbol v}^{(\lambda)}(\vec k)$ of this basis represents a complex classical state where only one of the couples $ (\Gamma_{s}(\vec k),\Pi_{s}(\vec k))$ is initially displaced from equilibrium.

Using (\ref{creatopt}), we obtain that the annihilation operators associated with this choice are
\be \hat a_{1}(\vec k)=\sqrt{\frac{a^2\, \mathcal{V}_0}{8\, \kappa}} \left(\sqrt{k}\, \hat{ \Gamma}_{0} (\vec k)+i\, \frac{4 \kappa}{a^2}\frac{1}{\sqrt{k}}\, \hat{ \Pi}_{0}(\vec k)\right)\, , \ee
\be \hat a_{2}(\vec k)=\sqrt{\frac{a^2\, \mathcal{V}_0}{8\, \kappa}} \left(\sqrt{k}\, \hat{ \Gamma}_{+2} (\vec k)+i\, \frac{4 \kappa}{a^2}\frac{1}{\sqrt{k}}\, \hat{ \Pi}_{+2}(\vec k)\right)\, , \ee
\be \hat a_{3}(\vec k)=\sqrt{\frac{a^2\, \mathcal{V}_0}{8\, \kappa}} \left(\sqrt{k}\, \hat{ \Gamma}_{-2} (\vec k)+i\, \frac{4 \kappa}{a^2}\frac{1}{\sqrt{k}}\, \hat{ \Pi}_{-2}(\vec k)\right)\, . \ee
We can see that $\hat a_{1}(\vec k)$ and $ \hat a^{\dagger}_{1}(\vec k)$ respectively  annihilate and create quanta associated with the field  $ \hat \Gamma_{0}(\vec k)$ and do not modify the quantum state associated with the degrees of freedom of $ \hat \Gamma_{\pm2}(\vec k)$, and vice versa. This  also implies that the vacuum state can be expressed as the tensor product $|0\rangle_0\otimes|0\rangle_{+2}\otimes |0\rangle_{-2}$ of the vacuum of each degree of freedom (recall that this is the state at time $t_0$; time evolution will be described in the next section).

From (\ref{fopk}), we obtain that the field operators in Fourier space at the initial time take the form 
\be \hat \Gamma_{0}(\vec k) = \sqrt{\frac{4\kappa }{a^2\, \mathcal{V}_0}} \, \frac{1}{\sqrt{2\, k}} \left(  \hat a_{1}(\vec k)+    \hat a^{\dagger}_{1}(-\vec k)\right)\, \, , \ee
\be \hat \Gamma_{+2}(\vec k) = \sqrt{\frac{4\kappa }{a^2\, \mathcal{V}_0}} \, \frac{1}{\sqrt{2\, k}} \left(  \hat a_{2}(\vec k)+    \hat a^{\dagger}_{2}(-\vec k)\right)\, \, , \ee
\be \hat \Gamma_{-2}(\vec k) = \sqrt{\frac{4\kappa }{a^2\, \mathcal{V}_0}} \, \frac{1}{\sqrt{2\, k}} \left(  \hat a_{3}(\vec k)+    \hat a^{\dagger}_{3}(-\vec k)\right)\, \, . \ee
and the momentum operators
\be \hat \Pi_{0} (\vec k)=-i\, a\sqrt{\frac{k}{8\kappa\mathcal{V}_0}} \left(  \hat a_{1}(\vec k)-    \hat a^{\dagger}_{1}(-\vec k)\right)\,  , \ee
\be \hat \Pi_{+2} (\vec k)=-i\,a\sqrt{\frac{k}{8\kappa\mathcal{V}_0}} \, \left(  \hat a_{2}(\vec k)-    \hat a^{\dagger}_{2}(-\vec k)\right)\,  , \ee
\be \hat \Pi_{-2} (\vec k)=-i\, a\sqrt{\frac{k}{8\kappa\mathcal{V}_0}}\left(  \hat a_{3}(\vec k)-    \hat a^{\dagger}_{3}(-\vec k)\right)\,  . \ee
The power spectra [for  field operators $\hat \Gamma_{s}(\vec k)$ only] are
\be  \mathcal{P}_{s s'}=\, \hbar \, \kappa\, \frac{k^2}{a^2\pi^2} \, \delta_{s,s'}\, .\ee
In this last expression we have restored $\hbar$ in order to show explicitly the  quantum nature of $\mathcal{P}_{s s'}$. Note also that the fiducial volume ${\mathcal V}_0$ introduced in our calculations does not  appear in these physical observables. The presence of the Kronecker delta reveals the absence of correlations at the initial time between $\hat \Gamma_{0}$, $\hat \Gamma_{+2}$, and $\hat \Gamma_{-2}$  in the vacuum state  we have chosen. However, because these fields are coupled in the physical Hamiltonian, the time evolution  will generate such correlations. Therefore, at later times, we should expect nonvanishing off-diagonal components in $ \mathcal{P}_{s s'}$. This happens because, in general, the time evolution of any of the basis modes  ${\boldsymbol v}^{(\lambda)}(\vec k)$ will have nonzero values in all six components.

\section{Dynamics: $\mathcal{S}$-matrix and generation of entanglement}\label{sec.5}

Dynamics is simpler to write in the Heisenberg picture. The Heisenberg operators are  obtained  from (\ref{fopk})  simply by applying  time evolution  to each element $ {\boldsymbol v}^{(\lambda)}_S(\vec k)$  of the basis functions, namely
\be\hat V_{S} (\vec k,t)=\sum_{\lambda=1}^3 \left[ {\boldsymbol v}^{(\lambda)}_S(\vec k,t)\, \hat a_{\lambda}(\vec k)\, +{\bar {\boldsymbol v}}^{(\lambda)}_S(-\vec k,t)\, \hat a^{\dagger}_{\lambda}(-\vec k)\right] \, , \ee
where ${\boldsymbol v}^{(\lambda)}_S(\vec k,t)$ denotes the solution to the classical Hamilton's equation with initial data  ${\boldsymbol v}^{(\lambda)}_S(\vec k)$. With this, the power spectra at any time are
\be \label{FPS} \mathcal{P}_{ss'}(\vec k,t)={\mathcal V}_0\, \frac{k^3}{2\pi^2} \,   \sum_{\lambda=1}^3\, \left[\boldsymbol{v}^{(\lambda)}_s(\vec k,t)\,\bar {\boldsymbol{v}}^{(\lambda)}_{s'}(\vec k,t)\right] \, , \ee
where again, we are focusing here on the power spectra of field operators and not momenta. This expression is exact, in the sense that it is not the result of any perturbative expansion in the shears $\sigma_i$. To evaluate the right-hand side, all we need is to solve the set of coupled, second-order ordinary differential equations (\ref{eqginper}) with appropriate initial data, a task that is always possible to do using numerical algorithms. 

It is interesting to study the evolution also in the Schr\"{o}dinger picture, since it illuminates complementary aspects of the dynamics, particularly regarding the generation of quantum entanglement between the different  perturbations.  In order to write the evolution operator that implements the dynamics, we first need to specify a final Fock space. It is common in this context to use the label $in$ for the initial vacuum and Fock space, and $out$ for the late time counterparts.

The time evolution operator is  a unitary map from the Fock space $\mathcal{F}_{in}$ to $\mathcal{F}_{out}$, known also as the $\mathcal{S}$-matrix, and denoted by $\mathcal{S}_{(in,out)}$ \cite{waldbook}. It is common to build $\mathcal{S}_{(in,out)}$ from the standard textbook expression in terms  of the time-ordered exponential of the Hamiltonian, $T\left[ \exp (-i/\hbar \int_{t_{in}}^{t_{out}}\hat H(t')\, dt')\right]$ and use it as the starting point for a perturbative expansion. However, it is more convenient to express $\mathcal{S}_{(in,out)}$ in terms of the so-called Bogoliubov coefficients $\alpha_{\lambda\lambda'}(k)$ and $ \beta_{\lambda \lambda'}(k)$ (see also Appendix \ref{appB}). If we denote $\boldsymbol{v}_{in}^{(\lambda)}(\vec{k},t)$ and $\boldsymbol{v}_{out}^{(\lambda)}(\vec{k},t)$ as the three orthonormal vectors that define the bases  defining the $in$ and $out$ vacua, respectively, these Bogoliubov coefficients are 
\be \alpha_{\lambda\lambda'}(\vec k):=\langle  \boldsymbol v_{out}^{(\lambda')}(\vec k,t_{out}) ,\boldsymbol v_{in}^{(\lambda)}(\vec k,t_{out})\rangle \, , \hspace{0.4cm} \beta_{\lambda\,\lambda'}(\vec k) :=-\langle \bar{\boldsymbol v}_{out}^{(\lambda')}(\vec k,t_{out}) ,\boldsymbol v_{in}^{(\lambda)}(\vec k,t_{out})\rangle\, ,
\ee
i.e.\ $\alpha_{\lambda \lambda'}$ and $\beta_{\lambda \lambda'}$ ``measure" the  positive- and negative-norm components of  the $in$ modes with respect to the $out$ basis, respectively. In terms of these coefficients, the $\mathcal{S}$-matrix takes the form of a generalized squeezing operator, and its action on the $in$ vacuum produces
\be \label{Sinout2} \mathcal{S}_{(in,out)}|in\rangle=  N\, \bigotimes_{\vec k} \exp{\Big[\sum_{\lambda,\lambda'=1}^3  {\rm V}_{\lambda\lambda'}(\vec k) \, \h  a^{out\, \dagger}_{\lambda}(\vec k)\  \h a^{out \, \dagger}_{\lambda'}(-\vec k)\Big]} \, |out \rangle \, ,\ee
where $N$ is a normalization factor and $V_{\lambda\lambda'}:=\sum_{\lambda''=1}^3 \frac{1}{2} \,\bar \beta_{\lambda'' \lambda}(\vec k)\, \bar\alpha^{-1}_{\lambda'\lambda''}(\vec k)$, where $\alpha^{-1}_{\lambda\lambda'}$ is the inverse of the matrix $\alpha_{\lambda\lambda'}$ (the properties of these coefficients ensure that $\alpha_{\lambda\lambda'}$ is invertible). 

One can  prove from the properties of  $\alpha_{\lambda\lambda'}$ and $\beta_{\lambda\lambda'}$ (see Appendix \ref{appB})   that $V_{\lambda\lambda'}$ is symmetric. Expression (\ref{Sinout2}) is commonly  interpreted by saying that the evolution of the state $|in\rangle$ from $t_{in}$ to $t_{out}$ results in ``the exponential of a two-particle state'' in  $\mathcal{F}_{out}$. More precisely, we can better understand this result by expanding the exponential (\ref{Sinout2}): 
\bea \label{explicit} \mathcal{S}_{(in,out)}|in\rangle&=&N \,\bigotimes_{\vec k} \Big[|out_{\vec k}\rangle+V_{11} \, |1_{\vec k}1_{-\vec k}\rangle_1|0\rangle_{2}|0\rangle_{3}+ V_{12} \,\Big( |1_{\vec k}\rangle_1|1_{-\vec k}\rangle_{2}|0\rangle_{3}+|1_{-\vec k}\rangle_1|1_{\vec k}\rangle_{2}|0\rangle_{3}\Big)\nonumber \\ &+& V_{13} \,\Big( |1_{\vec k}\rangle_1|0\rangle_{2}|1_{-\vec k}\rangle_{3}+|1_{-\vec k}\rangle_1|0\rangle_{2}|1_{\vec k}\rangle_{3}\Big)+\cdots\Big]\, , \ea
where states in the right-hand side belong to $\mathcal F_{out}$, and the subscript $\lambda=1,2,3$ in the quantum states indicates that they correspond to excitations created by $a^{out\, \dagger}_{\lambda}(\vec k)$ over the $out$ vacuum state $|out_{\vec k}\rangle=|0_{\vec k}\rangle_1|0_{\vec k}\rangle_{2}|0_{\vec k}\rangle_{3}$ for the Fourier mode $\vec k$ ($|out\rangle = \bigotimes_{\vec k} |out_{\vec k}\rangle$). We see from this expression that the result of the evolution is the product of linear combination of states containing $2^N$ particles, with $N\in \mathbb{N}$. Furthermore, some of these pairs are made of quanta associated with different degrees of freedom, and hence they show the existence of  quantum  entanglement in the final state. Note also that the entanglement only takes place between quanta with wave numbers $\vec{k}$ and $-\vec{k}$. This is a consequence of the homogeneity of the Bianchi I geometry, that implies momentum conservation. One can then interpret Eq. (\ref{explicit}) by saying that the evolution has created pairs of entangled quanta with opposite wave numbers.

The previous discussion is generic, in the sense that it is valid regardless of the  choice of basis vectors one uses to define the out-Fock space $\mathcal F_{out}$. But if $t_{out}$ is chosen to be the end of inflation, because at that time the Universe is isotropic, the natural choice  of $\mathcal{F}_{out}$ is the product of the Fock spaces for scalar and tensor perturbations constructed from the familiar Bunch-Davies vacua. With this choice, $a^{out\, \dagger}_{\lambda}$ with $\lambda=1,2,3$ creates quanta of the scalar, and tensor perturbations with helicity $+2$ and $-2$, respectively. The final state (\ref{explicit}) contains then correlations between scalar and tensor quanta. These are  the  same correlations described by  the power spectra ${\mathcal{P}}_{ss'}(\vec k)$. 

If the offdiagonal couplings ${\cal U}_{\mu\mu'}$ in the Hamiltonian (\ref{Ham2}) were zero, then the Bogoliubov coefficients, and consequently the matrix $V_{\lambda\lambda'}$,  would become diagonal. The action of the $\cal S$-matrix on the vacuum in that situation would then be 
\be  \label{eq:vac-fact}\mathcal{S}_{(in,out)}|in\rangle=  N\, \bigotimes_{\lambda=1}^3 \left( \bigotimes_{\vec k}  \exp{\Big[   V_{\lambda\lambda}(\vec k) \, \h  a^{out\, \dagger}_{\lambda}(\vec k)\  \h a^{out \, \dagger}_{\lambda}(-\vec k)\Big]} \, |out \rangle \right)\, . \ee
The right-hand side is  a product state that  contains no correlations or entanglement between different degrees of freedom.

The main take-home points of this analysis are twofold: (i) Anisotropies in the early Universe produce primordial spectra that are  in general anisotropic. This fact  manifests itself  in that  the spectra $\mathcal{P}_{ss'}(\vec k,t)$ depend on the direction of $\vec k$. (ii) Anisotropies generate quantum entanglement, or correlations, between scalar and tensor perturbations, as well as among the two tensor modes. As a consequence, either the nondiagonal spectra, $\mathcal{P}_{ss'}(\vec k,t)$ for $s\neq s'$ is nonzero, or, if we work in the Schr\"odinger picture, the form of the final state is the one given in (\ref{Sinout2}) rather than \eqref{eq:vac-fact}. [The existence of entanglement can also be evaluated by writing the density matrix associated with the final state and by computing the entanglement entropy between the degrees of freedom associated with the  three fields $\hat \Gamma_s$ (see Appendix \ref{appB}.)] It is also important to emphasize that these features are not necessarily washed out by the fact that the Universe isotropizes at late time. A large expansion will  certainly red-shift all wave numbers, including those containing  anisotropies and entanglement, and the question of whether they are observable in the CMB depends on the details of the model. In general, anisotropic  effects are expected to be  larger for the longest wavelengths we can observe. 

There exist however one difficulty that prevents us from making concrete predictions about the effects of anisotropies in the CMB, and it is the lack of a preferred initial state in Bianchi I spacetimes in classical general relativity. In the literature of quantum field theory in curved spacetimes, it is known that the notion of adiabatic vacuum can be used to provide a preferred choice of vacuum, at least for short distances or wavelengths,  relative  to the radius of curvature of the spacetime (which is proportional to the Hubble radius in most models). In isotropic FLRW spacetimes, the wavelength of any mode grows monotonically in time in an expanding universe. If there was a phase of inflation during which the Hubble  radius remained constant,  there is a time at which  the modes that we can probe in the CMB had all arbitrarily small wavelength. So for them  there exists a preferred initial state. This is not always true in  Bianchi I geometries, as pointed out in \cite{ppu-BI1,ppu-BI2}. There, even if the universe expands---in the sense that volume grows  in time and the mean Hubble rate is positive---directional Hubble rates can be negative, and hence wavelengths of modes pointing in such directions would decrease in time. This means that, in the  presence of  anisotropies, one cannot guarantee that all the modes that we observe in the CMB were in an adiabatic regime at some early time, and consequently there is no unambiguous way of defining an initial vacuum state. This is to say, the predictions for anisotropies are subject to the choice of initial state, and no universal statement can be made about the power spectra or any other observable quantity unless one introduces extra ingredients in the theory  to  single out a  preferred choice.  We show this fact explicitly in the next section.

\section{Example}\label{sec.6}

This section  illustrates the general analysis presented above with a concrete example. We consider a scenario for the early Universe in which the expansion is initially dominated by anisotropies, followed by a phase of slow-roll inflation.  We will follow the evolution of cosmic perturbations and compute the primordial power spectra of scalar and tensor perturbations. We first obtain the evolution of the Bianchi I geometry following Sec. \ref{sec.2}, and then we evolve perturbations thereon. 

\begin{enumerate}

\item{\bf Evolution of the background fields}

 As explained in Sec. \ref{sec.2}, we first obtain the evolution of the mean scale factor $a(t)$ and the scalar field $\phi(t)$. We consider initial data at a time $t_0=0$ given by $a(0) =1$, $H(0) =3{.}5\times 10^{-5}$, $\phi(0)=3{.}3$, and $\Sigma=7{.}67\times 10^{-5}$, all in Planck units. Then, the Hamiltonian constraint  (\ref{cons}) determines $\dot \phi(0)$ up to a sign, that we choose to be positive. For the scalar field potential $V(\phi)$ we use the simple quadratic form $V(\phi)=\frac{1}{2}m^2\phi^2$, with $m$ obtained from observations \cite{planck-inf}, $m=1.28\times 10^{-6}$, again in Planck units. We obtain the solution to Eqs. (\ref{aver}) with this initial data, and plot in Fig. \ref{fig:1} the time evolution of the kinetic and the potential energy of the scalar field $\phi(t)$, together with the evolution of the shear $\sigma^2(t)=\Sigma^2/a^6(t)$. These are the three terms in the right-hand side of the  Friedmann equation (\ref{cons}). We see in  Fig. \ref{fig:1} that the solution we have chosen is  dominated by the shear at early times. But the cosmic  expansion makes the shear lose relative relevance,  until  finally the potential energy dominates, the Universe enters in a phase of slow-roll inflation, and it quickly  isotropizes. On the other hand, if we evolve backwards in time, we find the big bang singularity at $t=-5.3\times 10^{3}$ Planck times.

\begin{figure}
\centering
\includegraphics[width=0.6\textwidth]{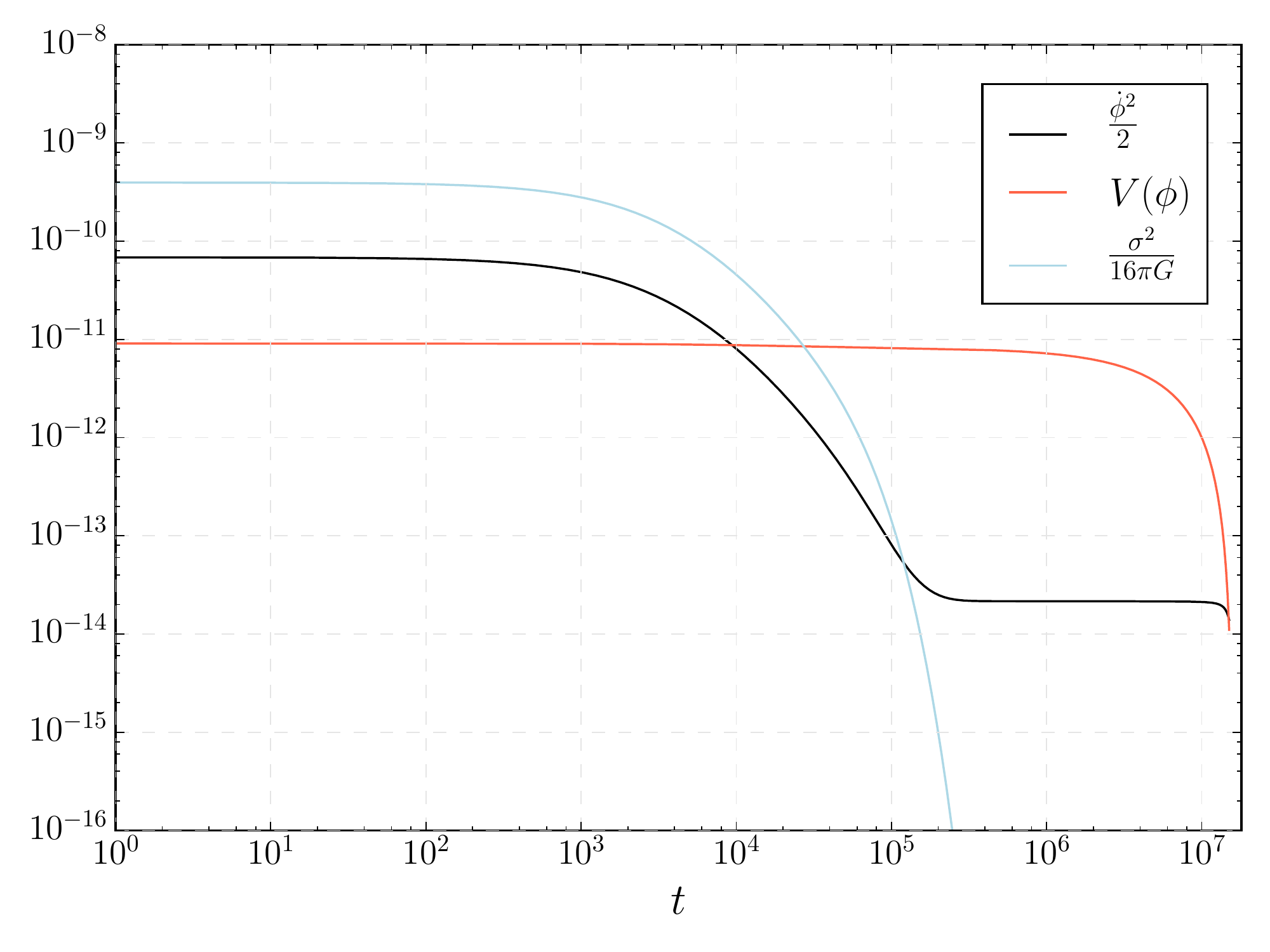} 
\caption{Evolution of the kinetic and potential energy densities of the scalar field $\phi(t)$, and  the shear $\sigma^2(t)$. The Universe is initially dominated by the shear. During the forward evolution $\sigma^2(t)$ falls off as $1/a^6(t)$ and the potential energy gains relative relevance until it dominates. At that time the Universe starts expanding in an accelerated way and inflation begins.}
\label{fig:1}
\end{figure}

Next,  the evolution of the  shears $\sigma_i(t)$ is given by Eqs. (\ref{shearsev}). To obtain the solution to these equations, we first need to specify  the value of the angle $\Psi$ that indicates the way the total shear $\sigma$ is distributed among the three principal directions. Notice that, since $\sigma_1+\sigma_2+\sigma_3=0$, the three components cannot have the same sign. We choose $\Psi=\pi/4$ in this example, and plot in Fig. \ref{fig:2} the evolution of the directional scale factors $a_i(t)$. We fix the freedom in the value of the directional scale factors by choosing $a_1(t_{\rm end})=a_2(t_{\rm end})=a_3(t_{\rm end})$, where $t_{\rm end}$ is the time when inflation ends. Hence the three scale factors $a_i(t)$ and their derivatives agree at late times, but they differ significantly in the  earliest stages of evolution. For our choice of  $\Psi$ the scale factor $a_2$ is initially contracting ($H_2<0$), while $a_1$ and $a_3$ are expanding. This implies that the wavelength of Fourier modes of perturbations with wave number $\vec k$ that point in the direction of $a_2$ will initially contract while the mean scale factor $a(t)$ expands. Therefore, these wavelengths grow when propagated  back in time, and they will not  generically find an adiabatic regime, no matter how far to the past we go  \cite{ppu-BI1,ppu-BI2}. As discussed before, the absence of an adiabatic regime for cosmological perturbations is a generic feature of anisotropic spacetimes. We illustrate below with a simple example that this fact translates into an ambiguity in the predictions for the primordial power spectra.

\begin{figure}
\centering
\includegraphics[width=0.6\textwidth]{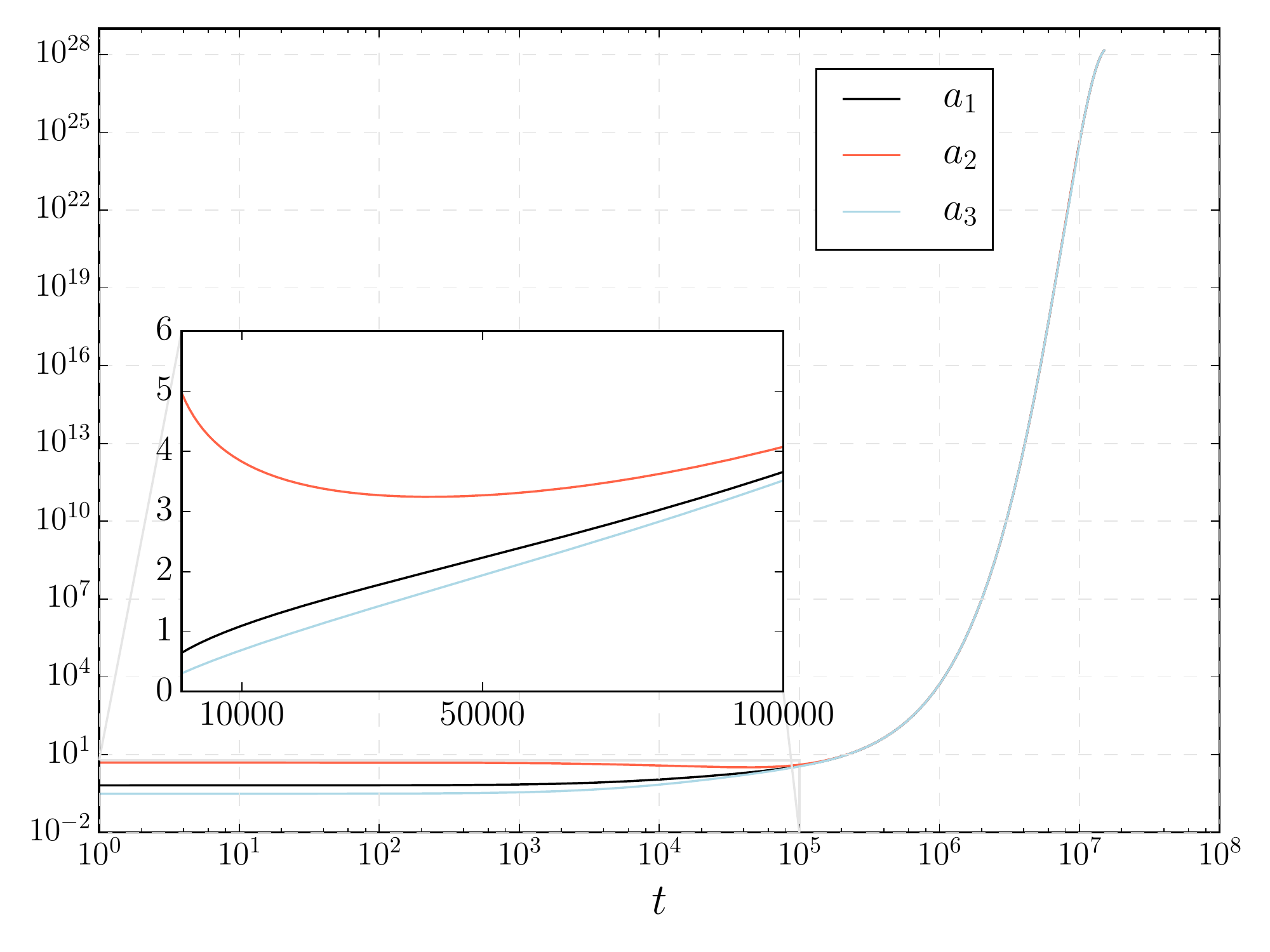} 
\caption{Evolution of the directional scale factors $a_i(t)$. At late times, when the Universe enters in a phase of accelerated expansion, the three $a_i(t)$ and their derivatives quickly approach each other (we have used the freedom in rescaling the coordinates to make the value of all $a_i$  equal at late times). At early times the three $a_i(t)$  are very different. In our example, the scale factor $a_2$ bounces when we go backwards in time, while $a_1$  and $a_3$ go to zero and reach the big bang singularity in a finite amount of proper time.}
\label{fig:2}
\end{figure}

\item {\bf Initial state for perturbations}

Let us start by thinking about states in the  Schr\"odinger  picture. For the initial state of perturbations  at $t_0$, we start by choosing the same one we used  in the example at the end of Sec. \ref{sec.4}, and that is specified in Eq. (\ref{exp}).    As explained there, since each of the three basis vectors ${\boldsymbol v}^{(\lambda)}(\vec k)$ only contain a nonzero entry in the ``direction'' of the field $\Gamma_{s}$, the vacuum state they define is the product  of a vacuum for each field, $|0\rangle_1\otimes|0\rangle_2\otimes |0\rangle_3$. It is obvious that this state  does not contain correlations between scalar and tensor modes. We call this state the  ``instantaneous Minkowski vacuum,'' because it corresponds to the state that one would choose  in  Minkowski spacetime. (In the terminology of adiabatic states \cite{parker-toms}, this is a zeroth-order adiabatic vacuum. It is also possible to build states of higher order in the adiabatic expansion, see e.g.\ \cite{ana,hyb-adiab}.)
As emphasized before, in Bianchi I spacetimes there is no sense in which this initial state is preferred with respect to any other. Therefore, the form of  the power spectra given below contains  information not only about the spacetime geometry on which perturbations propagate upon, but also about our choice of initial state.  To illustrate this point with a concrete example, we will also consider the same Schr\"odinger state  but now at a different time, more concretely 4500 Planck seconds before $t_0$. We call this vacuum state $\widetilde{|0\rangle}$. We will show below that the  power spectrum of ${|0\rangle}$ and $\widetilde{|0\rangle}$ at the end of inflation are quite different. Since there is no preferred time to specify the initial state, this simple example illustrates well the ambiguity in the physical predictions.

\item{\bf Evolution of  perturbations and observables}

We will discuss  here evolution in both the Heisenberg and Schr\"odinger pictures. In order to obtain the evolution of the operator fields $\hat \Gamma_{s}$ in the Heisenberg picture, all we need is  the time evolution of the  basis elements ${\boldsymbol v}^{(\lambda)}(\vec k)$, and to plug the result in  (\ref{FPS}). This requires us to solve the equations of motion (\ref{perteqH}) using (\ref{exp}) as initial data at $t_0$. At late times, the basis element ${\boldsymbol v}^{(\lambda)}(\vec k,t)$ will contain in general nonzero values in all six components. 

We compute the power spectra of the comoving curvature perturbation
\be
\hat{\mathcal{R}}(\v k) = \frac{1}{\sqrt{4\kappa}} \left(\frac{H}{\dot \phi}\right)\hat\Gamma_0(\v k)\, ,
\ee
 and the two tensor perturbations $\hat \Gamma_{\pm 2}$. Concretely, the power spectra involving the comoving curvature perturbations $\hat{\mathcal{R}}(\v k)$, are related to the spectra $\mathcal{P}_{ss'}$ defined above by 
 \be {\mathcal P}_{\mathcal{R}}(\v k) = \frac{1}{4\kappa} \left(\frac{H}{\dot \phi}\right)^2 \, {\mathcal P}_{00}(\v k) \, , \ \ \ \  {\rm and } \ \ \ \ {\mathcal P}_{\pm 2\mathcal{R}}(\v k) = \frac{1}{\sqrt{4\kappa}} \left(\frac{H}{\dot \phi}\right) \, {\mathcal P}_{\pm 20}(\v k) \, .\ee
Figure \ref{fig:3} shows the result for all the spectra. Since the direction dependence of power spectra is quantified better in the harmonic space, we have presented the results for the multipolar components ${\cal P}_{ss'}^{LM}$. These plots contain two main messages: (1) Power spectra are anisotropic, in the sense that they depend strongly on the direction of the wave number $\vec k$. (2) There exist significant cross-correlations between scalar and tensor modes, as well as between the two tensor modes, that fall off approximately as $1/k$. These two facts find their origin in the anisotropic phase of the Universe before the beginning of inflation, and make manifest that, even though the background spacetime isotropizes, perturbations maintain memory of that phase. More concretely, the effects of the anisotropic phase on the correlation functions are larger for infrared scales (large angular correlations). 
\begin{figure}[!]
\centering
\includegraphics[width=0.49\textwidth]{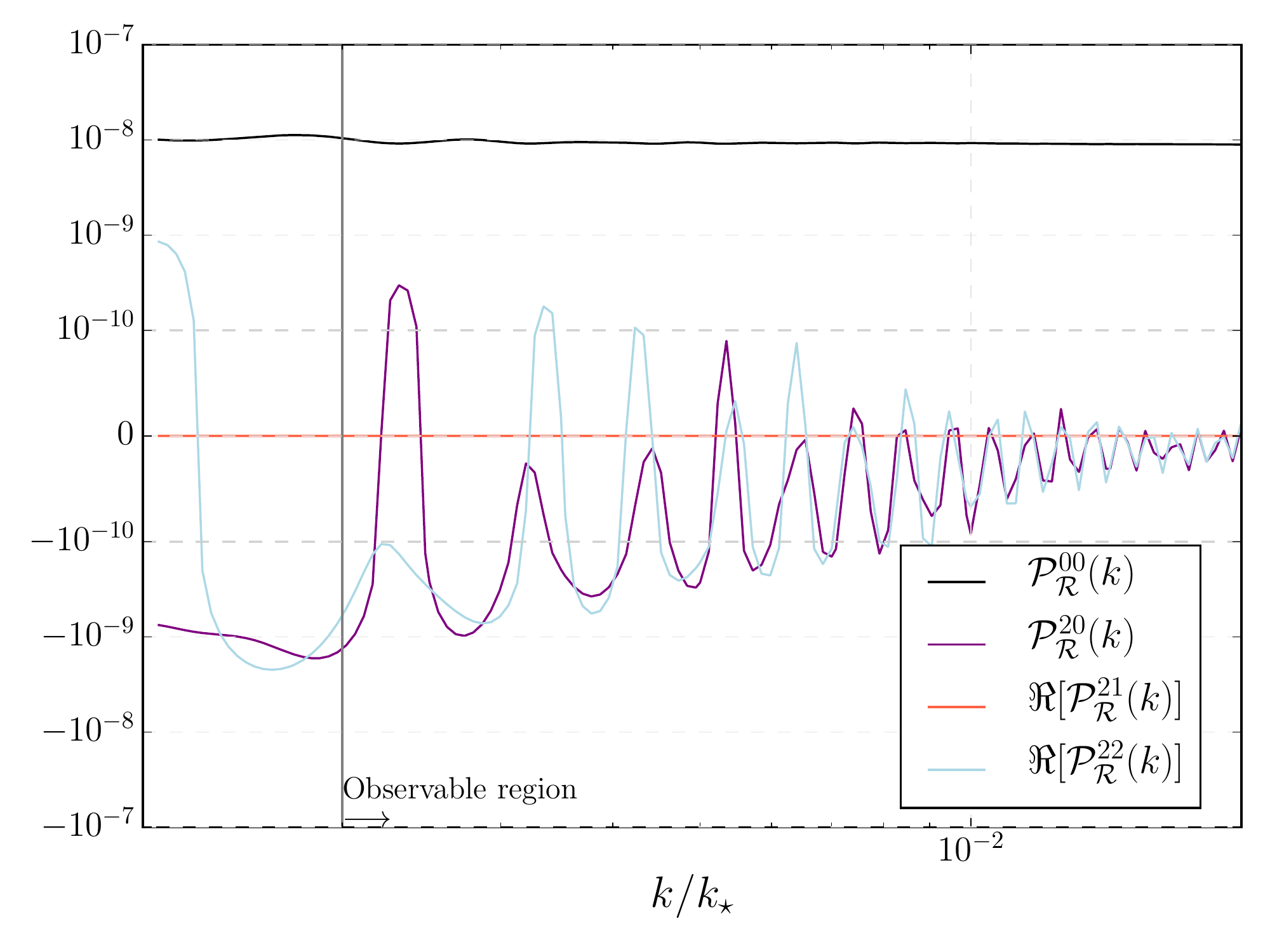}
\includegraphics[width=0.49\textwidth]{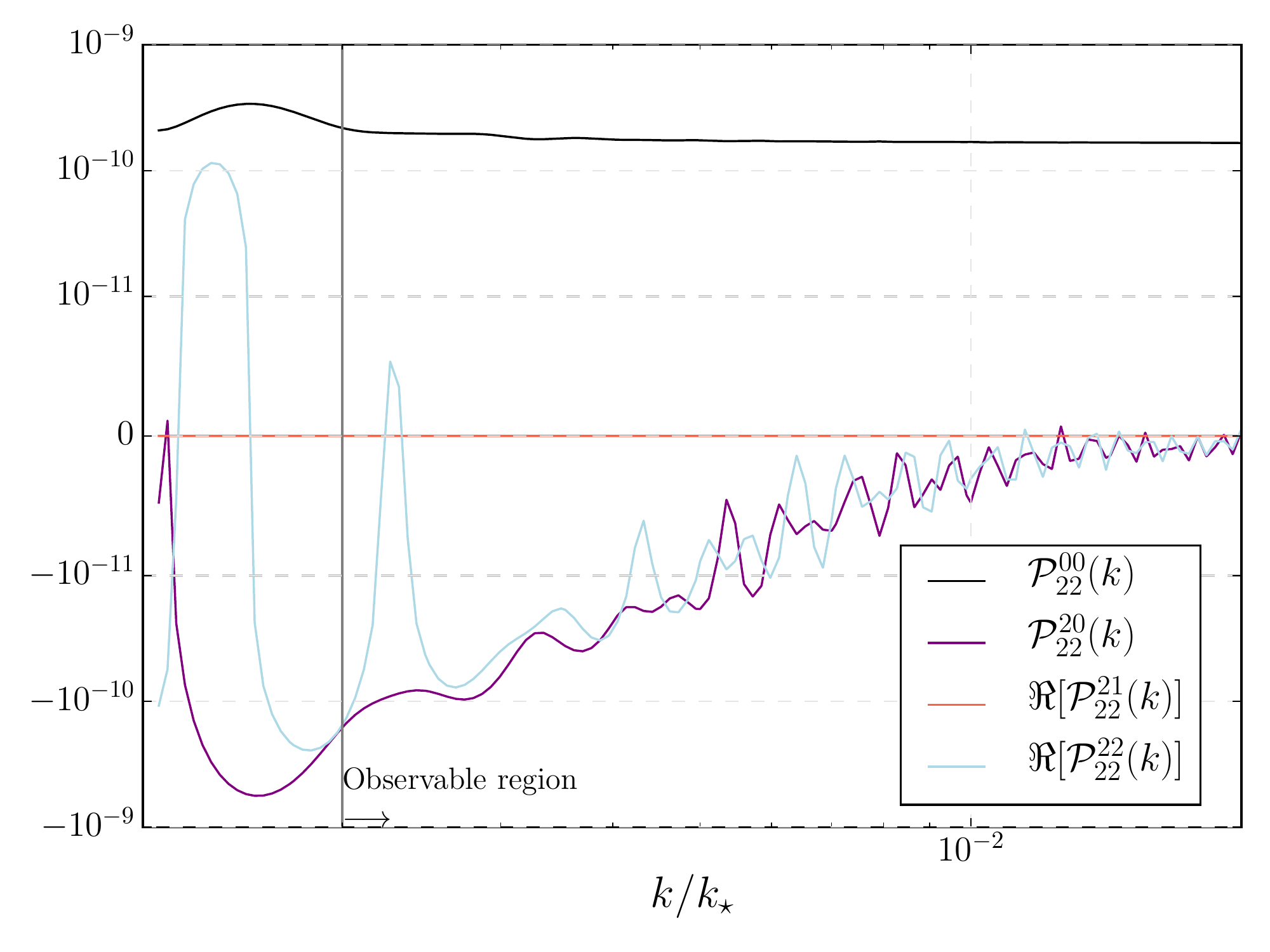}\\
\includegraphics[width=0.49\textwidth]{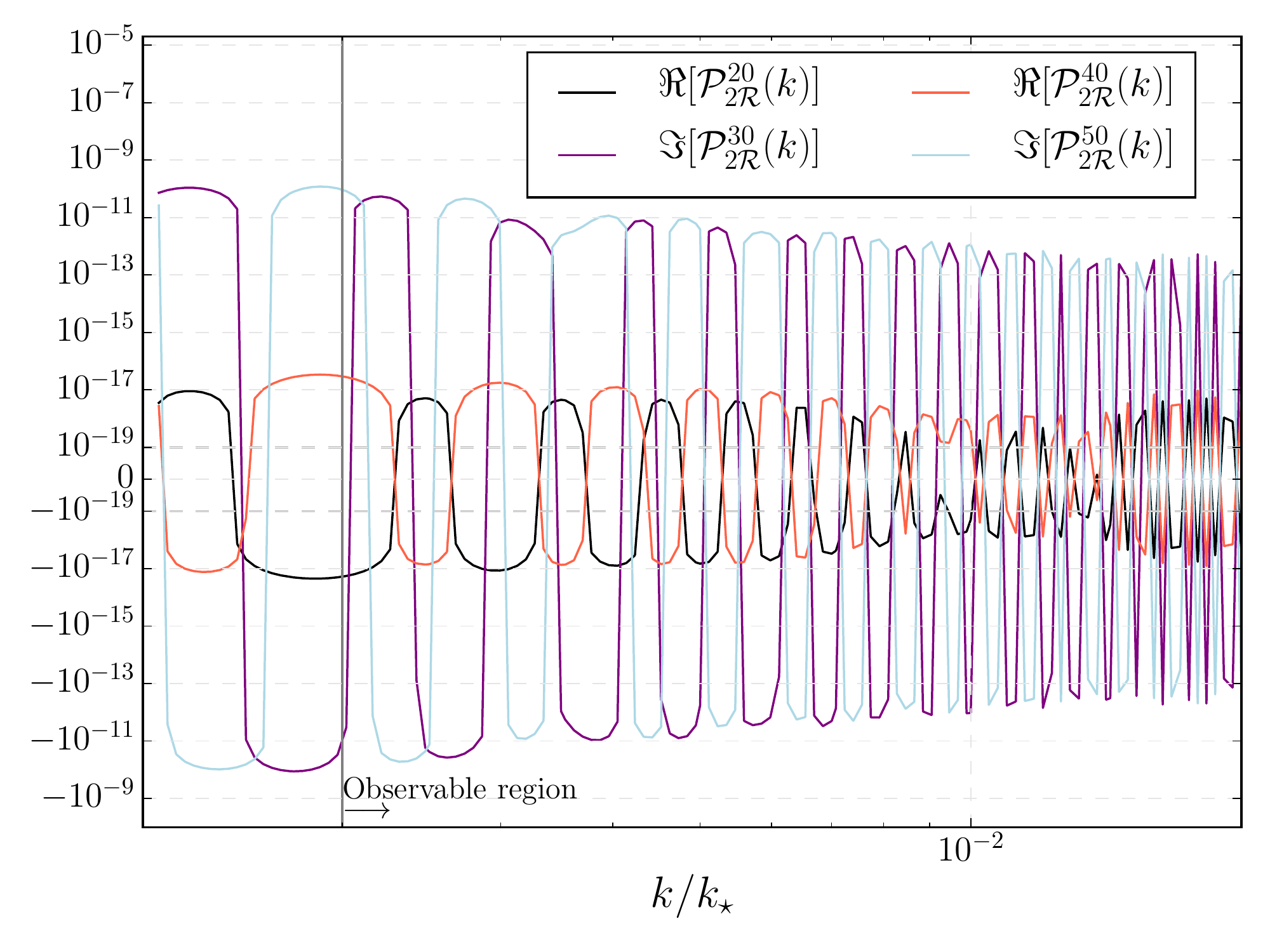}
\includegraphics[width=0.49\textwidth]{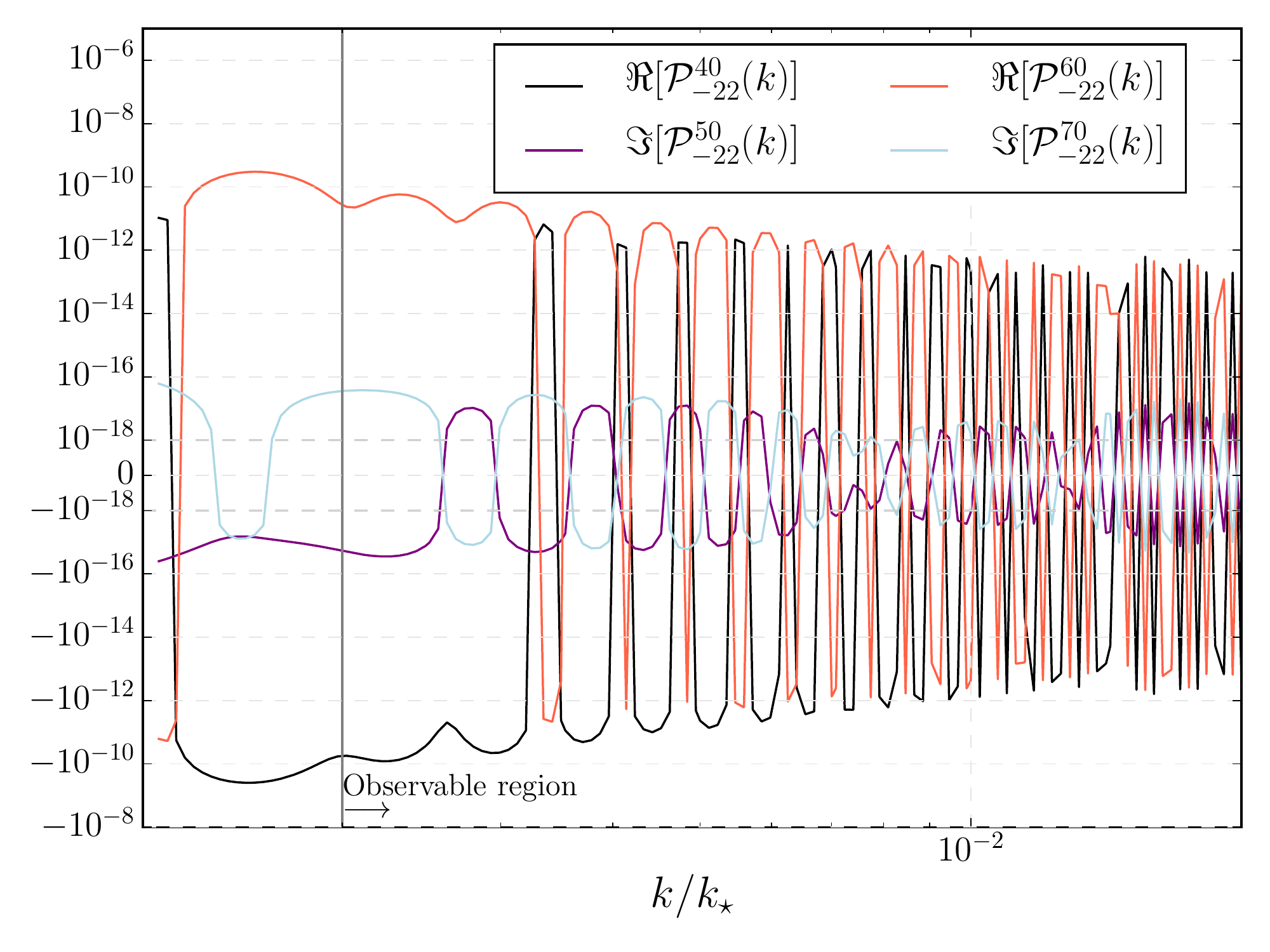} 
\caption{Multipoles $\mathcal{P}^{LM}_{ss'}(k)$ resulting from the decomposition of the primordial power spectra $\mathcal{P}_{ss'}(\vec k)$ in spin-weighted spherical harmonics.  Departure from isotropy is encoded in  multipoles with $L>0$. These anisotropic features are significantly larger for infrared scales. We recover nearly scale invariant and isotropic power spectra for large $k$. $k_{\star}$ is a reference scale, and it corresponds to a wave number whose physical value today is $0.05\, {\rm Mpc}^{-1}$.}
\label{fig:3}
\end{figure}

 However,  as advertised above, the results  in Fig. \ref{fig:3}  depend on the choice of vacuum state.  Let us consider  the vacuum state defined by the initial data for the basis modes (\ref{exp}), but now imposed at  $\tilde t_0=t_0-4.5\times 10^{3}$ Planck times, rather than $t_0$. The new initial time  $\tilde t_0$ is  far enough from the big bang singularity for the semiclassical approximation to be valid. Using the initial data  (\ref{exp}) at the new  initial time gives rise to different basis functions $\tilde {\boldsymbol v}^{(\lambda)}(\vec k,t)$, and consequently  to a different Heisenberg state $\widetilde{|0\rangle}$. Figure\ \ref{fig:4}  shows the lowest multipoles of the scalar and tensor power spectra computed from this state,  $\widetilde{{\cal P}}_{ss'}^{LM}(k)$, and shows that it differs substantially from ${{\cal P}}_{ss'}^{LM}(k)$. In order to remove the ambiguity in the physical predictions, one needs to introduce additional physical input. As an example, we  argue in Ref.\ \cite{aos2} that in models of quantum cosmology  where the big bang singularity is replaced by a cosmic bounce,  the ambiguity disappears,  since all Fourier modes relevant for the CMB start in an adiabatic regime in the the prebounce contracting phase. 
\begin{figure}[!]
\centering
\includegraphics[width=0.49\textwidth]{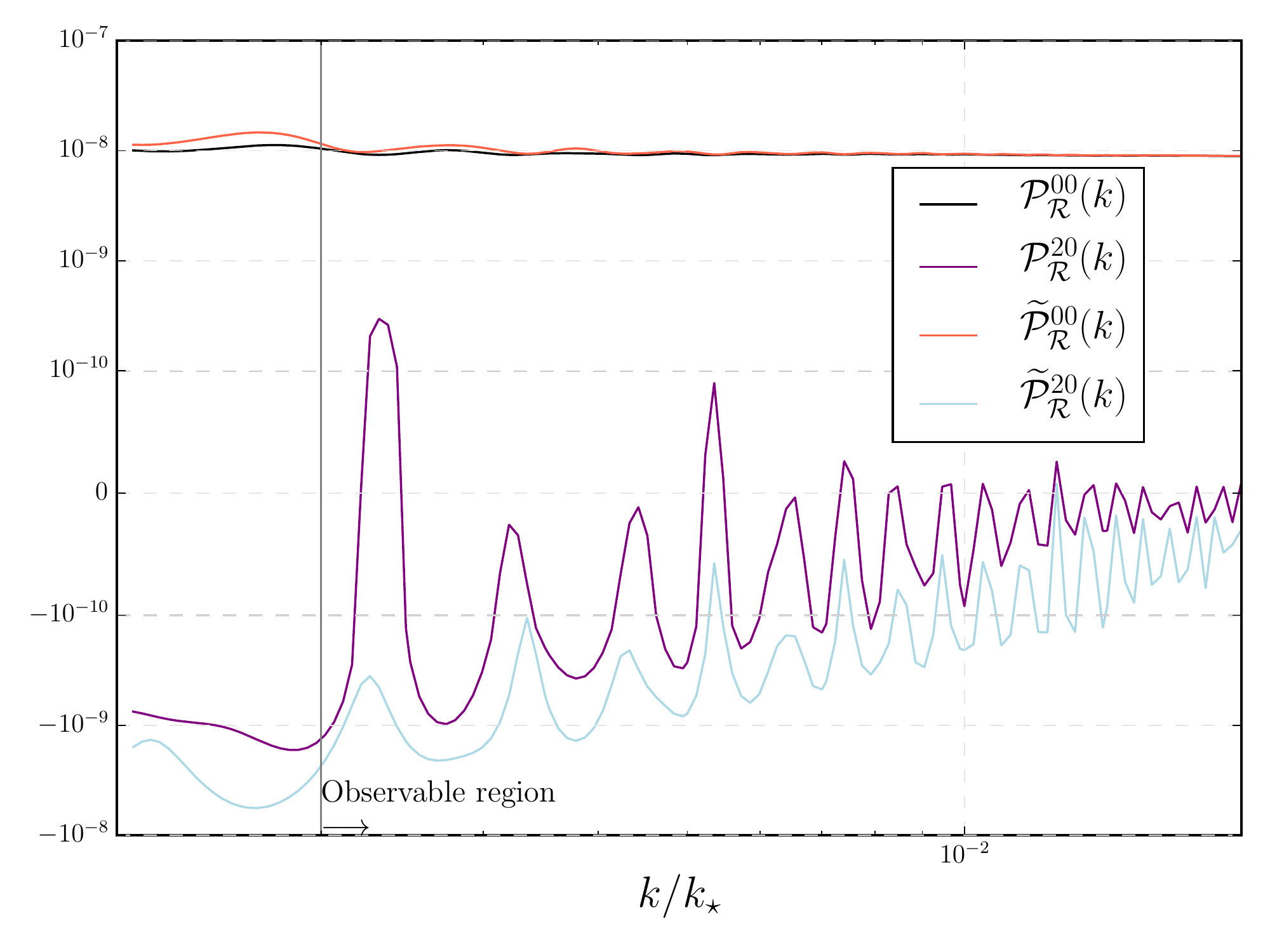}
\includegraphics[width=0.49\textwidth]{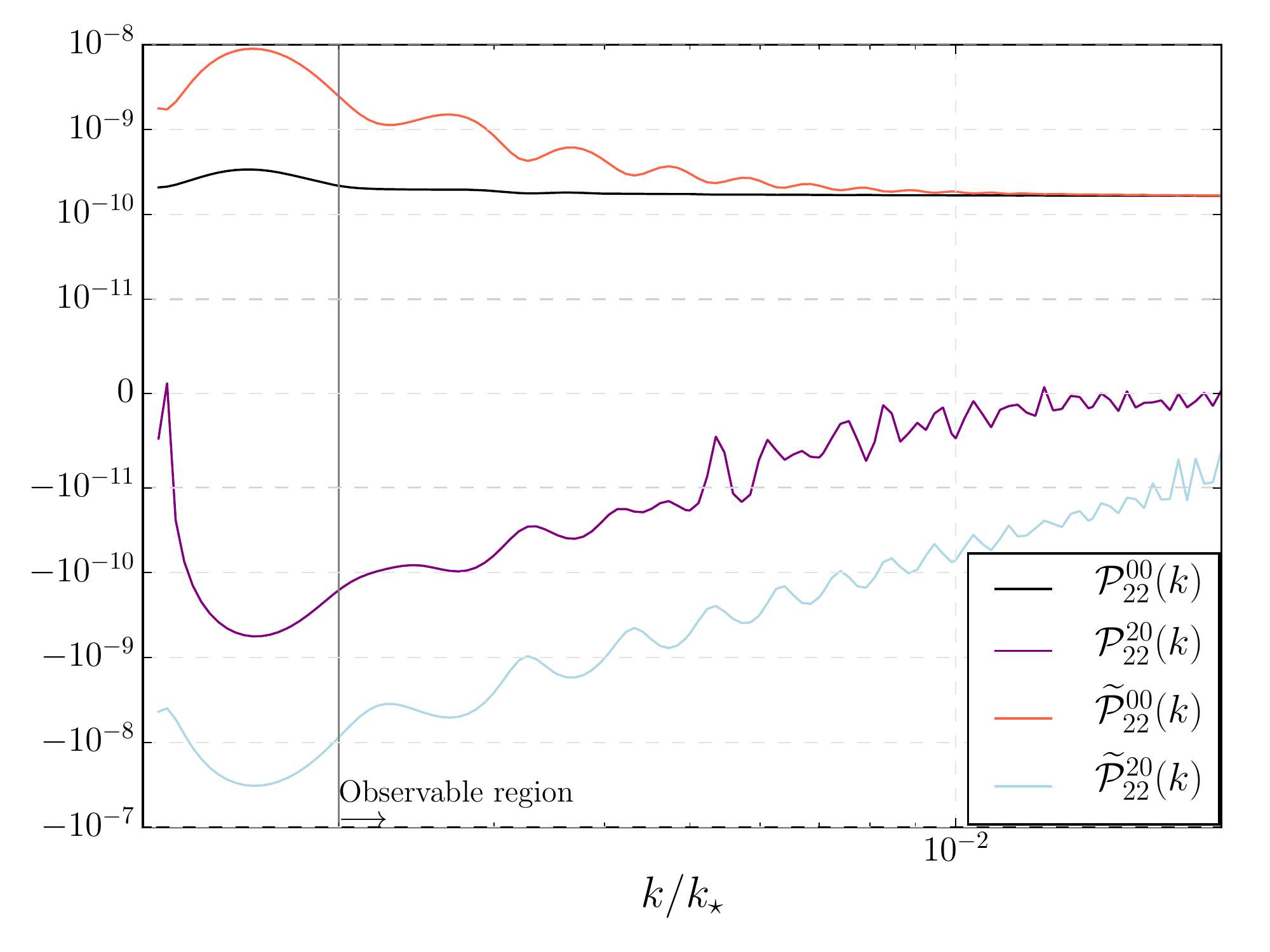}
\caption{Comparison between multipoles $L=0$ and $L=2$ of $\mathcal{P}^{LM}_{ss'}(k)$  and $\widetilde{\mathcal{P}}^{LM}_{ss'}(k)$, obtained from the two vacua considered in this section. For illustrative purposes we show in this plot only the results for  scalar perturbations (left) and one of the two tensor modes (right).  We observe significant  differences, especially in tensor modes and at infrared scales (precisely because they are more sensitive to anisotropies).}
\label{fig:4}
\end{figure}

\end{enumerate}

To describe the evolution in the Schr\"odinger picture, we  need to provide a reference state at late times that plays the role of the ``out" vacuum. Since the inflationary phase makes the Universe highly isotropic, it is natural to  use the familiar Bunch-Davies vacuum there. Such state is given by  the positive-negative norm decomposition defined by using the following  basis elements 
\bea \label{BD} {\boldsymbol v}^{(1)}_{BD}(\vec k)&=&\left(\Gamma^{\rm BD}_{\beta}(k,\eta),0,0;\frac{a^2}{4\kappa}\frac{d}{d\eta}\Gamma^{\rm BD}_{\beta}(k,\eta),0,0\right)\Big |_{\eta_{end}} \, , \nonumber \\
{\boldsymbol v}^{(2)}_{BD}(\vec k)&=&\left(0,\Gamma^{\rm BD}_{\nu}(k,\eta),0;0,\frac{a^2}{4\kappa}\frac{d}{d\eta}\Gamma^{\rm BD}_{\nu}(k,\eta),0\right)\Big |_{\eta_{end}} \, ,\nonumber \\
{\boldsymbol v}^{(3)}_{BD}(\vec k)&=&\left(0,0,\Gamma^{\rm BD}_{\nu}(k,\eta);0,0,\frac{a^2}{4\kappa}\frac{d}{d\eta}\Gamma^{\rm BD}_{\nu}(k,\eta)\right)\Big |_{\eta_{end}} \, ,
\ea
where
\be\Gamma^{\rm BD}_{\beta}(k,\eta)\equiv \sqrt{\frac{4\, \kappa}{ a^2\, \mathcal{V}_0}}\sqrt{\frac{\eta\, \pi}{4}} \,H^{(1)}_{\beta}(-k\, \eta)\, ,\ee
$\eta$ corresponds to conformal time,   and $\eta_{end}$ denotes the end of inflation. $H^{(1)}_{\beta}(x)$ is a Hankel function, and  $\beta =3/2+2\, \epsilon +\delta$, and $\nu =3/2+\epsilon$, where $\epsilon$ and $\delta$ are the standard slow-roll parameters. The ``out'' vacuum state is therefore the familiar tensor product of the Bunch-Davies vacuum for scalar and tensor modes. 

With this, the mode functions defining our initial vacuum ${\boldsymbol v}^{(\lambda)}(\vec k)$, after they are evolved until the end of inflation can be written in  terms of the Bunch-Davies modes and their conjugates via the Bogoliubov coefficients $\alpha_{\lambda\lambda'}$ and $\beta_{\lambda\lambda'}$ as

\be {\boldsymbol v}^{(\lambda)}(\vec k,\eta_{\rm end})=\sum_{\lambda'=1}^3\alpha_{\lambda\lambda'}\, {\boldsymbol v}^{(\lambda')}_{BD}(\vec k)+\beta_{\lambda\lambda'}\, \bar{{\boldsymbol v}}_{BD}^{(\lambda)}(\vec k) \, .\ee
We show here the value of some of these coefficients for the example considered in this section. For $\v k$ pointing in the principal direction of the scale factor $a_1$, and for $k/k_\star=2\times 10^{-3}$, we obtain  
\bea 
\alpha_{11} = & 6{.}49\times 10^{-1} - 1{.}01\,i \, , \hspace{0.5cm} &\beta_{11}\,= \,6{.}84\times 10^{-1} -2{.}98\times 10^{-3}\,i, \nonumber \\
\alpha_{12} = & 1{.}37\times 10^{-1} + 6{.}55\times 10^{-2}\,i \, , \hspace{0.5cm} &\beta_{12}\,=\,-3{.}52\times 10^{-3}\,+\,4{.}71\times 10^{-2}\,i \, ,  \nonumber \\
\alpha_{13} = & -3{.}82\times 10^{-13} - 4{.}09\times 10^{-13}\,i \, , \hspace{0.5cm}& \beta_{13}\,= \, 1{.}76\times 10^{-13}\,-\,9{.}12\times 10^{-14}\,i , \nonumber \\
\alpha_{21} = & 1{.}36\times 10^{-1} + 6{.}72\times 10^{-1}\,i \, , \hspace{0.5cm} &\beta_{21}\,=\, -3{.}47\times 10^{-3}\,+\,4{.}74\times 10^{-2}\,i \, ,  \nonumber \\
\alpha_{22} = & 3{.}37\times 10^{-1} - 1{.}16\times 10^{0}\,i \, , \hspace{0.5cm}& \beta_{22}\,=\, 6{.}91\times 10^{-2}\,-\,1{.}02\times 10^{-1}\,i \, ,  \nonumber \\
\alpha_{23} = & 1{.}63\times 10^{-12} + 3{.}61\times 10^{-12}\,i \, , \hspace{0.5cm} &\beta_{23}\,=\, -2{.}04\times 10^{-12}\,-1\,1{.}90\times 10^{-13}\,i\, ,   \nonumber \\
\alpha_{31} = & -4{.}83\times 10^{-13} - 2{.}68\times 10^{-13}\,i\, , \hspace{0.5cm}& \beta_{31}\,=\, 2{.}61\times 10^{-14}\,-\,8{.}42\times 10^{-14}\,i\, ,   \nonumber \\
\alpha_{32} = & 2{.}09\times 10^{-12} + 3{.}01\times 10^{-12}\,i \, , \hspace{0.5cm} &\beta_{32}\,=\, -1{.}38\times 10^{-12}\,-\,1{.}76\times 10^{-13}\,i\, ,   \nonumber \\
\alpha_{33} = & 1{.}00 + 7{.}05\times 10^{-2}\,i \, , \hspace{0.5cm} &\beta_{33}\,=\, -3{.}48\times 10^{-2}\,-\,9{.}50\times 10^{-2}\,i \, . \ea
Hence, the value of these  coefficients contain information about the evolution of the initial vacuum state to the end of inflation in a particular direction. More explicitly, from them we can compute the coefficients $V_{\lambda\lambda'}(\v k):=\sum_{\lambda''=1}^3 \frac{1}{2} \,\bar \beta_{ \lambda''\lambda}(\v k)\, \bar\alpha^{-1}_{\lambda'\lambda''}(\v k)$. In this particular case (i.e. $\v k$ pointing in the  direction of $a_1$), they are
\begin{align} 
&V_{11}=\,(1{.}53\,-\, 2{.}37\,i)\times 10^{-1} \, , \,\,  V_{22}=\,(1{.}19\,-\, 2{.}60\,i)\times 10^{-1} \, , \,\, V_{33}=(1{.}39\,+\,4{.}84\,i)\times 10^{-2} \,, \nonumber \\
&V_{12}=(1.34\,-\,0{.}96\,i)\times 10^{-2} \, , \,\, V_{13}=(-2{.}70\,-\,9{.}48\,i)\times 10^{-14}\,, \,\, V_{23}=(-2{.}15\,+\,9{.}74\,i)\times 10^{-13} \, . 
\end{align}
Substituting them in  expression (\ref{Sinout2}), we obtain the explicit form of the evolution of the initial state written in terms of excited states over the Bunch-Davies vacuum. We can explicitly see that the ``in'' vacuum evolves to an {\em excited and entangled} state between scalar and tensor perturbations at the end of inflation, and all details about this entanglement (entanglement entropy, mutual information, etc.) can be now straightforwardly computed using the coefficients $V_{\lambda\lambda'}(\v k)$. 

\section{Conclusions}

This paper contains a detailed derivation of the classical and quantum theory of gauge invariant linear cosmological perturbations in Bianchi I spacetimes from a Hamiltonian viewpoint. At the classical level, the problem of isolating the gauge invariant degrees of freedom  and their dynamics in phase space reduces to solving a Hamilton-Jacobi-like equation for the generating function of a canonical transformation. Among the possible choices, we consider a particular set of gauge invariant fields that reduce  to the familiar scalar and tensor perturbations commonly used in the isotropic limit. The presence of  anisotropies introduces terms in the physical Hamiltonian that {\em couple} these fields among themselves. These couplings  introduce subtleties in the quantization process, but as long as one is restricted to linear perturbations, the formulation of the quantum theory and the derivation of its physical predictions can be done in an exact manner, without relying on any perturbative expansion on the anisotropies.  We have described in detail this quantum theory from a canonical viewpoint, and spelled out the time evolution of quantum perturbations both in the Heisenberg and the Schr\"odinger pictures. In the latter, the couplings in the  Hamiltonian induce entanglement  in the quantum state of scalar and tensor modes, as well as for tensor modes with different polarizations.

Therefore, if an anisotropic phase existed  in the early Universe before inflation, one should expect the quantum state of cosmic perturbations at the onset of the slow-roll era to be anisotropic, and to contain nontrivial entanglement between the different types of perturbations. These two features can be imprinted in the CMB through anisotropic power spectra and cross-correlations between scalars and tensors modes.  Some of the phenomenological consequences of entanglement between scalar and tensors perturbations in inflation have been discussed in the literature (see e.g.\ \cite{holmanentaglement,collins});  the framework constructed in this paper provides a concrete mechanism to generate the entanglement postulated in these works. We have  developed the tools needed to explicitly compute all aspects of  this entanglement, both in the Heisenberg and the Schr\"odinger pictures.

One of the advantages of (and partially the motivation for) the Hamiltonian formulation presented in this paper is that it is suitable to be applied to theories of canonical quantum gravity. We show a concrete example in a companion paper \cite{aos2}, where we use our formalism on a quantum Bianchi I spacetime, as predicted by loop quantum cosmology, where the big bang singularity is replaced by a cosmic bounce \cite{awe-bi,mmp-bi,gs-bi}. Such anisotropic bounce connects two  isotropic FLRW spacetimes in the past and future. In that scenario perturbations find an adiabatic regime in the remote past, which makes a preferred initial quantum state for perturbations available. Therefore, that setting offers a clean scenario where concrete predictions arising from an anisotropic phase of the  Universe can be made.

\acknowledgments
We have benefited from  discussions with Abhay Ashtekar,  Mar Bastero-Gil, Brajesh Gupt, Guillermo A. Mena Marug\'an, Jorge Pullin, Parampreet Singh and Edward Wilson-Ewing. This work is supported by the NSF CAREER Grant No. PHY-1552603, Project. No. FIS2017-86497-C2-2-P of MICINN from Spain, and from funds of the Hearne Institute for Theoretical Physics.  V.S. was also supported by Louisiana State University and Inter-University Centre for Astronomy and Astrophysics during different stages of this work. Portions of this research were conducted with high performance computational resources provided by Louisiana State University (http://www.hpc.lsu.edu). 

\appendix
\section{TOTAL HAMILTONIAN FOR PERTURBATIONS: FOURIER EXPANSION}\label{appA}
This appendix provides further details, omitted in the main text, about the SVT decomposition of perturbations on Bianchi I spacetimes. Let us first recall that the linearized scalar and vector constraints of general relativity take the following general form (see Sec. \ref{sec.2} for the definitions of the different quantities that appear in this equation)
 \begin{align}\nonumber
&\mathbb S^{(1)}(\v x) = \frac{2\kappa}{\sqrt{h}}\left[2\pz_{ij}\delta\pi^{ij}-{\pz_{i}}{}^i\delta{\pi_{j}}^j+\delta h_{ij}\left(2{\pz_{k}}^i\pz^{jk}-\pz^{ij}{\pz_{k}}{}^k\right)-\frac{1}{2}\hz^{ij}\delta h_{ij}\left({\pz_{kl}}\pz^{kl}-\frac{1}{2}{\pz_{k}}{}^k{\pz_{l}}{}^l\right)\right]\\
  &+\frac{\sqrt{h}}{2\kappa}\left(\hz^{ij}\hz^{kl}-\hz^{ik}\hz^{jl}\right)\delta h_{ij,k,l}+\hz^{ij}\delta h_{ij} \left( -\f{\pp^2}{4\,\sqrt{h}} + \f{1}{2}\sqrt{h}\, V(\phi)\right)+ \sqrt{h}\, V_{\phi}\,\dph + \f{{\pp}\, \dpp}{\sqrt{h}}\, ,\\
  & \mathbb V^{(1)}_i(\v x) = \pz^{jk}\l( \delta h_{jk,i} - 2\,\delta h_{ij,k}   \r) - 2 h_{ij}\,\delta\pi^{jk}_{\,\,,k} + \pi_{\phi}\, \delta\phi_{,i}\, ,
\end{align}
 where a comma indicates coordinate derivative, e.g.\ $\delta h_{ij,k}\equiv \partial_k h_{ij}$. We now Fourier expand the perturbations  $\delta h_{ij}$, $\delta \pi^{ij}$, $\dpp$, $\delta\phi$ as in (\ref{Fexph}) and (\ref{Fexph2}), and furthermore carry out the SVT decomposition as defined in (\ref{eq:dh-to-gamma}). This decomposition must be implemented in the phase space as  a time-dependent canonical transformation, since the matrices $A^{(n)}_{ij}$ depend on time via $\hz_{ij}$ and the orthonormal vectors $(\h k, \h x, \h y)$. Concretely, the time derivatives of  $\hz_{ij}$ and $(\h k, \h x, \h y)$, denoted as $(\partial_t)$ and understood as their Poisson bracket with the background Hamiltonian $ \mathcal H_{_{\rm BI}}$, are
\begin{align}
\frac{1}{N}\partial_t{\hz}_{ij} & =\frac{4\kappa}{\sqrt{\hz}}\left(\pz_{ij}-\frac{1}{2}\hz_{ij}\pz\right), \\
  \frac{1}{N}\partial_t{\h k}_i & = \frac{2 \kappa}{\sqrt{\hz}} \h k_j \h k_k (\pz^{jk} - \frac{1}{2}\hz^{jk} \pz_l{}^l) \h k_i,\\
  \frac{1}{N}\partial_t{\h x}_i & = \frac{4 \kappa}{\sqrt{\hz}} (\pz_i{}^{j} - \frac{1}{2}\hz_i{}^{j} \pz_k{}^k) \h x_j+R_{xx}\, \h x_i +R_{xy}\, \h y_i,\\
  \frac{1}{N}\partial_t{\h y}_i & = \frac{4 \kappa}{\sqrt{\hz}} (\pz_i{}^{j} - \frac{1}{2}\hz_i{}^{j} \pz_k{}^k) \h y_j+R_{yy}\, \h y_i +R_{yx}\, \h x_i,
\end{align}
where $N$ is the lapse function and 
\begin{align}
  R_{xx} &= -\frac{2 \kappa}{\sqrt{\hz}} (\pz^{ij} - \frac{1}{2}\hz^{ij} \pz_k{}^k) \h x_i\h x_j,\\
  R_{yy} &= -\frac{2 \kappa}{\sqrt{\hz}} (\pz^{ij} - \frac{1}{2}\hz^{ij} \pz_k{}^k) \h y_i\h y_j,\\
  R_{xy} &=R_{yx}=-\frac{2 \kappa}{\sqrt{\hz}} (\pz^{ij} - \frac{1}{2}\hz^{ij} \pz_k{}^k) \h x_i\h y_j.
\end{align}
These equations can be easily obtained from the definition of $\h k$, the orthonormality conditions of $(\h k, \h x, \h y)$, the equations of motion of the background variables, and the extra condition $R_{xy} = R_{yx}$, that introduces convenient simplifications (see Refs.\ \cite{ppu-BI1,ppu-BI2} for additional details). It is also convenient to compute the time derivative of the comoving wave number
\begin{equation}
\frac{1}{N}\partial_t k =  -\frac{2 \kappa}{\sqrt{\hz}} k \, \h k_i \h k_j (\pz^{ij} - \frac{1}{2}\hz^{ij} \pz_k{}^k) \, .
\end{equation}
From these quantities, it is straightforward to obtain the time derivatives of the matrices ${A}^{{(n)}}_{ij}$. For the canonical transformation that implements the SVT decomposition, we adopt a mode-by-mode  type 3 generating function, which  depends on new configuration variables $\gamma_n$ and old momenta $\delta \tilde\pi^{ij}$. More explicitly 
\begin{equation}
g(\v k) = - \delta \tilde\pi^{ij}(\v k) \sum_{n=1}^6A^{(n)}_{ij}(\v k) \gamma_n(\v k). 
\end{equation}
New momenta are defined as
\begin{equation}
\pi_n(\v k) = - \frac{\partial g(\v k)}{\partial \gamma_n(\v k)}.
\end{equation}
As we see,  $g(\v k)$ depends on the time-dependent matrices $A^{(n)}_{ij}(\v k)$. This fact will be important to obtain the Hamiltonian for the new variables. Let us now focus on the linear constraints $\mathbb S^{(1)}(\v x)$ and $\mathbb V^{(1)}_i(\v x)$. In terms of the new canonical variables $\gamma_\alpha(\vec k)$ and $\pi_\alpha(\vec k)$ (we have also incorporated the perturbations of the scalar field) they take the form:
\begin{align}
  &\tilde{\mathbb{S}}^{(1)}(\v k) = \frac{\gamma_0}{\sqrt{4\kappa}}a^3\,V_{\phi}+ \frac{\gamma_1}{\sqrt{3}}\, \l(-\frac{a^3|\v k|^2}{\kappa}-\frac{\kappa p^2_a}{24 a}-\frac{3 \pp^2}{4 a^3}+\frac{3}{2}a^3 V+\f{a^3}{4\,\kappa}\,\sigma^2\r) + 
 \frac{\gamma_2}{\sqrt{6}\kappa}\, \biggl( a^3\,|\v k|^2 + \f{\kappa a p_a }{\sqrt{6}} \sigma_{(2)}+ a^3 \sigma_{(2)}^2\nonumber\\
& + \frac{1}{2}a^3\sigma_{(3)}^2+\frac{1}{2}a^3\sigma_{(4)}^2-a^3\sigma_{(5)}^2-a^3\sigma_{(6)}^2 \biggr) + 
 \frac{\gamma_3}{\sqrt{2}\kappa}\,\l( \f{\kappa a p_a\,}{3 \sqrt{2}}\sigma_{(3)} + \f{a^3 \sigma_{(2)}\sigma_{(3)}}{\sqrt{3}} + a^3 \sigma_{(3)}\sigma_{(5)} + a^3 \sigma_{(4)}\sigma_{(6)}\r) \nonumber\\
&+\frac{\gamma_4}{\sqrt{2}\kappa}\,\l( \f{\kappa a p_a\,}{3 \sqrt{2}}\sigma_{(4)} + \f{a^3 \sigma_{(2)}\sigma_{(4)}}{\sqrt{3}} - a^3 \sigma_{(4)}\sigma_{(5)} + a^3 \sigma_{(3)}\sigma_{(6)}\r) + \frac{\gamma_5}{\sqrt{2}\kappa}\,\biggl( \f{\kappa a p_a\,}{3 \sqrt{2}}\sigma_{(5)} - \f{2a^3 \sigma_{(2)}\sigma_{(5)}}{\sqrt{3}} \nonumber\\
&+ \frac{1}{2}a^3 \sigma_{(3)}^2 - \frac{1}{2}a^3 \sigma_{(4)}^2\biggr)
 + \frac{\gamma_6}{\sqrt{2}\kappa}\,\biggl( \f{\kappa a p_a\,}{3 \sqrt{2}}\sigma_{(6)} - \f{2a^3 \sigma_{(2)}\sigma_{(6)}}{\sqrt{3}} + a^3 \sigma_{(3)}\sigma_{(4)} \biggr)\, + \frac{2 \sqrt{\kappa} \pp}{a^{3}}\pi_0-\, \f{\kappa\,p_a}{\sqrt{3}\,a^2}\,\pi_1 
 \nonumber\\
&+\, 2\sigma_{(2)}\pi_2\, +\, 2\sigma_{(3)}\pi_3\, +\, 2\sigma_{(4)}\pi_4\,  +\, 2\sigma_{(5)}\pi_5 +\, 2\sigma_{(6)}\pi_6\, ,\label{eq:S1}\\
& \h k^i\,\tilde{\mathbb{V}}_i^{(1)}(\v k)\,=\,i\,|\v k|\,\biggl[ \gamma_0\frac{\pp}{\sqrt{4\kappa}}+\gamma_1 \l(\f{a p_a}{6\sqrt{3}}\, -\, \f{\sqrt{2}a^3\,\sigma_{(2)}}{3\kappa}\r)\, 
 -\, \gamma_2\, \l( \sqrt{\f{2}{3}}\frac{ap_a}{3}\, +\, \f{a^3\,\sigma_{(2)}}{6\,\kappa} \r) +\, \f{a^3\,\sigma_{(5)}}{2\,\kappa}\,\gamma_5 \nonumber\\
&
 +\, \f{a^3\,\sigma_{(6)}}{2\,\kappa}\,\gamma_6\,-\, \f{2}{\sqrt{3}}\,\pi_1-\, 2\sqrt{\frac{2}{3}}\,\pi_2\, \biggr],\label{eq:Vk}\\
& \h x^i\,\tilde{\mathbb{V}}_i^{(1)}(\v k)\, =\, i\,|\v k|\,\biggl[ \f{a^3\,\sigma_{(3)}}{\sqrt{6}\kappa}\,\gamma_1\, -\, \f{a^3\,\sigma_{(3)}}{2\sqrt{3}\,\kappa}\, \gamma_2\, 
+\, \l( \f{ap_a}{3\,\sqrt{2}}\, +\, \f{a^3\, \sigma_{(2)}}{\sqrt{3}\,\kappa} \r)\, \gamma_3\, 
+\, \f{a^3\,\sigma_{(3)}}{2\,\kappa}\,\gamma_5\, +\, \f{a^3\,\sigma_{(4)}}{2\,\kappa}\,\gamma_6\nonumber\\ 
& +\, \sqrt{2}\,\pi_3\biggr],\label{eq:Vx}\\
& \h y^i\,\tilde{\mathbb{V}}_i^{(1)}(\v k)\, =\, i\,|\v k|\,\biggl[ \f{a^3\,\sigma_{(4)}}{\sqrt{6}\kappa}\,\gamma_1\, -\, \f{a^3\,\sigma_{(4)}}{2\sqrt{3}\,\kappa}\, \gamma_2\, 
+\, \l( \f{ap_a}{3\,\sqrt{2}}\, +\, \f{a^3\, \sigma_{(2)}}{\sqrt{3}\,\kappa} \r)\, \gamma_4\, 
-\, \f{a^3\,\sigma_{(4)}}{2\,\kappa}\,\gamma_5\, +\, \f{a^3\,\sigma_{(3)}}{2\,\kappa}\,\gamma_6\nonumber\\ 
& +\, \sqrt{2}\,\pi_4\biggr]\label{eq:Vy}.
\end{align}
With these expressions, one can check the following algebra of the linearized constraints 
\begin{align}
\{\tilde{\mathbb{S}}^{(1)}(\v k),\h k^i\,\tilde{\mathbb{V}}_i^{(1)}(\v k')\}&=-i\,|\v k|\,\delta_{\v k,-\v k'}\,\mathbb S^{(0)}\;\mathring{\approx} \;0,\nonumber \\ 
\{\tilde{\mathbb{S}}^{(1)}(\v k),\h x^i\,\tilde{\mathbb{V}}_i^{(1)}(\v k')\}&=0,\nonumber \\
\{\tilde{\mathbb{S}}^{(1)}(\v k),\h y^i\,\tilde{\mathbb{V}}_i^{(1)}(\v k')\}&=0,\nonumber \\
\{\tilde{\mathbb{V}}_i^{(1)}(\v k),\,\tilde{\mathbb{V}}_j^{(1)}(\v k')\}&=0. 
\end{align}
Here, the symbol $\;\mathring{\approx} \;0$ means that we evaluate the background quantities on shell. These expressions show that the linear constraints form a first class system. From (\ref{eq:S1})--(\ref{eq:Vy}), it is trivial to obtain the Poisson brackets between the canonical variables $\gamma_n(\vec k)$ and $\pi_n(\vec k)$ and the linearized constraints (for instance, $\{\gamma_1,\tilde{\mathbb{S}}^{(1)}\}$ is given by the coefficient multiplying $\pi_1$ in $\tilde{\mathbb{S}}^{(1)}$). These Poisson brackets indicate the way all these variables change under the gauge transformations generated by the constraints; i.e.\ none of them  are gauge invariant. 

Next, we obtain  the Fourier transform of the second-order scalar constraint $\tilde{\mathbb{S}}^{(2)}(\v k)$. But we must keep in mind that, since we are dealing with a time-dependent canonical transformation, we must add the time derivative of the generating function $g(\v k)$.  The result is the following second-order Hamiltonian for $\gamma_\alpha(\vec k)$ and $\pi_\alpha(\vec k)$:

\begin{align} \label{S2}
 &\int d^3x \, {\mathbb{S}}^{(2)}(\v x)=\sum_{\vec k}|\gamma_0^2 |\bigg(\frac{a^{3}V_{\phi\phi}}{8 \kappa}+\frac{a^{3} |\v k|^2}{8 \kappa}\bigg)+|\gamma_1^2 |\bigg(-\frac{a^{3} |\v k|^2}{12 \kappa}+\frac{\kappa p_a^{2}}{288 a}+\frac{5 \pp^{2}}{16 a^{3}}-\frac{a^{3} \sigma^{2}}{48 \kappa}\nonumber \\
  &
  +\frac{1}{8} a^{3} V\bigg)+|\gamma_2^2 |\bigg(-\frac{a^{3} |\v k|^2}{24 \kappa}  +\frac{5\kappa p_a^{2}}{144 a}+\frac{\pp^{2}}{8 a^{3}}+\frac{a p_a \sigma_{(2)}}{3\sqrt{6}}+\frac{a^{3} \sigma_{(2)}^2}{8 \kappa}-\frac{a^{3} \sigma_{(3)}^{2}}{24 \kappa}-\frac{a^{3} \sigma_{(4)}^{2}}{24 \kappa}+\frac{5 a^{3} \sigma_{(5)}^{2}}{24 \kappa}\nonumber \\
  &
  +\frac{5 a^{3} \sigma_{(6)}^{2}}{24 \kappa}-\frac{1}{4} a^{3} V\bigg)+|\gamma_3^2 |\bigg(\frac{5 \kappa p_a^{2}}{144 a} +\frac{\pp^{2}}{8 a^{3}}+\frac{ a p_a \sigma_{(2)}}{6 \sqrt{6}}-\frac{a^{3} \sigma_{(2)}^{2}}{6 \kappa}+\frac{a^{3} \sigma^{2}}{8 \kappa}+\frac{a p_a \sigma_{(5)}}{6 \sqrt{2}}+\frac{a^{3} \sigma_{(2)} \sigma_{(5)}}{2 \sqrt{3} \kappa} \nonumber\\
  &
  -\frac{1}{4} a^{3} V\bigg)+|\gamma_4^2 |\bigg(\frac{5 \kappa p_a^{2}}{144 a} +\frac{\pp^{2}}{8 a^{3}}+\frac{ a p_a \sigma_{(2)}}{6 \sqrt{6}}-\frac{a^{3} \sigma_{(2)}^{2}}{6 \kappa}+\frac{a^{3} \sigma^{2}}{8 \kappa}-\frac{a p_a \sigma_{(5)}}{6 \sqrt{2}}-\frac{a^{3} \sigma_{(2)} \sigma_{(5)}}{2 \sqrt{3} \kappa}-\frac{1}{4} a^{3} V\bigg)\nonumber\\
  &
  +|\gamma_5^2 |\bigg(\frac{a^{3} |\v k|^2}{8 \kappa}+\frac{5 \kappa p_a^{2}}{144 a}+\frac{\pp^{2}}{8 a^{3}}-\frac{ a p_a \sigma_{(2)}}{3 \sqrt{6}}+\frac{5a^{3} \sigma_{(2)}^{2}}{24 \kappa}+\frac{a^{3} \sigma_{(3)}^{2}}{8 \kappa}+\frac{a^{3} \sigma_{(4)}^{2}}{8 \kappa}+\frac{a^{3} \sigma_{(5)}^{2}}{8 \kappa}-\frac{a^{3} \sigma_{(6)}^{2}}{8 \kappa}-\frac{1}{4} a^{3} V\bigg)\nonumber\\
  &
  +|\gamma_6^2 |\bigg(\frac{a^{3} |\v k|^2}{8 \kappa}+\frac{5 \kappa p_a^{2}}{144 a}+\frac{\pp^{2}}{8 a^{3}}-\frac{ a p_a \sigma_{(2)}}{3 \sqrt{6}}+\frac{5a^{3} \sigma_{(2)}^{2}}{24 \kappa}+\frac{a^{3} \sigma_{(3)}^{2}}{8 \kappa}+\frac{a^{3} \sigma_{(4)}^{2}}{8 \kappa}-\frac{a^{3} \sigma_{(5)}^{2}}{8 \kappa}+\frac{a^{3} \sigma_{(6)}^{2}}{8 \kappa}-\frac{1}{4} a^{3} V\bigg)\nonumber\\
  & -\frac{3\kappa}{a^3}|\pi_1^2 |+\frac{2\kappa}{a^3} \left(|\pi_0^2 |+|\pi_1^2 |+|\pi_2^2 |+|\pi_3^2 |+|\pi_4^2 |+|\pi_5^2 |+|\pi_6^2 |\right)+\frac{\sqrt{3}}{4\sqrt{\kappa}}a^3V_\phi \Re[\gamma_0 \bar\gamma_1 ]+\Re[\gamma_1 \bar\gamma_2 ]\bigg(\frac{a^{3} |\v k|^2}{6 \sqrt{2} \kappa}\nonumber\\
&
  -\frac{a p_a \sigma_{(2)}}{12 \sqrt{3}} -\frac{a^{3} \sigma_{(2)}^{2}}{6 \sqrt{2} \kappa}-\frac{a^{3} \sigma_{(3)}^{2}}{12 \sqrt{2} \kappa}-\frac{a^{3} \sigma_{(4)}^{2}}{12 \sqrt{2} \kappa}  +\frac{a^{3} \sigma_{(5)}^{2}}{6 \sqrt{2} \kappa}+\frac{a^{3} \sigma_{(6)}^{2}}{6 \sqrt{2} \kappa}\bigg)+\Re[\gamma_1 \bar\gamma_3 ]\bigg(-\frac{ a p_a  \sigma_{(3)}}{12 \sqrt{3}}-\frac{a^{3} \sigma_{(2)} \sigma_{(3)}}{6 \sqrt{2} \kappa}\nonumber\\
  &
  -\frac{a^{3} \sigma_{(3)} \sigma_{(5)}}{2 \sqrt{6}{\kappa}}-\frac{a^{3} \sigma_{(4)} \sigma_{(6)}}{2 \sqrt{6} \kappa}\bigg)+\Re[\gamma_1 \bar\gamma_4 ]\bigg(-\frac{ a p_a  \sigma_{(4)}}{12 \sqrt{3}}-\frac{a^{3} \sigma_{(2)} \sigma_{(4)}}{6 \sqrt{2} \kappa}+\frac{a^{3} \sigma_{(4)} \sigma_{(5)}}{2 \sqrt{6} \kappa}-\frac{a^{3} \sigma_{(3)} \sigma_{(6)}}{2 \sqrt{6} \kappa}\bigg)\nonumber \\
  &
  +\Re[\gamma_1 \bar\gamma_5 ]\bigg(-\frac{a^{3} \sigma_{(3)}^{2}}{4 \sqrt{6} \kappa}+\frac{a^{3} \sigma_{(4)}^{2}}{4 \sqrt{6} \kappa}-\frac{a p_a \sigma_{(5)}}{12 \sqrt{3}}+\frac{a^{3} \sigma_{(2)} \sigma_{(5)}}{3 \sqrt{2} \kappa}\bigg)+\Re[\gamma_1 \bar\gamma_6 ]\bigg(-\frac{a^{3} \sigma_{(3)} \sigma_{(4)}}{2 \sqrt{6} \kappa}-\frac{a p_a \sigma_{(6)}}{12 \sqrt{3}}\nonumber \\
  &
  +\frac{a^{3} \sigma_{(2)} \sigma_{(6)}}{3 \sqrt{2} \kappa}\bigg)+\Re[\gamma_2 \bar\gamma_3 ]\bigg(\frac{a p_a \sigma_{(3)}}{3 \sqrt{6}}+\frac{a^{3} \sigma_{(2)} \sigma_{(3)}}{3 k}-\frac{a^{3} \sigma_{(3)} \sigma_{(5)}}{2 \sqrt{3} k}-\frac{a^{3} \sigma_{(4)} \sigma_{(6)}}{2 \sqrt{3} k}\bigg)+\Re[\gamma_2 \bar\gamma_4 ]\bigg(\frac{a p_{a} \sigma_{(4)}}{3 \sqrt{6}}\nonumber \\
  &
  +\frac{a^{3} \sigma_{(2)} \sigma_{(4)}}{3 \kappa}+\frac{a^{3} \sigma_{(4)} \sigma_{(5)}}{2 \sqrt{3} \kappa}-\frac{a^{3} \sigma_{(3)} \sigma_{(6)}}{2 \sqrt{3} \kappa}\bigg)+\Re[\gamma_2 \bar\gamma_5 ]\bigg(\frac{a^{3} \sigma_{(3)}^{2}}{2 \sqrt{3} \kappa}-\frac{a^{3} \sigma_{(4)}^{2}}{2 \sqrt{3} \kappa}-\frac{1}{3} \sqrt{\frac{2}{3}} a p_a \sigma_{(5)}\nonumber\\
  &
  -\frac{a^{3} \sigma_{(2)} \sigma_{(5)}}{6 \kappa}\bigg)+\Re[\gamma_2 \bar\gamma_6 ]\bigg(\frac{a^{3} \sigma_{(3)} \sigma_{(4)}}{\sqrt{3} \kappa}-\frac{1}{3} \sqrt{\frac{2}{3}} a p_a \sigma_{(6)}-\frac{a^{3} \sigma_{(2)} \sigma_{(6)}}{6 \kappa}\bigg)+\Re[\gamma_3 \bar\gamma_4 ]\bigg(\frac{a p_{a} \sigma_{(6)}}{3 \sqrt{2}}\nonumber \\
  &
  +\frac{a^{3} \sigma_{(2)} \sigma_{(6)}}{\sqrt{3} \kappa}\bigg)+\Re[\gamma_3 \bar\gamma_5 ]\bigg(\frac{a p_{a} \sigma_{(3)}}{3 \sqrt{2}}-\frac{\mathrm{a}^{3} \sigma_{(2)} \sigma_{(3)}}{2 \sqrt{3} \kappa}-\frac{\mathrm{a}^{3} \sigma_{(4)} \sigma_{(6)}}{2 \kappa}\bigg)+\Re[\gamma_3 \bar\gamma_6 ]\bigg(\frac{a p_{a} \sigma_{(4)}}{3 \sqrt{2}}-\frac{\mathrm{a}^{3} \sigma_{(2)} \sigma_{(4)}}{2 \sqrt{3} \kappa}\nonumber\\
  &
  +\frac{\mathrm{a}^{3} \sigma_{(4)} \sigma_{(5)}}{2 \kappa}\bigg)+\Re[\gamma_4 \bar\gamma_5 ]\bigg(-\frac{a p_{a} \sigma_{(4)}}{3 \sqrt{2}}+\frac{a^{3} \sigma_{(2)} \sigma_{(4)}}{2 \sqrt{3} \kappa}+\frac{a^{3} \sigma_{(3)} \sigma_{(6)}}{2 \kappa}\bigg)+\Re[\gamma_4 \bar\gamma_6 ]\bigg(\frac{a p_a \sigma_{(3)}}{3 \sqrt{2}}\nonumber\\
  &-\frac{a^{3} \sigma_{(2)} \sigma_{(3)}}{2 \sqrt{3} \kappa}-\frac{a^{3} \sigma_{(3)} \sigma_{(5)}}{2 \kappa}\bigg)+\frac{a^{3} \sigma_{(5)} \sigma_{(6)}}{2 \kappa}\Re[\gamma_5 \bar\gamma_6 ]-\Re[\gamma_1 \bar\pi_0 ]\frac{\sqrt{3\kappa} \pp}{a^{3}}-\Re[\gamma_1 \bar\pi_2 ]\frac{\sigma_{(2)}}{\sqrt{3}}-\Re[\gamma_1 \bar\pi_3 ]\frac{\sigma_{(3)}}{\sqrt{3}}\nonumber\\
  &
  -\Re[\gamma_1 \bar\pi_4 ]\frac{\sigma_{(4)}}{\sqrt{3}}-\Re[\gamma_1 \bar\pi_5 ]\frac{\sigma_{(5)}}{\sqrt{3}}-\Re[\gamma_1 \bar\pi_6 ]\frac{\sigma_{(6)}}{\sqrt{3}} -\Re[\gamma_2 \bar\pi_1 ]\frac{\sigma_{(2)}}{\sqrt{3}}+\Re[\gamma_2 \bar\pi_3 ]\frac{2\sqrt{2}\sigma_{(3)}}{\sqrt{3}}+\Re[\gamma_2 \bar\pi_4 ]\frac{2\sqrt{2}\sigma_{(4)}}{\sqrt{3}}\nonumber\\
  &
  -\Re[\gamma_2 \bar\pi_5 ]\frac{\sqrt{2}\sigma_{(5)}}{\sqrt{3}}-\Re[\gamma_2 \bar\pi_6 ]\frac{\sqrt{2}\sigma_{(6)}}{\sqrt{3}}-\Re[\gamma_3 \bar\pi_1 ]\frac{\sigma_{(3)}}{\sqrt{3}}-\Re[\gamma_3 \bar\pi_2 ]\frac{\sqrt{2}\sigma_{(3)}}{\sqrt{3}}-\Re[\gamma_3 \bar\pi_4 ]\l( \frac{\sigma_{(6)}}{\sqrt{2}} - \sqrt{2}\sigma_6 \r)\nonumber
  \end{align}
  \begin{align}
  &
  +\Re[\gamma_3 \bar\pi_5 ]\sqrt{2}\sigma_{(3)}+\Re[\gamma_3 \bar\pi_6 ]\sqrt{2}\sigma_{(4)}-\Re[\gamma_4 \bar\pi_1 ]\frac{\sigma_{(4)}}{\sqrt{3}}-\Re[\gamma_4 \bar\pi_2 ]\frac{\sqrt{2}\sigma_{(4)}}{\sqrt{3}}-\Re[\gamma_4 \bar\pi_3 ]\l(\frac{\sigma_{(6)}}{\sqrt{2}} - \sqrt{2} \sigma_6\r)\nonumber\\
  &
  -\Re[\gamma_4 \bar\pi_5 ]\sqrt{2}\sigma_{(4)}+\Re[\gamma_4 \bar\pi_6 ]\sqrt{2}\sigma_{(3)}
  -\Re[\gamma_5 \bar\pi_1 ]\frac{\sigma_{(5)}}{\sqrt{3}}-\Re[\gamma_5 \bar\pi_2 ]\frac{\sqrt{2}\sigma_{(5)}}{\sqrt{3}}-\Re[\gamma_6 \bar\pi_1 ]\frac{\sigma_{(6)}}{\sqrt{3}}-\Re[\gamma_6 \bar\pi_2]\frac{\sqrt{2}\sigma_{(6)}}{\sqrt{3}} \nonumber\\
  &
  +\Re[\gamma_1 \bar\pi_1]\frac{\kappa p_a}{6 a^2}
  + \Re[\gamma_2 \bar\pi_2]\l( \frac{2 \kappa p_a}{3 a^2} + \sqrt{\frac{2}{3}}\sigma_{(2)}\r)
  +\Re[\gamma_3\bar\pi_3]\l( \frac{2 \kappa p_a}{3 a^2} + \sqrt{\frac{2}{3}} \sigma_{(2)} -\frac{\sigma_{(2)}}{\sqrt{6}} - \frac{\sigma_{(5)}}{\sqrt{2}} + \sqrt{2}\sigma_{(5)} \r) \nonumber\\
  &
  + \Re[\gamma_4\bar\pi_4]\l(  \frac{2 \kappa p_a}{3 a^2} + \sqrt{\frac{2}{3}} \sigma_{(2)} -\frac{\sigma_{(2)}}{\sqrt{6}} + \frac{\sigma_{(5)}}{\sqrt{2}} - \sqrt{2}\sigma_{(5)} \r)
  + \Re[\gamma_5\bar\pi_5]\l( \frac{2 \kappa \pi_a}{3 a^2} - \sqrt{\frac{2}{3}}  \sigma_{(2)} \r) \nonumber\\
  &
  + \Re[\gamma_6 \bar\pi_6]\l(\frac{2 \kappa \pi_a}{3 a^2} - \sqrt{\frac{2}{3}}  \sigma_{(2)} \r).
\end{align}

It is an interesting exercise to compute the time evolution of the  linear constraints

\begin{align}
  &\frac{1}{N}\frac{d}{dt}\left(\tilde{\mathbb{S}}^{(1)}(\v k)\right) \;\mathring{\approx}\; \{\tilde{\mathbb{S}}^{(1)}(\v k),\mathbb{S}^{(2)}\}+\frac{1}{N}\partial_t\,\tilde{\mathbb{S}}^{(1)}(\v k) = i k^i\,\tilde{\mathbb{V}}_i^{(1)}(\v k),\\
  &\frac{1}{N}\frac{d}{dt}\left(\h k^i\,\tilde{\mathbb{V}}_i^{(1)}(\v k)\right) \;\mathring{\approx}\; \{\h k^i\,\tilde{\mathbb{V}}_i^{(1)}(\v k),\mathbb{S}^{(2)}\}+\frac{1}{N}\partial_t\,\left(\h k^i\,\tilde{\mathbb{V}}_i^{(1)}(\v k)\right) =  \sqrt{2}\sigma_{(4)}\,\h y^i\,\tilde{\mathbb{V}}_i^{(1)}(\v k)\nonumber \\
  &+\h k^i\,\tilde{\mathbb{V}}_i^{(1)}(\v k)\left(\frac{\kappa p_a}{6a^2}-
  \sqrt{\frac{2}{3}}\sigma_{(2)}+\sqrt{2}\sigma_{(3)}\right), \\
    &\frac{1}{N}\frac{d}{dt}\left(\h x^i\,\tilde{\mathbb{V}}_i^{(1)}(\v k)\right) \;\mathring{\approx}\; \{\h x^i\,\tilde{\mathbb{V}}_i^{(1)}(\v k),\mathbb{S}^{(2)}\}+\frac{1}{N}\partial_t\,\left(\h x^i\,\tilde{\mathbb{V}}_i^{(1)}(\v k)\right) =  -\frac{\sigma_{(6)}}{\sqrt{2}}\,\h y^i\,\tilde{\mathbb{V}}_i^{(1)}(\v k)\nonumber \\
  &+\h x^i\,\mathbb V_i^{(1)}(\v k)\left(\frac{\kappa p_a}{6a^2}+
  \frac{\sigma_{(2)}}{\sqrt{6}}-\frac{\sigma_{(5)}}{\sqrt{2}}\right),\\
  &\frac{1}{N}\frac{d}{dt}\left(\h y^i\,\tilde{\mathbb{V}}_i^{(1)}(\v k)\right) \;\mathring{\approx}\; \{\h y^i\,\tilde{\mathbb{V}}_i^{(1)}(\v k),\mathbb{S}^{(2)}\}+\frac{1}{N}\partial_t\,\left(\h y^i\,\tilde{\mathbb{V}}_i^{(1)}(\v k)\right) =  -\frac{\sigma_{(6)}}{\sqrt{2}}\,\h x^i\,\tilde{\mathbb{V}}_i^{(1)}(\v k)\nonumber \\
  &+\h y^i\,\tilde{\mathbb{V}}_i^{(1)}(\v k)\left(\frac{\kappa p_a}{6a^2}+
  \frac{\sigma_{(2)}}{\sqrt{6}}+\frac{\sigma_{(5)}}{\sqrt{2}}\right).
\end{align}
We see that the right-hand sides of these equations are linear combinations of the constraints themselves, and hence vanish on-shell, as expected from a system of first class constraints. 

\section{DECOUPLING GAUGE INVARIANT VARIABLES\label{B}}

In this appendix we provide further information about the canonical transformation introduced in Eq.\ (\ref{CT}). In Eq. \eqref{eq:new-Pi} we provided expressions for the new conjugate momenta $\Pi_{\alpha}$ for $\alpha=3,4,5,6$. We complement that information with the form of the new pure gauge configuration variables $\Gamma_{\alpha}$ for $\alpha=3,4,5,6$ in terms of old ones, namely,
\begin{align}
 & \Gamma_{3}(\v k) = \sqrt{\frac{3}{2}}\frac{a^2 |\v k|}{\kappa p_a + \sqrt{6} a^2 \sigma_{(2)}} \l( \gamma_2 - \sqrt{2} \gamma_1 \r)\\
 & \Gamma_{4}(\v k) = -\sqrt{\frac{3}{2}} \frac{1}{2 |\v k| \l(\kappa p_a + \sqrt{6} a^2 \sigma_{(2)}\r)} \l( \kappa p_a \gamma_2 + 2\sqrt{3} a^2 \sigma_{(2)} \gamma_1 \r)\\
 & \Gamma_5 = -\frac{\gamma_3}{\sqrt{2}} + \frac{\sqrt{3} a^2 \sigma_{(3)}}{\kappa p_a + \sqrt{6} a^2 \sigma_{(2)}} \l( \gamma_2 - \sqrt{2} \gamma_1 \r)\\
 & \Gamma_6 = -\frac{\gamma_4}{\sqrt{2}} + \frac{\sqrt{3} a^2 \sigma_{(4)}}{\kappa p_a + \sqrt{6} a^2 \sigma_{(2)}} \l( \gamma_2 - \sqrt{2} \gamma_1 \r)
\end{align}
On the other hand, we also wrote in Eqs. (\ref{eq:G2}) the form of the gauge invariant  variables $\Gamma_0$, $\Gamma_1$ and $\Gamma_2$. We write here their   conjugate momenta (also gauge invariant)
\begin{align}
 & \Pi_0 = \pi_0 + \frac{3 \pp^2}{4 a \l(\kappa p_a + \sqrt{6} a^2 \sigma_{(2)}\r)}\gamma_0 - \biggl( \frac{3 \sqrt{3 \kappa} \pp^3}{2 a^2 \l(\kappa p_a + \sqrt{6} a^2 \sigma_{(2)}\r)^2} + \frac{3 a^2 \pp \sigma_{(2)}}{2 \sqrt{2 \kappa} \l(\kappa p_a + \sqrt{6} a^2 \sigma_{(2)}\r)} \nonumber\\ 
 & + \frac{3\sqrt{3} a^4 \pp}{2 \sqrt{\kappa} \l(\kappa p_a + \sqrt{6} a^2 \sigma_{(2)}\r)^2}\l( \sigma_{(5)}^2 + \sigma_{(6)}^2 \r) + \frac{\sqrt{3} a^5 V_{\phi}}{2 \sqrt{\kappa} \l(\kappa p_a + \sqrt{6} a^2 \sigma_{(2)}\r)} \biggr) \gamma_1 
 + \biggl( \sqrt{\frac{3 \kappa}{2}} \frac{3 \pp^3}{2 a^2 \l(\kappa p_a + \sqrt{6} a^2 \sigma_{(2)}\r)^2}\nonumber\\
 &  + \sqrt{\frac{3}{2 \kappa}} \frac{3 a^4 \pp}{2 \l(\kappa p_a + \sqrt{6} a^2 \sigma_{(2)}\r)^2}\l( \sigma_{(5)}^2 + \sigma_{(6)}^2\r) - \sqrt{\frac{3}{2 \kappa}} \frac{\kappa \pp p_a - 2 a^5 V_{\phi}}{4 \l(\kappa p_a + \sqrt{6} a^2 \sigma_{(2)}\r)}\biggr) \gamma_2 - \frac{3 a^2 \pp \sigma_{(5)}}{4\sqrt{\kappa} \l(\kappa p_a + \sqrt{6} a^2 \sigma_{(2)}\r)} \gamma_5\nonumber\\
 & - \frac{3 a^2 \pp \sigma_{(6)}}{4\sqrt{\kappa} \l(\kappa p_a + \sqrt{6} a^2 \sigma_{(2)}\r)}\gamma_6\\
& \Pi_1 = \pi_5 - \frac{3 a^2 \pp \sigma_{(5)}}{4\sqrt{\kappa} \l(\kappa p_a + \sqrt{6} a^2 \sigma_{(2)}\r)} \gamma_0 + \biggl( \sqrt{\frac{3}{2}}\frac{a^5 }{2 \kappa \l(\kappa p_a + \sqrt{6} a^2 \sigma_{(2)}\r)} \l( \sigma_{(3)}^2 - \sigma_{(4)}^2 \r) - \frac{3 \sqrt{3} a \pp^2 \sigma_{(5)}}{2 \l(\kappa p_a + \sqrt{6} a^2 \sigma_{(2)}\r)^2} \nonumber\\
& - \frac{3 \sqrt{3} a^7 \sigma_{(5)}}{2 \kappa \l(\kappa p_a + \sqrt{6} a^2 \sigma_{(2)}\r)^2} \l( \sigma_{(5)}^2 + \sigma_{(6)}^2 \r)  - \frac{3 a^5 \sigma_{(2)} \sigma_{(5)}}{2 \sqrt{2} \kappa \l(\kappa p_a + \sqrt{6} a^2 \sigma_{(2)}\r)} + \frac{a^3 \sigma_{(5)}}{2 \sqrt{3} \kappa} \biggr) \gamma_1 + \biggl( \sqrt{\frac{3}{2}} \frac{3 a \pp^2 \sigma_{(5)}}{2 \l(\kappa p_a + \sqrt{6} a^2 \sigma_{(2)}\r)^2} \nonumber\\
& - \frac{\sqrt{3}a^5}{4 \kappa \l(\kappa p_a + \sqrt{6} a^2 \sigma_{(2)}\r)} \l( \sigma_{(3)}^2 - \sigma_{(4)}^2 \r) + \sqrt{\frac{3}{2}} \frac{3 a^7 \sigma_{(5)}}{2 \kappa \l(\kappa p_a + \sqrt{6} a^2 \sigma_{(2)}\r)^2} \l( \sigma_{(5)}^2 + \sigma_{(6)}^2 \r) + \frac{3 a^5 \sigma_{(2)} \sigma_{(5)}}{4 \kappa \l(\kappa p_a + \sqrt{6} a^2 \sigma_{(2)}\r)} \nonumber\\
& - \frac{5 a^3 \sigma_{(5)}}{4 \sqrt{6} \kappa} \biggr) \gamma_2 + \frac{a^3 \sigma_{(3)}}{2 \sqrt{2} \kappa} \gamma_3 - \frac{a^3 \sigma_{(4)}}{2 \sqrt{2} \kappa} \gamma_4 + \biggl( \frac{a p_a}{6} - \frac{a^3 \sigma_{(2)}}{2 \sqrt{6}\kappa} - \frac{3 a^5 \sigma_{(5)}^2}{4 \kappa \l(\kappa p_a + \sqrt{6} a^2 \sigma_{(2)}\r)}\biggr) \gamma_5 \nonumber\\
& - \frac{3 a^5 \sigma_{(5)} \sigma_{(6)}}{4 \kappa \l(\kappa p_a + \sqrt{6} a^2 \sigma_{(2)}\r)} \gamma_6 
\end{align}
\begin{align}
& \Pi_2 = \pi_6 - \frac{3 a^2 \pp \sigma_{(6})}{4 \sqrt{\kappa} \l(\kappa p_a + \sqrt{6} a^2 \sigma_{(2)}\r) } \gamma_0 - \biggl( \frac{3\sqrt{3} a \pp^2 \sigma_{(6)} }{2 \l(\kappa p_a + \sqrt{6} a^2 \sigma_{(2)}\r)^2 } 
+ \frac{3 a^5 \sigma_{(2)} \sigma_{(6)}}{2\sqrt{2} \kappa \l(\kappa p_a + \sqrt{6} a^2 \sigma_{(2)}\r)} 
\nonumber\\
& - \sqrt{\frac{3}{2}} \frac{a^5 \sigma_{(3)} \sigma_{(4)}}{\kappa \l(\kappa p_a + \sqrt{6} a^2 \sigma_{(2)}\r)} 
+ \frac{3 \sqrt{3} a^7 \sigma_{(6)}}{2 \kappa \l(\kappa p_a + \sqrt{6} a^2 \sigma_{(2)}\r)^2}\l( \sigma_{(5)}^2 + \sigma_{(6)}^2 \r) - \frac{a^3 \sigma_{(6)}}{2 \sqrt{3} \kappa}\biggr) \gamma_1 
\nonumber\\
& + \biggl( \sqrt{\frac{3}{2}}\frac{3 a \pp^2 \sigma_{(6)}}{2 \l(\kappa p_a + \sqrt{6} a^2 \sigma_{(2)}\r)^2 } + \frac{3 a^5 \sigma_{(2)} \sigma_{(6)}}{4 \kappa \l(\kappa p_a + \sqrt{6} a^2 \sigma_{(2)}\r)} - \frac{\sqrt{3} a^5 \sigma_{(3)} \sigma_{(4)}}{2 \kappa \l(\kappa p_a + \sqrt{6} a^2 \sigma_{(2)}\r)} \nonumber \\
& + \sqrt{\frac{3}{2}}\frac{3 a^7 \sigma_{(6)}}{2 \kappa \l(\kappa p_a + \sqrt{6} a^2 \sigma_{(2)}\r)^2 } \l(\sigma_{(5)}^2 + \sigma_{(6)}^2 \r) - \frac{5 a^3 \sigma_{(6)}}{4 \sqrt{6} \kappa} \biggr) \gamma_2 + \frac{a^3 \sigma_{(4)}}{2 \sqrt{2} \kappa} \gamma_3 + \frac{a^3 \sigma_{(3)}}{2 \sqrt{2} \kappa} \gamma_4 
\nonumber\\
& - \frac{3 a^5 \sigma_{(5)} \sigma_{(6)}}{4 \kappa \l(\kappa p_a + \sqrt{6} a^2 \sigma_{(2)}\r)} \gamma_5 
+ \biggl( \frac{a p_a}{6} - \frac{a^3 \sigma_{(2)}}{2 \sqrt{6} \kappa} - \frac{3 a^5 \sigma_{(6)}^2}{4 \kappa \l(\kappa p_a + \sqrt{6} a^2 \sigma_{(2)}\r)} \biggr) \gamma_6
\end{align}
As a check, one can easily see that these variables satisfy the canonical Poisson algebra
\bea \label{Gamma-comm}\{\Gamma_\alpha(\v k), \Pi_\beta (\v{k}')\}&=&
\mathcal{V}_0^{-1} \, \delta_{\alpha\beta} \, \delta_{\vec{k},-\vec{k}'}\, , \nonumber \\
\{ \Gamma_\alpha(\v k),\Gamma_\beta (\v{k}')\}&=&0
\, , \nonumber \\
\{ \Pi_\alpha(\v k), \Pi_\beta (\v{k}')\}&=&0\, .
\ea

The total Hamiltonian for the perturbations $\mathcal{H}_{\rm total}=\int d^3x N\, \mathbb{S}^{(2)}(\vec x)$ can now be written in terms of these new variables, starting from Eq.\ (\ref{S2}) (again, one needs to perform a time-dependent canonical transformation). One obtains
\begin{equation}
\mathcal{H}_{\rm total} = \mathcal{H_{\rm pert}} + \frac{N(t)}{2\, a(t) }\, \sum_{\vec k}\sum_{\alpha,\alpha'=3}^6 \tilde{\cal U}_{\alpha\alpha'}\Gamma_\alpha(\vec k)\bar\Gamma_{\alpha'}(\vec k)+\sum_{\vec k}\sum_{\alpha=3}^6\, \Lambda_\alpha(\vec k)\, \Pi_{\alpha}(\vec k),
\end{equation}
where  $\Lambda_\alpha(\vec k)$ are functions of the perturbations of the lapse and shift, that also depend linearly on $\Gamma_{\alpha}(\vec k)$ and $\Pi_{\alpha}(\vec k)$ with $\alpha=3,4,5,6$. But note that $\Lambda_\alpha(\vec k)$ are multiplying the linearized constraints, so they are Lagrange multipliers and, furthermore, they do not affect the dynamics of the gauge invariant variables, since the constraints vanish on-shell. The term $\mathcal{H_{\rm pert}}$ was defined in (\ref{Ham2}) and it only involves gauge invariant variables. Hence, this expression for $\mathcal{H}_{\rm total}$ shows explicitly that the dynamics of the gauge invariant degrees of freedom $\Gamma_{\alpha},\Pi_{\alpha}$ for $\alpha=0,1,2$ decouples from pure gauge ones. This is why  in Sec. \ref{sec.3.B} we restricted our attention to the term  $\mathcal{H_{\rm pert}}$.

\section{FOCK QUANTIZATION OF TWO HARMONIC OSCILLATORS WITH A TIME-DEPENDENT COUPLING : A PEDAGOGICAL EXAMPLE}\label{appB}

This appendix summarizes the Hamiltonian formulation of classical and quantum theories of two coupled harmonic oscillators, with spring ``constants'' that depend on time. This system has many similarities with the evolution of cosmological perturbations in Bianchi I  spacetimes discussed in the main body of this article, although the phase space of the latter is infinite dimensional. Hence, the goal of this appendix is to serve as a pedagogical introduction to the Fock quantization techniques of coupled linear systems used in this paper, in the simpler situation of a finite dimensional model. 

\subsection{Classical theory}

Consider two point masses $m_1$ and $m_2$, each of them attached to a spring, with time-dependent spring constants $k_1(t)$ and $k_2(t)$ respectively, and joined together by another spring with constant $k_c(t)$, also  time dependent. The phase space $\mathbb{V}$ of this system  is four dimensional. Elements $v$ of $\mathbb{V}$ are characterized by the values of two pairs of canonically conjugated variables $v_a=(x_1,x_2,p^1,p^2)$, where the index $a$ runs from 1 to 4. The basic  Poisson brackets are
\be \label{cc} \{v_a,v_b\}=\Omega_{ab}\ , \  {\rm with}\ \  \Omega_{ab}=\left( {\begin{array}{cc} 0& \mathbb{I}_{2\times2}   \\ -\mathbb{I}_{2\times2}& 0 \end{array} } \right)\, , \ee or, written in components
\be \label{cc} \{x_i,x_j\}=0\, ,\hspace{1cm} \, \{p^i,p^j\}=0\, ; \hspace{1cm} \{x_i,p^j\}=\delta_{i}^{j}\, . \ee
Dynamics in $\mathbb{V}$ is generated by the Hamiltonian 
\be \label{ham-osc} H(t)=\frac{1}{2}\, p^i p^j\, M^{-1}_{ij}+\frac{1}{2}\,x_i x_j \, K^{ij}(t),\ee
where \be M_{ij} =\left( {\begin{array}{cc} m_1& 0   \\ 0& m_2 \end{array} } \right)\, , \hspace{1cm} K^{ij} =\left( {\begin{array}{cc} k_1(t)+k_{\rm c}(t)& -k_{\rm c}(t)  \\ -k_{\rm c}(t) & k_2(t)+k_{\rm c}(t) \end{array} } \right) \, .\ee
Hamilton's equations are
\bea \dot x_i&=&\{x_i,H \}=M^{-1}_{ij} p^j\, ,\\ \dot p^i &=&\{p^i,H \}=-K^{ij} x_j  \, .\nonumber \ea
More explicitly
\bea \dot x_1&=&p_1/m_1 \, ,\\ \nonumber
\dot p_1&=&-(k_1+k_{\rm c})\, x_1+k_{\rm c}\, x_2 \, ,\\ \nonumber
 \dot x_2&=&p_2/m_2 \, ,\\ \nonumber
 \dot p_2&=&k_c x_1-(k_2+k_{\rm c})\, x_2 \, .\ea
These equations can be combined into  second-order differential equations 
\be \label{eom} \ddot x_i(t)+{\Lambda_i}^j(t) \, x_j(t)=0 \, ,\ee
where ${\Lambda_i}^j(t)=M_{ik}K^{k j}(t)$.
If ${\Lambda_i}^j(t)$ were time independent, these equations could be easily decoupled, and both the classical and quantum theories would reduce to the study of two independent oscillators. But in the time-dependent situation one cannot diagonalize simultaneously  ${\Lambda_i}^j(t)$ and the differential operator $\delta_{i}^{\, j} \frac{d^2}{dt^2}$ by means of usual transformations local in time. However, in spite of the coupling between the point masses, the equations of motion are  {\em linear}, and consequently the space of solutions is a vector space (i.e.\ linear combinations of solutions are  solutions).  It is this linear structure that makes it possible to quantize the system in an exact  way.

In the remainder of this subsection we will write the classical theory in a convenient form that will serve as  starting point to build a Fock quantization in the next subsection.  First, consider the complex phase space $\mathbb{V}_{\mathbb{C}}$, constructed by taking all possible linear combination with complex coefficients of elements of $\mathbb{V}$. It turns out that $\mathbb{V}_{\mathbb{C}}$ carries a natural {\em product} on it, which originates from the symplectic structure of the Hamiltonian theory, and plays a central role in the quantization of the system. Given two  elements of  $\mathbb{V}_{\mathbb{C}}$, $v^{(1)}=(\vec{x}^{(1)},\vec{p}^{(1)})$ and $v^{(2)}=(\vec{x}^{(2)},\vec{p}^{(2)})$, their product is the complex number given by
\be \label{prod} \langle v_{(1)}, v_{(2)}\rangle\equiv i \, \alpha^{-1} \, v_a^{(1)}\,  \bar{v}_b^{(2)} \, \Omega^{ab}= i \alpha^{-1}\,  (\bar x^{(1)}_{i}\,   p^{i\, (2)} -\bar p^{ i\, (1)}\,  x_i^{(2)}) \, ,
\ee
(sum over repeated indices is understood). In this expression, $\alpha $ is an arbitrary parameter with dimensions of action, and it is introduced to make this product dimensionless. Physical predictions will be insensitive to the choice of $\alpha $. Although  it is natural to fix $\alpha=\hbar$, we prefer not to make such a choice, because that would make  unclear how to take the classical limit in several expressions below, since this limit corresponds to $\hbar \to 0$ with fixed $\alpha$. 

We will now take advantage of the product (\ref{prod}) to  describe in more detail the classical theory and, in the next section,  to  quantize it. It is easy to check that (\ref{prod}) satisfies  all properties of a Hermitian inner product, except that it is not positive definite in $\mathbb{V}_{\mathbb{C}}$. Therefore, the obvious candidate for Hilbert space of the quantum theory, namely the Cauchy completion of the vector space $\mathbb{V}_{\mathbb{C}}$ with the product $\langle \cdot,\cdot \rangle$, is not a viable choice. The standard way to proceed is to notice that $\mathbb{V}_{\mathbb{C}}$ can always be written as the direct sum of two subspaces $\mathbb{V}_{\mathbb{C}}=\mathbb{V}_{\mathbb{C}}^+\oplus \mathbb{V}_{\mathbb{C}}^-$,  satisfying  that  $\langle \cdot,\cdot \rangle$ is positive definite when restricted to  $\mathbb{V}_{\mathbb{C}}^+$, and negatively definite in $\mathbb{V}_{\mathbb{C}}^-$.\footnote{\label{Mink}A pedagogical mathematical analogy is used to consider  the  Minkowski spacetime $M_2$ in two spacetime dimensions, and think about different ways of writing  $M_2$ as a  direct sum of two mutually orthogonal one-dimensional  subspaces,  $M_2=M_+\oplus M_-$, with $M_+$  spacelike and $M_-$ timelike, so the Minkowski metric is positive and negative definite when restricted to them, respectively. Familiarity with special relativity tells us that there are infinitely many different choices for $M_+$, as many as  inertial reference frames.} It is convenient to choose $\mathbb{V}_{\mathbb{C}}^-$ to be the complex conjugate of $\mathbb{V}_{\mathbb{C}}^+$. It is the subspace $\mathbb{V}_{\mathbb{C}}^+$ that will be used to build the Hilbert space of the quantum theory. 

A convenient practical way to make a choice of $\mathbb{V}_{\mathbb{C}}^+$  is to choose a set $\{ {\boldsymbol v}^{(\lambda)} \}$, with $\lambda=1,2$, of two  orthogonal elements of $\mathbb{V}_{\mathbb{C}}$ of positive norm (and equal 1 for convenience). $\mathbb{V}_{\mathbb{C}}^+$ arises then as the subspace spanned by $\{ {\boldsymbol v}^{(\lambda)}  \}$; the conjugate set, $\{ \bar{\boldsymbol v}^{(\lambda)} \}$, spans $\mathbb{V}_{\mathbb{C}}^-$, and therefore both sets together form a complete  basis of $\mathbb{V}_{\mathbb{C}}$. Once this choice has been made, any element $v$ of our  physical, real phase space $\mathbb{V}$ can be written in a unique manner in terms of this basis (since $\mathbb{V}$ is a subspace of $\mathbb{V}_{\mathbb{C}}$)
\be \label{gb} v=(\vec{x},\vec{p})=\sum_{\lambda=1}^2 a_{\lambda}\, {\boldsymbol v}^{(\lambda)} +\bar a_{\lambda}\, \bar {\boldsymbol v}^{(\lambda)} \, , \ee
where $a_{\lambda}$ are complex coefficients. These coefficients  can be then determined by projecting $v$ on the basis element $ v_{\lambda}$
\be \label{a} a_{\lambda}=\langle  {\boldsymbol v}^{(\lambda)} ,v \rangle \, . \ee
Then, using (\ref{a}), the canonical Poisson brackets for $x_i$ and $p^j$ (\ref{cc}) imply\footnote{For the inverse to also be  true, i.e.\ for the algebra of creation and annihilation operators to imply the canonical Poisson brackets, the basis vectors  $ v^{(\lambda)}_{a}$ must also satisfy the condition:
\be\label{exctcond2} \frac{1}{\alpha} \,\sum_{\lambda=1}^2 \left(\boldsymbol v^{(\lambda)}_{a}\bar{\boldsymbol v}^{(\lambda)}_{b}-\bar{\boldsymbol v}^{(\lambda)}_{a}\boldsymbol v^{(\lambda)}_{b}\right)=i\, \Omega_{ab}\, , \ee where 
\be \Omega_{ab}=\left( {\begin{array}{cc} 0& \mathbb{I}_{2\times2}   \\ -\mathbb{I}_{2\times2}& 0 \end{array} } \right)\, .\ee}

\bea \label{acc} \{a_{\lambda}, a_{\lambda'} \} &=&\frac{i}{\alpha} \langle  {\boldsymbol v}^{(\lambda)} , \bar{\boldsymbol v}^{(\lambda')} \rangle=0\,, \\
 \{a_{\lambda}, \bar a_{\lambda'} \} &=&-\frac{i}{\alpha} \, \langle {\boldsymbol v}^{(\lambda)} ,{\boldsymbol v}^{(\lambda')} \rangle=-\frac{i}{\alpha}\, \delta^{\lambda,\lambda'}\, .\nonumber\ea
(Note that $a_{\lambda}$ is dimensionless.) An important fact to keep in mind in this construction is that there is ambiguity in the choice of $\mathbb{V}_{\mathbb{C}}^+$: there are (infinitely)  many different ways of splitting  $\mathbb{V}_{\mathbb{C}}$ into a direct sum of two subspaces with the  properties mentioned above. If the Hamiltonian is time independent, the  symmetry under time translations of the system provides a natural choice of $\mathbb{V}_{\mathbb{C}}^+$, commonly called the positive frequency subspace. But this choice is not available in a general time-dependent situation.\footnote{This issue has important consequences in a field theory with infinitely many degrees of freedom, where the Stone-von Newman theorem does not apply. For a finite number of harmonic oscillators, different choices of $\mathbb{V}_{\mathbb{C}}^+$ give rise to Hilbert spaces that are all unitarily equivalent, although the state that we call ``the vacuum'' depends on the choice.}

We will now discuss the classical dynamics. Time evolution from time $t_0$ to $t$ will map  each of the basis elements $\boldsymbol{v}^{(\lambda)}\in \mathbb{V}_{\mathbb{C}}^+$ to another element ${\boldsymbol v}^{(\lambda)}(t):=E_{t,t_0} {\boldsymbol v}^{(\lambda)} $ of $\mathbb{V}_{\mathbb{C}}$, where $E_{t,t_0}$ is the canonical map implementing the Hamiltonian flow in phase space. Then, we can substitute  ${\boldsymbol v}^{(\lambda)}(t)$ in Eq.\ (\ref{gb}) to obtain the  evolution of an arbitrary element of the real phase space $v\in \mathbb{V}$ 
\be  \label{evol} v(t)=(\vec{x}(t),\vec{p}(t))=\sum_{\lambda=1}^2 a_{\lambda}\,   {\boldsymbol v}^{(\lambda)}(t) +\bar a_{\lambda}\,  \bar {\boldsymbol v}^{(\lambda)}(t) \, . \ee
 
As an example, consider  the positive norm subspace $\mathbb{V}_{\mathbb{C}}^{+}$ spanned by 

\be \label{bvti} {\boldsymbol v}^{(1)}=(\left( {\begin{array}{c} \frac{1}{\sqrt{2 w_1(t_0)\,  m_1/\alpha}}  \\ 0  \end{array} } \right) ,\left( {\begin{array}{c}   \frac{-i\, w_1(t_0)\,  m_1}{\sqrt{2 w_1(t_0)\,  m_1/\alpha}} \\ 0 \end{array} } \right) ) \, ; \hspace{0.1cm}  {\boldsymbol v}^{(2)}=(\left( {\begin{array}{c}0\\  \frac{1}{\sqrt{2 w_2(t_0) \, m_2/\alpha}}   \end{array} } \right) ,\left( {\begin{array}{c} 0\\  \frac{-i\, w_2(t_0)\,  m_2}{\sqrt{2 w_1(t_0) m_2/\alpha}}  \end{array} } \right) )\ee
where $t_0$ is a chosen instant of time and $w_i(t)\equiv \sqrt{k_i(t)/m_i}$. These two basis vectors, together with their conjugates, provide a complete basis  in $\mathbb{V}_{\mathbb{C}}$. It is straightforward to show  the orthonormality relations $\langle {\boldsymbol v}^{(1)}, {\boldsymbol v}^{(1)}\rangle= \langle  {\boldsymbol v}^{(2)}, {\boldsymbol v}^{(2)}\rangle=1$, $\langle  {\boldsymbol v}^{(1)}, {\boldsymbol v}^{(2)}\rangle=\langle  {\boldsymbol v}^{(1)}, \bar{\boldsymbol v}^{(1)}\rangle=\langle  {\boldsymbol v}^{(1)}, \bar {\boldsymbol v}^{(2)}\rangle=\langle  {\boldsymbol v}^{(2)}, \bar {\boldsymbol v}^{(2)}\rangle=0$, as well as properties (\ref{exctcond2}). If the two oscillators were decoupled and the spring constants were time independent,  $ {\boldsymbol v}^{(1)}$ and $ {\boldsymbol v}^{(2)}$ in (\ref{bvti}) would be the initial data for positive frequency solutions for which only the first or second oscillator is excited, respectively: \bea  {\boldsymbol v}^{(1)}(t)&:=&E_{t,t_0}  {\boldsymbol v}^{(1)}=(\left( {\begin{array}{c} \frac{e^{-i\, w_1 t}}{\sqrt{2 w_1 m_1/\alpha}}  \\ 0  \end{array} } \right),\left( {\begin{array}{c} \frac{-i\, w_1m_1\, e^{-i\, w_1 t}}{\sqrt{2 w_1 m_1/\alpha}}  \\ 0  \end{array} } \right))\, , \nonumber \\   {\boldsymbol v}^{(2)}(t)&:=&E_{t,t_0}  {\boldsymbol v}^{(1)}=(\left( {\begin{array}{c}0\\  \frac{e^{-i\, w_2 t}}{\sqrt{2 w_2 m_2/\alpha}}  \end{array} } \right),\left( {\begin{array}{c} 0\\ \frac{-i\, w_2 \, m_2\, e^{-i\, w_2 t}}{\sqrt{2 w_2 m_2/\alpha}}   \end{array} } \right))\, . \ea But in the time-dependent case under consideration, the form of $ {\boldsymbol v}^{(1)}(t)$ and $ {\boldsymbol v}^{(2)}(t)$ is more complicated, and will generically contain excitations in both oscillators, even if only one of them was initially excited.

\subsection{Quantum  theory} 

Now that we have written the classical theory in a convenient way, the quantization is straightforward. Given a   positive-negative norm decomposition, $\mathbb{V}_{\mathbb{C}}=\mathbb{V}_{\mathbb{C}}^+\oplus \mathbb{V}_{\mathbb{C}}^-$, the one-particle Hilbert  space $\mathfrak{h}$ is simply given by $\mathbb{V}_{\mathbb{C}}^+$ equipped with the Hermitian inner product $\langle \cdot,\cdot \rangle$. The Hilbert space of the theory is then the symmetric Fock space $\mathcal{F}$ constructed from $\mathfrak{h}$ (see e.g. Appendix A of \cite{waldbook} for details of this construction).\footnote{In textbooks, it is more common to use the space of square integrable functions in the configuration space to build  the Hilbert  space of a  finite set of harmonic oscillators. We use here a different representation, namely a Fock representation based on the classical phase space. Both representations are, of course, unitarily equivalent, and hence describe the same physics. The Fock approach is however convenient in quantum field theory, due to the infinite number of degrees of freedom of the system.} The position and momentum operators at the initial time $t_0$ are represented in  $\mathcal{F}$ as

\be \label{qgb} \hat V=(\hat{\vec{x}},\hat{\vec{p}})=\sum_{\lambda=1}^2  \hat a_{\lambda}\, {\boldsymbol v}^{(\lambda)} +\hat a_{\lambda}^{\dagger}\, \bar {\boldsymbol v}^{(\lambda)} \, .\ee
The commutation relations are obtained from the Poisson brackets of the classical theory via the Dirac replacement rule $\{\cdot ,\cdot\}\rightarrow  [\cdot,\cdot]/(i \hbar)$. Therefore
 \be \label{qcc1} [\hat V_a,\hat V_b]=i\hbar \, \Omega_{ab}\, , \ee
or more explicitly
\be \label{qcc} [\h x_i, \h x_j]=0\, ;\hspace{1cm} \, [\h p^i,\h p^j]=0\, ; \hspace{1cm} [\h  x_i,\h p^j]=i\hbar \, \delta_{i}^{j}\, . \ee
And from (\ref{acc}) we have
\bea [\h a_{\lambda}, \h a_{\lambda'} ] &=&- \frac{\hbar}{\alpha}\,  \langle {\boldsymbol v}^{(\lambda)} ,\bar {\boldsymbol v}^{(\lambda')} \rangle=0\\
 \lbrack \hat a_{\lambda}, \hat a^{\dagger}_{\lambda'} \rbrack&=& \frac{\hbar}{\alpha} \,  \langle{\boldsymbol v}^{(\lambda)} ,{\boldsymbol v}^{(\lambda')} \rangle= \frac{\hbar}{\alpha} \, \delta_{\lambda,\lambda'}\nonumber \, .\ea
These commutation relations reveal that $\hat a_{\lambda}$ and  $\hat a^{\dagger}_{\lambda}$ are creation and annihilation operators. With the choice $\alpha=\hbar$, we recover the textbook expression $\lbrack \hat a_{\lambda}, \hat a^{\dagger}_{\lambda'} \rbrack=\delta_{\lambda\lambda'}$. Now, the state $|0\rangle$ that is annihilated by the operators $\hat a_{\lambda}$ is called the Fock vacuum. A basis of the Fock space is obtained by acting repeatedly on $|0\rangle$ with the creation operators $a^{\dagger}_{\lambda}$: $|n_1, n_2 \rangle\equiv \left(\frac{\alpha}{\hbar}\right)^{\frac{n_1}{2}\frac{n_2}{2}}(n_1! n_2!)^{-1/2}\, (\hat a^{\dagger}_{1})^{n_1}(\hat a^{\dagger}_{2})^{n_2}|0\rangle$, for all integers $n_1$ and $n_2$.  It should be obvious from this construction that the notion of vacuum depends on our initial choice of positive norm subspace $\mathbb{V}_{\mathbb{C}}^+$, since the definition of annihilation operators $\hat a_{\lambda}$ rests on that choice. 

Let us now consider quantum evolution. Given  initial  and final times, $t_0$ and $t>t_0$, dynamics can be implemented either in the Heisenberg or Schr\"odinger pictures. Formally, time evolution  is generated by the standard time-ordered exponential $\hat U_{t,t_0}=T\left[ \exp (-i/\hbar \int_{t_0}^{t}\hat H(t')\, dt')\right]$, where $\hat H(t)$ is the quantum Hamiltonian obtained from Eq.\ \eqref{ham-osc}. This unitary operator $\hat U_{t,t_0}$ is the starting point of the perturbative expansion for small coupling constant $k_c\ll k_1,k_2$, obtained by truncating the exponential at a suitable order in powers of $k_c$.

However, if one looks for exact solutions for general values of the coupling $k_c$, it is more convenient to proceed in a different way, which in fact is closer to what is commonly done in quantum field theories in curved spacetimes. In the Heisenberg picture, where states do not evolve in time, the evolution of position and momentum operators can be obtained from the classical expression  (\ref{evol}) by simply substituting  $a_{\lambda}$ and $\bar a_{\lambda}$ by the associated operators or, equivalently, by substituting the basis vectors ${\boldsymbol v}^{(\lambda)}$ in (\ref{qgb}) by the classical  solutions ${\boldsymbol v}^{(\lambda)}(t)=E_{t,t_0}{\boldsymbol v}^{(\lambda)}$ 
\be  \hat V(t)=(\hat{\vec{x}}(t),\hat{\vec{p}}(t))=\sum_{\lambda=1}^2  \hat a_{\lambda}\, {\boldsymbol v}^{(\lambda)}(t) +\hat a_{\lambda}^{\dagger}\, \bar{\boldsymbol v}^{(\lambda)}(t) \, .\ee
Therefore, to evolve the position and momentum operator we just need the solution to the classical equations of motion (\ref{eom}) for each basis  vector ${\boldsymbol v}^{(\lambda)}$. No perturbative expansion is required in this calculation, and therefore the result is valid for arbitrary values of the coupling $k_c$. 

In the Schr\"odinger picture, the evolution of the Fock vacuum can be written as\footnote{It would be incorrect to identify the unitary operator $ \hat U_{t,t_0}$ with the nonunitary operator written in the right-hand side of this equation. Rather, this expression only tells us the result of acting with  $ \hat U_{t,t_0}$ on the vacuum.}
\be \label{evolop} \hat U_{t,t_0}|0\rangle= N\, \exp{\big[\frac{\alpha}{\hbar}\sum_{\lambda,\lambda'=1}^2 V_{\lambda\lambda'} a^{\dagger}_{\lambda} a^{\dagger}_{\lambda'}\big]} \, |0\rangle \, ,\ee
where $N^2=\left(\sum_{n,m=0}^{\infty} |\Delta_{nm}|^2 n!m!\right)^{-1}$, with $$\Delta_{nm}:=\sum_{n_1,n_2,n_3}\frac{1}{n_1!n_2!n_3!} \, (V_{11})^{n_1} (V_{22})^{n_2} (2V_{12})^{n_3} \, \delta_{2n_1+n_3,n} \, \delta_{2n_2+n_3,m}\, ,$$
and $V_{\lambda\lambda'}(t,t_0):=\sum_{\lambda''} \frac{1}{2}\bar \beta_{\lambda''\lambda}(t,t_0)\bar \alpha^{-1}_{\lambda'\lambda''}(t,t_0)$. In these expressions, $\alpha_{\lambda \lambda'}(t,t_0)$ and $\beta_{\lambda\lambda'}(t,t_0)$ are the Bogoliubov coefficients\footnote{Note that these coefficients encode the classical dynamics, in the sense that they provide the relation between  ${\boldsymbol v}^{(\lambda)}(t)$ and initial data ${\boldsymbol v}^{(\lambda)}(t_0)$: ${\boldsymbol v}^{(\lambda)}(t)=\sum_{\lambda'} \alpha_{\lambda\lambda'} (t,t_0)\, {\boldsymbol v}^{(\lambda')}(t_0)+\beta_{\lambda\lambda'}(t,t_0)\,  \bar{\boldsymbol v}^{(\lambda')}(t_0)$.} $\alpha_{\lambda\lambda'} (t,t_0):=\langle {\boldsymbol v}^{(\lambda')}(t_0) ,{\boldsymbol v}^{(\lambda)}(t)\rangle$ and $\beta_{\lambda\lambda'} (t,t_0):=-\langle \bar{\boldsymbol v}^{(\lambda')}(t_0) ,{\boldsymbol v}^{(\lambda)}(t)\rangle$. 
They satisfy the following properties:
\begin{align}\label{eq:bogo1}
\sum_{\lambda''}\alpha_{\lambda\lambda''}\bar\alpha_{\lambda'\lambda''}-\beta_{\lambda\lambda''}\bar\beta_{\lambda'\lambda''}&=\delta_{\lambda\lambda'},\\
\label{eq:bogo2}
\sum_{\lambda''}\alpha_{\lambda\lambda''}\beta_{\lambda'\lambda''}-\beta_{\lambda\lambda''}\alpha_{\lambda'\lambda''}&=0.%\\
\end{align}
In addition, $\bar \alpha^{-1}_{\lambda'\lambda''}(t,t_0)$ is the $\lambda' \lambda''$ component of the inverse of matrix $\bar \alpha(t,t_0)$ [Eqs.\ \eqref{eq:bogo1} and \eqref{eq:bogo2} guarantee that this matrix is invertible]. Furthermore, from Eq.\ \eqref{eq:bogo2}, one can easily prove that the matrix $V_{\lambda\lambda'}$ is symmetric, $V_{\lambda\lambda'}=V_{\lambda'\lambda}$.

The state (\ref{evolop}) is an excited state, and  has a quite interesting structure. These details are further discussed in the next subsection in a concrete scenario of direct relevance for the main body of this paper.

\subsection{The in and out representations and the $\mathcal{S}$-matrix}

Consider now the example in which the following two conditions hold: \begin{enumerate} 

\item  The spring ``constants''  $k_1(t)$ and $k_2(t)$ are indeed constant $k_1(t)=k^{\rm in}_1$ and $k_2(t)=k^{\rm in}_2$ in the past until $t=t_{\rm in}$, then vary smoothly till $t=t_{\rm out}$, and then become constant again $k_1(t)=k^{\rm out}_1$ and $k_2(t)=k^{\rm out}_2$ to the future of $t_{\rm out}$.

\item  The coupling between the oscillators $k_c(t)$ vanishes to the past of $t_{\rm in}$ and to the future of $t_{\rm out}$, but it is nonzero in between. 
\end{enumerate}
Then, before $t_{\rm in}$ and after $t_{\rm out}$ the two oscillators are time independent and uncoupled, although their initial and final spring constants are different. We are concerned now with describing the evolution of the system from an initial time $t_1<t_{\rm in}$ to a final instant $t_2>t_{\rm out}$. Note that since the Hamiltonian is time independent in the past and in the future, we have two natural quantum representations, the {\it in} and {\it out}, that are selected by the time translational symmetry in each asymptotic region. We will denote the associated Fock space as $\mathcal{F}_{in}$ and $\mathcal{F}_{out}$, respectively. The vacuum state in $\mathcal{F}_{in}$, $|in\rangle$, is the preferred notion of vacuum (ground state of the Hamiltonian) to the past of $t_{\rm in}$ and, similarly, the vacuum  state in $\mathcal{F}_{out}$, $|out\rangle$, is the ground state of the Hamiltonian to the future of $t_{\rm out}$. We want to answer the following question: if the system is prepared at $t_1$ in the $|in\rangle$ state, and then evolved to $t_2$, how does the evolved state look when compared to $|out\rangle$? Note that this question is slightly different from the discussion on time evolution around Eq.  (\ref{evolop}); now we want to express the evolved state in the {\it out} Fock space. The operator providing this evolution is known as the $\mathcal{S}$-matrix, and we will denote  it as $\mathcal{S}_{(in,out)}$. Its action on  $|in\rangle$ produces 
\be \label{inout} \mathcal{S}_{(in,out)}|in\rangle=  N\, \exp{\big[\frac{\alpha}{\hbar}\sum_{\lambda,\lambda'=1}^2  V_{\lambda\lambda'} \, \h  a^{out\, \dagger}_{\lambda} \h a^{out \, \dagger}_{\lambda'}\big]} \, |out \rangle \, , \ee
where, as before, $V_{\lambda\lambda'}:=\sum_{\lambda''} \frac{1}{2}\, \bar \beta_{\lambda'' \lambda } \, \bar \alpha^{-1}_{\lambda' \lambda''}$,  but  the Bogoliubov coefficients  that appear in this equation are now given by
\be \alpha_{\lambda\lambda'}:=\langle {\boldsymbol v}_{out}^{(\lambda')}(t_2) ,{\boldsymbol v}_{in}^{(\lambda)}(t_2)\rangle \, , \hspace{0.5cm} \beta_{\lambda\lambda'} :=-\langle \bar{\boldsymbol v}_{out}^{(\lambda')}(t_2) ,{\boldsymbol v}_{in}^{(\lambda)}(t_2)\rangle\, .\ee
Equation (\ref{inout}) tells us that the ground state at early times evolves to a state which is quite different from the vacuum in the {\it out} region. Expanding the exponential in (\ref{inout}) one can see that the evolved state is made of linear combinations of states containing an {\em even number of excitations} at late times
\be \mathcal{S}_{(in,out)}|in\rangle=N\, \Big (|out\rangle+\sqrt{2!}\, V_{11}\, |2_1\rangle+\sqrt{2!}\, V_{22}\, |2_2\rangle+2V_{12}\, |1_11_2\rangle+ \frac{\sqrt{3!}\, }{2!}4\, V_{11}V_{12}\, |3_11_2\rangle+... \Big )\, ,\ee
where $|n_1m_2\rangle$ indicates a state in $\mathcal{F}_{out}$ with $n$ excitations  in the first oscillator and $m$ in the second. This result is commonly interpreted by saying that  the evolution has created pairs of excitations.  For a general coupling $k_c(t)$, this state cannot be written as the product of two states each belonging to the Hilbert space of one of the oscillators, and hence the two oscillators become {\em entangled} quantum mechanically at late times. Since there is no entanglement in the initial state $|in\rangle$, this entanglement can be entirely attributed to the coupling between the oscillators at intermediate stages of the evolution. 
Recall now that a density matrix represents a pure state if and only if it is idempotent, i.e.\ its square is itself (or equivalently if the trace of the density matrix squared is equal to one).

One way of showing explicitly the existence of entanglement between the two oscillators in the final state is by following the textbook recipe: Think about oscillator 1 and oscillator 2 as two subsystems.  Build the density matrix $\rho$ for the pure state (\ref{inout})
\be \rho=\mathcal{S}_{(in,out)}|in\rangle \langle in| \mathcal{S}^{\dagger}_{(in,out)}\, .\ee
Now, trace-out from  $\rho$ the degrees of freedom of one of the subsystems, say oscillator 1
\be \label{red} \rho_{\rm red}:={\rm Tr}_1[\rho]=N^2 \sum_{n_2,m_2,k=0}^{\infty} \,k! \sqrt{n_2!}\sqrt{m_2!}  \, \Delta_{kn_2}\, \bar\Delta_{km_2} \, |n_2\rangle \, \langle m_2| \, . \ee
The square of this reduced density matrix,  $\rho_{\rm red}^2$, has trace different from one for a generic coupling $k_c(t)$, and hence it represents a mixed state. An equivalent way of accounting for this entanglement is by  simply computing   the Von Neumann entropy of $\rho_{\rm red}$, which agrees with the entanglement entropy between the two oscillators (since the initial state is a pure state).  On the other hand, in the absence of coupling, $k_{\rm c}(t)=0$ for all $t$, one finds that the Bogoliubov coefficients $\beta_{12}$ and   $\beta_{21}$ vanish, and the final state  becomes  a product state 
\be \mathcal{S}_{(in,out)}|in\rangle=  N\, \left(\exp{\big[ \frac{\alpha}{\hbar}\, V_{11} \, \h  a^{out\, \dagger}_{1} \h a^{out \, \dagger}_{1}\big]} \otimes \exp{\big[ \frac{\alpha}{\hbar}\, V_{22} \, \h  a^{out\, \dagger}_{2} \h a^{out \, \dagger}_{2}\big]}\right) \,  |out \rangle \, .\ee
The reduced density matrix represents then a pure state, and the two oscillators are unentangled, as expected. 
 
The existence  of entanglement can also be understood by computing the correlation functions of this theory. In the ``in" vacuum they are
\be \langle in|\hat V_{(a}\hat V_{b)}|in\rangle =\frac{\hbar}{\alpha}\, \sum_{\lambda=1}^2\left({\boldsymbol v}_{in\, (a}^{(\lambda) }\bar{\boldsymbol v}_{ in\, b)}^{(\lambda)}\right)\, , \ee
where the brackets around indices indicates symmetrization (the  antisymmetric part is state independent and completely determined by the canonical commutation relations). The time evolution of this expression is more easily computed using the Heisenberg picture, and it only requires one to evolve the ``in" modes in the right-hand side. The entanglement between the two oscillators is manifest in the time evolution of the cross-correlation
\be \langle in| \hat x_{1}(t)\, \hat x_{2}(t)|in\rangle\, , \ee 
which turns out to be equal to zero for early times $t<t_{in}$, but it generically becomes different from zero at late times  if the coupling $k_c(t)$ is different from zero at some intermediate time.

%%%%%%%%%%%%%%%%%%%%%%%%%%%%%%%%%%%%%%%%%%%%%%%%%%%%%%%%%%%%%%%%%%%%%%%%%%%%%%%

\end{document}